
\topskip 24pt
\overfullrule=0pt
\magnification=\magstep1
\hsize=5.75truein
\hoffset=.375truein
\baselineskip=14pt
\rightline {CU-TP-662 and RU-95-2-B}
\vskip20pt
\centerline {\bf NONCOMPACT LATTICE FORMULATION OF GAUGE THEORIES}
\vskip40pt
\centerline {R. Friedberg, T. D. Lee}
\vskip8pt
\centerline {Columbia University, New York, N.Y. 10027}
\vskip16pt
\centerline {Y. Pang}
\vskip8pt
\centerline {Columbia University, New York, N.Y. 10027}
\vskip2pt\centerline {and Brookhaven National Laboratory, Upton, N.Y. 11973}
\vskip16pt\centerline {H. C. Ren}
\vskip8pt
\centerline{The Rockefeller University, New York, N.Y. 10021}
\vskip46pt
\centerline {\bf ABSTRACT}
\vskip20pt We expand the gauge field in terms of a suitably constructed
complete
set of
 Bloch wave functions, each labeled by a band designation $\,n\,$ and a wave
number $\,\vec K\,$ restricted to the Brillouin zone.  A noncompact formulation
of lattice QCD (or QED) can be derived by restricting the expansion only to the
$\,0^{th}$-band ($\,n = 0\,$) functions, which are simple continuum
interpolations of discrete values associated with sites or links on a lattice.
The exact continuum theory can be reached through the inclusion of all $\,n =
0\,$ and $\,n
\ne 0\,$ bands, without requiring the lattice size $\,\ell \to 0\,$.  This
makes
it possible, at a nonzero $\,\ell\,$, for the lattice coupling $\,g_\ell\,$ to
act as the renormalized continuum coupling.  All physical results in the
continuum are, of course, independent of $\,\ell\,$.
\vskip10pt\hrule
\vskip10pt
\centerline{This research was supported in part by the U.S. Department of
Energy.}
\vfill\eject
\centerline{\bf 1.  INTRODUCTION}
\vskip18pt
	In a recent paper$^1$ (hereafter referred to as {\rm I}) we proposed a new
lattice
formulation of the continuum field theory.  Instead of the usual Fourier
series,
the continuum field operators are expanded in terms of a suitably chosen
complete set of orthonormal Bloch functions

$$\{\,f_n(\vec K\,|\,\vec r)\,\} \eqno(1.1)$$
\vskip4pt\noindent
where $\,\vec K\,$ denotes the Bloch wave number restricted to the Brillouin
zone,
and $\,n\,$ labels the different
bands, similar to the one-particle wave functions in a crystal (i.e.,
$\,e^{-i\vec K\cdot\vec r}\,f_n(\vec K\,|\,\vec r\,)\,$ has the periodicity of
the lattice).  The lattice approximation is then derived by either restricting
it
to only one band (say, $\,n = 0\,$), or to a few appropriately defined
low-lying
bands.  Since the inclusion of all bands {\it is} the original continuum
problem,
there is a natural connection between the lattice and the continuum in this
method.  By including the contributions due to more and more bands, one can
systematically arrive at the exact continuum solution from the lattice
approximation.  There is a large degree of freedom in choosing the Bloch
functions (1.1), as the original continuum theory has no crystal structure.
These extra degrees of freedom are analogous to gauge fixing; the final answer
to the continuum problem is independent of the particular choice of Bloch
functions (or the lattice structure).
\vskip4pt
	In paper {\rm I} we gave a few simple examples of such functions and
applied them to spin-0 and spin-${1\over2}$ fields.  We noted that this
approach
provides an effective means of removing the spurious lattice fermion solutions
for the Dirac equation.  The underlying reason is quite simple: After all, in
any crystal in nature all the electrons do move in a lattice and satisfy the
Dirac equation; yet there is not a single physical result that has ever been
entangled with a spurious fermion solution.  Consequently, it should not be
that
difficult to get rid of these unphysical elements.
\vskip4pt
On a discrete lattice, particles hop from point to point, whereas in a real
crystal the lattice structure is embedded in a continuum and electrons move
continuously from lattice cell to lattice cell.  In a discrete system, the
lattice functions are defined only on individual points (or links, as in the
case of the gauge field).  However, in a crystal, each Bloch wave $\,f_n(\vec
K\,|\,\vec r\,)\,$ is a continuous function in $\,\vec r\,$, and herein lies
one of the
essential differences.  In the case of gauge theories, there is an additional
complication, since the usual lattice gauge action$^{2,3}$ is compact, which
makes it
intrinsically different from the noncompact continuum action (except in the
limit of weak coupling or infinitesimal lattice size).  As we shall see, these
differences can be resolved by extending the lattice formulation developed in
I to both quantum
electrodynamics (QED) and quantum chromodynamics (QCD).  The result is a new
{\it noncompact} lattice formulation of gauge theories that can serve as the
first approximation of the continuum theory in a systematic way, {\it without}
requiring the lattice size  $\,\ell \to 0\,$.
\vskip4pt
The freedom in choosing any complete set of Bloch functions allows us to
explore a family of new functions that are convenient to use.   For higher
bands,
 we would like the  Bloch
functions to stay very close to the Fourier series of high wave numbers, so
that
 renormalization calculations can be carried out
analytically.  On the other hand, the zeroth band functions
$\,f_0(\vec K\,|\,\vec r\,)\,$ should be made
of simple continuum interpolations of the discrete values given at lattice
(or link) sites; this way, the restriction to zeroth band would naturally
resemble the usual lattice theory.   Furthermore, in order to retain the
``locality'' character of the original continuum theory in the one-band
approximation, we
 require $\,f_0(\vec K\,|\,\vec r\,)\,$ to satisfy, for any lattice site
$\,j\,$
located at $\,\vec r_j\,$,

$${1\over\sqrt\Omega}\;\sum_{\vec K}\,e^{-i\vec K\,\cdot\,\vec r_j}\,f_0(\vec K
|
\vec r\,)\;=\;0\hskip2em {\rm if} \hskip2em |\,\vec r - \vec r_j\,|\;>\;\nu\ell
\eqno(1.2)$$
\vskip4pt\noindent
with $\,\nu\,$ a finite number $\,O(1)\,$, $\,\ell\,$ the unit lattice size,
$\,\Omega\,$ the total volume $\,>> \ell^3\,$ and $\,\vec K\,$ being summed
over all wave numbers within the Brillouin zone.  This lattice locality
condition
rules out the simple Fourier series for $\,f_0(\vec K\,|\,\vec r\,)\,$, since
for $\,f_0(\vec K\,|\,\vec r\,) = \Omega^{-{1\over 2}}\,e^{i\vec K\cdot\vec
r}\,$
the corresponding sum (1.2) for a cubic lattice with $\,\Omega = L^3\,$ would
be

$${\sin {\pi\over\ell}\,(x - x_j)\;\sin {\pi\over\ell}\,(y - y_j) \;\sin
{\pi\over\ell}\,(z - z_j)\over L^3 \sin {\pi\over L}\,(x - x_j)\;\sin
{\pi\over L}\,(y - y_j) \;\sin  {\pi\over L}\,(z - z_j)} \eqno(1.3)$$
\vskip4pt\noindent
which is non-local.  Instead, condition (1.2) leads to
the concept of ``lump functions'' and ``link functions'', which will be
introduced in Section 2. \vskip4pt
For applications to gauge theories, there is also the problem of compatibility
between the continuum gauge-fixing condition and the band decomposition.  We
discuss QED first in the time-axial gauge in Section 3.1 and then in the
Coulomb
gauge in Section 3.2.  By using suitable combinations of these lump and link
functions we show that in both cases the continuum
gauge requirement,

$$A_4(\vec r, t)\;=\;0 \eqno(1.4)$$\noindent
for the time-axial gauge or
$$\vec\nabla \cdot \vec A\,(\vec r, t)\;=\;0 \eqno(1.5)$$
\vskip4pt\noindent
for the Coulomb gauge, can hold {\it within\ each\ $\,n^{th}\,$ band} at all
$\,\vec r\,$ and $\,t\,$, where
$\,A_\mu(\vec r, t)\,$ is the electromagnetic field operator.  Thus, as we
shall see, in the lattice
approximation (i.e., with $\,A_\mu(\vec r, t)\,$ restricted only to the $\,n
= 0\,$ band), we have the interesting situation of having the continuum
gauge-fixing
equation as well as its discrete realization.  For example, in the Coulomb
gauge and with the
restriction to $\,0^{th}\,$ band, $\,\vec A(\vec r, t)\,$ is a simple continuum
interpolation of its
discrete values $\,a_{ij}\,$, each of which is associated with a link
$\,\ell_{ij}\,$, where
$\,\ell_{ij}\,$ connects the $\,i^{th}\,$ lattice site to its nearest
neighboring site $\,j\,$.
The discrete version of $\,\vec\nabla \cdot \vec A = 0\,$ refers to Kirchoff's
law

$$\sum_j a_{ij}\;=\;0 \eqno(1.6)$$
\vskip4pt\noindent
which is valid at every $\,i\,$, with $\,j\,$ summed over all its nearest
neighbors.  As will
be shown by (3.37)-(3.39), (4.10) and (4.14)-(4.15) below, in our formulation
the discrete
Kirchoff law implies the validity of the continuum $\,\vec\nabla \cdot \vec A =
0\,$ within the
one-band approximation, and {\it vice versa}.  Such equivalent formulas can be
further generalized.
Take $\,\vec\nabla \cdot \vec A = 0\,$: in the continuum, this implies that
$\,\vec A\,$ equals the
curl of another vector function, say $\,\vec A = \vec\nabla \times \vec I\,$,
and therefore
$\,\vec A\,$ becomes invariant under the transformation $\,\vec I \to \vec I +
\vec\nabla \chi\,$.
We are pleasantly surprised to find that, within the same one-band
approximation in the Coulomb
gauge, each of these continuum equations has its exact equivalent discrete form
(see (4.24)-(4.29)
below), giving rise to a family of relations between the continuum and the
discrete.  Of course, a change
in the gauge has to induce a change in the corresponding Bloch functions, as
will be illustrated in
Section 3 for QED.
\vskip4pt
Similar considerations are also applicable to a generalized
Coulomb-like gauge where, instead of $\,\vec\nabla \cdot \vec A = 0\,$, we have
at all continuum $\,\vec r\,$

$$\int (\vec r\,|\,\vec \Gamma\,|\,\vec r\,') \cdot \vec A(\vec
r\,')\,d^3r'\,=\,0 \eqno(1.7)$$
\vskip4pt\noindent
with $\,\vec \Gamma\,$ a linear operator independent of $\,A_\mu(\vec r, t)\,$.
While QED is a relatively trivial model, the analysis given in Section 3
provides the necessary tools for the noncompact lattice formulation of
non-Abelian theories.
\vskip4pt
The extension to QCD in the Coulomb or Coulomb-like gauge is
 straightforward since the gauge condition, (1.5) or (1.7), is linear and
there are no additional supplementary conditions; hence, the group index is
external to the band decomposition. The details are given in Section 4.
Because
the noncompact lattice QCD formulation is the
$\,0^{th}$-band approximation of the continuum QCD, it is possible to regard
the
lattice QCD coupling constant
$\,g_\ell\,$ as the exact {\it renormalized} coupling constant of continuum
QCD,
without requiring the lattice size $\,\ell \to 0\,$ (i.e., at a
finite $\,\ell^{-1}\,$), as will be studied in Section 5.  We discuss how this
approach is related to the conventional renormalization procedure in continuum
QCD.  By taking advantage of the asymptotic freedom of the theory, we show
that,
in terms of the perturbation series in the Coulomb gauge, the bare coupling
$\,g_0\,$ and the conventional renormalized coupling $\,g_R\,$ of continuum QCD
can be readily expressed in terms of the lattice coupling $\,g_\ell\,$ in a
systematic way.
\vskip4pt
The extension to QCD in the time-axial gauge is more complicated: While the
gauge condition $\,A_4(\vec r, t) = 0\,$ being a linear equation offers no
problem in the band decomposition, the generalization of the supplementary
``Gauss Law''condition on the state vector ((3.19) below) becomes nonlinear in
QCD, which means that in solving this constraint, there would be additional
coupling between different bands which must be taken into account. The details
will not be analysed in this paper.
\vfill\eject
\centerline{\bf 2. MATHEMATICAL PRELIMINARIES}
\vskip8pt
In this section, we consider only the one-space-dimensional case.
\vskip16pt\noindent
{\it 2.1.  Lump Functions}
\vskip4pt
These functions are defined by

$$L_m(x)\;\equiv\;\int_{-\infty}^\infty\;{dk\over 2\pi}\;{(2 \sin {k\over
2})^m\over k^m}\;\times\;\cases{e^{ikx/\ell}&for $\,m\,$ even\cr
e^{ik(x - {\ell\over 2})/\ell}&for $\,m\,$ odd\cr}\eqno(2.1)$$
\vskip4pt\noindent
where $\,m\,$ is a positive integer $\,> 0\,$ and $\,x\,$ is the continuous
coordinate variable.  (When $\,m = 0\,$, $\,L_m(x)\,$ becomes the
$\,\delta$-function, which does not serve our particular purpose.)  As we shall
see, these functions are nonzero (shaped like a lump) only within $\,m\,$
lattice
cells, each of length $\,\ell\,$. When $\,m\,$ is even and $\,|\,x\,| >
{m\over 2}\;\ell\,$, we may deform the integration path in the complex
$\,k$-plane into a semi-circle of infinite radius encompassing the upper, or
lower, half-plane depending on whether $\,x\,$ is greater than $\,{m\over
2}\;\ell\,$, or less than $\,- {m\over 2}\;\ell\,$; this leads to zero for
(2.1).  Likewise, when $\,m\,$ is
odd the integral is also zero for $\,| x -
{\ell\over 2}\,| > {m\over 2}\;\ell\,$.  In the next section, by using these
lump
functions we can construct Bloch wave functions that satisfy the lattice
locality condition (1.2). From the definition, we see that for $\,m\,$ even

$$\ell\;{d\,L_m(x)\over dx}\;=\;- L_{m-1}(x) + L_{m-1}(x + \ell)
\eqno(2.2)$$\noindent
and for $\,m\,$ odd
$$\ell\;{d\,L_m(x)\over dx}\;=\;L_{m-1}(x) - L_{m-1}(x - \ell)\,.
\eqno(2.3)$$
\vskip4pt
The lump functions of lower order, $\,m = 1, 2\,$ and 3, are particularly
useful in later sections.  Because, in the application to gauge theories, there
will be several different kinds of indices, it is convenient to give these
functions special names.  We introduce
$$C(x)\;\equiv\;L_1(x)\;=\;\cases{1&for $0 < x < \ell$\cr
0&otherwise$\,$,\cr}$$
$$\Delta(x)\;\equiv\;L_2(x)\;=\;\cases{1 - |{x\over \ell} |&for $|x| <
\ell$\cr
0&otherwise\cr}\eqno(2.4)$$
\noindent
and
$$S(x) \equiv L_3(x) = \cases{{1\over 8}\,[(3 - 2\,|\,{x\over \ell}  - {1\over
2}\,|)^2 - 3(1 - 2\,|\,{x\over \ell}  - {1\over 2}\,|)^2]&for $|\,x -
{\ell\over
2}\,| < {\ell\over 2}$\cr
{1\over 8}\,(3 - 2\,|\,{x\over \ell}  - {1\over
2}\,|)^2&for
${\ell\over 2} < |\,x - {\ell\over 2}\,| < {3\ell\over 2}$\cr
0&otherwise$\,$.\cr}$$
\vskip4pt\noindent
Here, the notation $\,C(x)\,$ is chosen for its being a {\it constant}  within
$\,0 < x < \ell\,$, $\,\Delta(x)\,$ for being like a {\it triangle}  within
$\,- \ell < x < \ell\,$ and $\,{\cal S}(x)\,$ for its nonzero part being a
{\it second} order polynomial in $\,x\,$ (within unit intervals of $\,\ell\,$).
\vskip4pt
In general, the nonzero parts of all $\,L_m(x)\,$ are positive polynomials of
$\,(m - 1)^{th}\,$ order within each $\,\ell$-interval.  For $\,m > 1\,$,
$\,L_m(x)\,$ is continuous at all
$\,x\,$.  It follows from (2.2)-(2.3) that for $\,m > 2\,$ all
$\,d\,L_m(x)/dx\,$ are continuous functions; for $\,m > 3\,$, all
$\,d^2\,L_m(x)/dx^2\,$ are continuous, etc.  When $\,m\,$ is even $\,> 0\,$,
$\,L_m(x) = L_m(- x)\,$; it peaks at $\,x = 0\,$, decreases monotonically to
zero at $\,|\,x\,| = {m\over 2}\,\ell\,$ and remains zero for $\,|\,x\,| >
{m\over 2}\,\ell\,$.  When $\,m\,$ is odd $\,> 1\,$, $\,L_m(x - {\ell\over 2})
= L_m(- x + {\ell\over 2})\,$; its maximum is at $\,x = {\ell\over 2}\,$,
decreases monotonically to zero at $\,|\,x - {\ell\over 2}\,| = {m\over
2}\,\ell\,$ and remains zero for $\,|\,x - {\ell\over 2}\,| >
{m\over 2}\,\ell\,$.
\vskip4pt
Another useful property is, within the range $\,0 < x < \ell\,$,

$$\sum_{j=-(m - 2)/2}^{m/2} L_m(x - j\ell)\;=\;{\rm constant}\hskip2em{\rm
for}\hskip2em m\;\;{\rm even}\eqno(2.5)$$\noindent
and
$$\sum_{j=-(m - 1)/2}^{(m - 1)/2} L_m(x - j\ell)\;=\;{\rm
constant}\hskip2em{\rm
for}\hskip2em m\;\;{\rm odd}\,,\eqno(2.6)$$\vskip4pt\noindent
which can be readily proven by noting that their derivatives are both zero, on
account of (2.2)-(2.3).
\vskip26pt\noindent
{\it 2.2.  Bloch Functions}
\vskip4pt
Given an $\,m\,$, the corresponding (unnormalized) $\,0^{th}\,$ band Bloch
function is given by

$${\cal F}\,(K\,|\,x)\;=\;\sum_{j = - \infty}^\infty e^{ij\theta}\,L_m(x -
j\ell) \eqno(2.7)$$\noindent
with
$$\theta\;=\;K\ell\,,\eqno(2.8)$$\noindent
which satisfies
$$|\,\theta\,|\;\le\;\pi\,.\eqno(2.9)$$\noindent
Because
$${\cal F}(K\,|\,x + \ell)\;=\;e^{i\theta}\,{\cal
F}(K\,|\,x)\,,\eqno(2.10)$$ \vskip4pt\noindent
the passage from an infinite lattice to a finite one of size $\,N\ell\,$ and
with the periodic boundary condition can be readily derived by requiring

$$KN\ell\;=\;2\pi\; \times\;\;{\rm integer}\,.\eqno(2.11)$$
\vskip4pt\noindent
This convention will be used subsequently.  It follows from (2.5) and (2.6)
that, at $\,K = 0\,$, $\,{\cal F}(K\,|\,x)\,$ is a constant at all $\,x\,$.
Furthermore, the sum $\,N^{-1}\,\sum_K\,e^{-iKj\ell}$
\vskip1pt\noindent
$\times {\cal F}(K\,|\,x) = L_m(x
- j\ell)\,$ satisfies the lattice locality condition (1.2).
 \vskip4pt To construct
the higher-band Bloch functions, we follow one of the simple procedures given
in
{\rm I}.  Let

$$\psi_p(x)\;=\;(N\ell)^{-{1\over 2}}\;e^{ipx} \eqno(2.12)$$
\vskip4pt\noindent
be the usual plane-wave functions with $\,pN\ell = 2\pi\;\times\,$ any integer.
Resolve

$$p\;=\;K + {2\pi n\over \ell} \eqno(2.13)$$
\vskip4pt\noindent
where $\,n = \cdot\cdot\,, - 2, - 1, 0, 1, 2, \cdot\cdot\,$.  At a $\,K\,$
given by (2.8)-(2.9), write (2.12) as

$$\psi_n(x)\;=\;(N\ell)^{-{1\over 2}}\,e^{iKx}\,e^{i2\pi
nx/\ell}\,.\eqno(2.14)$$
\vskip4pt\noindent
The complete set of orthonormal Bloch functions, given by
$\,\{f_n(K\,|\,x)\}\,$, can be obtained by requiring, for $\,n = 0\,$

$$f_0(K\,|\,x)\;=\;c_m\; {\cal
F}(K\,|\,x)\,,\eqno(2.15)$$\vskip 4pt\noindent
with $\,c_m\,$ the normalization constant, and for $\,n  \ne 0\,$

$$f_n(K\,|\,x)\;=\;\psi_n(x) - {f_0(K\,|\,x) + \psi_0(x)\over 1 +
a_0}\;a_n\eqno(2.16)$$\noindent
where (for all $\,n\,$, including 0)

$$a_n\;\equiv\;\int_0^{N\ell} f_0(K\,|\,x)^*\;\psi_n(x)\,dx\eqno(2.17)$$
\noindent
and $\,^*\,$ the complex conjugation.  It follows then
$$\int_0^{N\ell} f_n(K\,|\,x)^*
f_{n'}(K'\,|\,x)\,dx\;=\;\delta_{KK'}\,\delta_{nn'} \eqno(2.18)$$
\vskip4pt\noindent
where $\,\delta_{KK'}\,$ and $\,\delta_{nn'}\,$ are Kronecker symbols.  By
using (2.1) and (2.7) we see that

$$|\,a_n\,|^2\;=\;|\,c_m\,|^2\;{2^m(1 - \cos \theta)^m\over
(\theta + 2\pi n)^{2m}}\,.\eqno(2.19)$$
  \vskip6pt\noindent
Since $\,\sum_{n=-\infty}^\infty |\,a_n\,|^2 = 1\,$, we may write (2.19) as

$$|\,a_n\,|^2\;=\;(\theta + 2\pi\,n)^{-2m}\;\left[ \sum_{n'=-\infty}^\infty
(\theta + 2\pi\,n')^{-2m}\right]^{-1}\,.$$
\noindent
Thus, when $\,\theta \to 0\,$,
$$|\,a_0\,|^2\;\to\;1 - 2\theta^{2m} \sum_{n=1}^\infty (2\pi\,n)^{-2m}\;=\;1 -
{\theta^{2m}\,B_m\over (2m)!}$$
\noindent
and, for $\,n \ne 0\,$, $|\,a_n\,|^2\;\to\;\theta^{2m}(2\pi\,n)^{-2m}\,$ where
$\,B_m\,$ denotes the Bernoulli numbers with $\,B_1 = {1\over 6}\,$, $\,B_2 =
{1\over 30}\,$, $\,B_3 = {1\over 42}\,$,  etc.
\vskip4pt
We note that, given $\,{\cal F}(K\,|\,x)\,$, there are many equally simple ways
to construct the complete set of Bloch functions, as discussed in {\rm I}.
\vskip20pt\noindent
{\bf Remarks:}
\vskip4pt\noindent
1.  When $\,m = 0\,$, $\,L_m(x)\,$ becomes $\,\delta(x/\ell)\,$, as mentioned
before; therefore, the
expansion (2.7) reduces to the usual expression for a lattice of discrete
points.
\vskip6pt\noindent
2.  In (2.19), we see that for nonzero $\,n\,$ and $\,m\,$ the factor
$\,(\theta
+ 2\pi\,n)^{-2m}\,$ is always less than $\,[\,(2\,|\,n\,| -
1)\,\pi\,]^{-2m}\,$.
In $\,D$-dimension, (2.19) will be changed into a product of $\,D\,$ such
factors, one for each space dimension.  Thus, the higher band Bloch functions
are
very close to the corresponding Fourier series of high wave numbers (see, e.g.,
Eq. (3.67) below).
 \vskip26pt\noindent
{\it 2.3.  Scalar Field (analysed in terms of $\,L_m(x)\,$ for $m = 2$)}
\vskip4pt
As an example, consider the Lagrangian for the free scalar field $\,\phi(x,
t)\,$, in units $\,\hbar = c = 1\,$:

$${\cal L}\;=\;\int_0^{N\ell}\,\left[\,\dot\phi^\dagger \dot\phi - {\partial
\phi^\dagger\over \partial x}\;{\partial\phi\over \partial x} -
\mu^2\,\phi^\dagger \phi\,\right]\,dx \eqno(2.20)$$
\vskip4pt\noindent
where $\,\mu\,$ is the mass, $\,^\dagger\,$ denotes the Hermitian conjugation
and the dot indicates the time derivative.  The restriction to the zeroth band
$\,n = 0\,$ by using the lump function of order $\,m = 2\,$ is identical to the
expansion

$$\phi(x, t)\;=\;\sum_j\,\phi_j(t)\,\Delta(x - j\ell)\eqno(2.21)$$
\vskip4pt\noindent
where $\,j\,$ goes over the $\,N\,$ lattice sites and $\,\Delta(x)\,$ is the
triangular function given by (2.4).  At the $\,j^{th}$ site $\,x = j\ell\,$,
$\,\phi =
\phi_j\,$ and the expansion (2.21) gives a simple linear interpolation,
converting the set of discrete values $\,\{\,\phi_j\,\}\,$ to a continuous
function.  (The use of piecewise flat but continuous functions such as
$\,\Delta(x - j\ell)\,$ for lattice formulation is the basis of discrete
mechanics.$^4$)
\vskip4pt
We may substitute (2.21) into $\,{\cal L}\,$ and derive the corresponding
lattice equation by setting to zero the variational derivative of the action
$\,\int\,{\cal L}\,dt\,$ with respect to $\,\phi_j(t)\,$.  The resulting
solution is

$$\phi_j(t)\;=\;\sum_K\,\left[ {3\over N\ell(2 + \cos \theta)} \right]^{1\over
2}\;e^{ij\theta}\,Q_K^0(t)\eqno(2.22)$$\noindent
where $\,Q_K^0(t)\,$ is the normal mode satisfying

$$\ddot Q_K^0\;=\;- (w^2(\theta) + \mu^2)\,Q_K^0 \eqno(2.23)$$\noindent
with
$$w^2(\theta)\;=\;{6(1 - \cos \theta)\over \ell^2(2 + \cos \theta)}
\eqno(2.24)$$
\vskip4pt\noindent
and $\,\theta\,$ given by (2.8)-(2.9).  Correspondingly,  the lattice
approximation (2.21) becomes
$$\phi(x, t)\;=\;\sum_K\,Q_K^0(t)\,f_0(K\,|\,x) \eqno(2.25)$$\noindent
where the normalized $\,f_0(K\,|\,x)\,$ is

$$f_0(K\,|\,x)\;=\;\left[ {3\over N\ell(2 + \cos \theta)} \right]^{1\over 2}\;
\sum_j\,e^{ij\theta}\,\Delta(x - j\ell)\,. \eqno(2.26)$$
\vskip4pt\noindent
Clearly, the full expansion in terms of all bands

$$\phi(x, t)\;=\;\sum_n \sum_K\,Q_K^n(t)\,f_n(K\,|\,x) \eqno(2.27)$$
\vskip4pt\noindent
yields the exact continuum result.  The details are given in I.
\vskip20pt\noindent
{\bf Remarks.}
\vskip6pt\noindent
{\bf 1.}  For  $\,K\,$ within the Brillouin zone (i.e., $\,|\,\theta\,| \le
\pi\,$), the above frequency $\,\omega(K)\;=\;\sqrt {w^2(\theta) + \mu^2}\,$ is
always {\it higher}\  than the exact continuum result $\,E(K) = \sqrt {K^2 +
\mu^2}\,$.  This is expected since the expansion (2.21) may be regarded as a
constraint imposed on the continuum problem; the corresponding optimal spectrum
can only serve as an upper bound.
\vskip6pt\noindent
{\bf 2.}  Instead of $\,\Delta(x - j\ell)\,$, in the expansion (2.21) for
$\,\phi(x, t)\,$ we may use another lump function $\,L_m(x - j\ell)\,$, but
with
$\,m > 2\,$, as a constraint on the continuum field.  (See, e.g., (2.28)
below.)
However, this does not work if
$\,m\,$ is
$\,< 2\,$:  When $\,m = 1\,$,  $\,L_1(x) = {\cal C}(x)\,$ and its derivative is
$\,\delta(x) - \delta(x - \ell)\,$, in accordance with (2.4); therefore, the
application of the corresponding $\,0^{th}$-band Bloch function to the
continuum scalar field Lagrangian $\,{\cal L}\,$ would lead to $\,\infty\,$.
(However, the $\,m = 1\,$ lump function can be used as a link function for the
gauge field, as will be discussed in Section 3.)  When $\,m = 0\,$, $\,L_0(x) =
\delta(x/\ell)\,$; the substitution of the corresponding expansion of
$\,\phi(x)\,$ into the continuum Lagrangian $\,{\cal L}\,$ would clearly also
give $\,\infty\,$.
\vskip6pt\noindent
{\bf 3.}  In this context, it may be useful to examine the usual discrete
lattice
of $\,N\,$ separate points.  The corresponding Lagrangian $\,{\cal L}_d\,$ for
a
discrete field $\,\psi_j\,$ would be, instead of (2.20),

$${\cal L}_d\;=\;\sum_{j=1}^N \;[\,\dot\psi_j^\dagger\,\dot\psi_j -
(\psi_{j+1}^\dagger - \psi_j^\dagger)\,(\psi_{j+1} - \psi_j)\,\ell^{-2}\,-
\mu^2\,\psi_j^\dagger\,\psi_j\,]\,.$$
\vskip4pt\noindent
In terms of its normal modes we have, in place of (2.22)-(2.24),

$$\psi_j(t)\;=\;\sum_K N^{-{1\over 2}}\,e^{ij\theta}\,q_K(t)\,,$$
$$\ddot q_K\;=\;- \omega_d(K)^2\,q_K$$\noindent
with
$$\omega_d(K)\;=\;\left[\,2(1 - \cos \theta)\,\ell^{-2} +
\mu^2\,\right]^{1\over
2}\,.$$
\vskip4pt\noindent
Note that the frequency $\,\omega_d(K)\,$ is {\it lower} than the continuum
result $\,E(K)\,$, which underscores the difference (for $\,\ell \ne 0\,$)
between a discrete lattice Lagrangian $\,{\cal L}_d\,$ and its continuum analog
$\,{\cal L}\,$.  When $\,\ell \to 0\,$, $\,\omega(K)\,$ and $\,\omega_d(K)\,$
both approach $\,E(K)\,$ with the deviations, $\,\omega(K)^2 - E(K)^2\,$ and
 $\,\omega_d(K)^2 - E(K)^2\,$, $\,O(\theta^4)\,$.
\vskip4pt
However, when $\,\ell \ne 0\,$, the usual discrete lattice Lagrangian $\,{\cal
L}_d\,$ is more an abstract model of the continuum theory, rather than a
systematic approximation of it.  This is different from the direction that we
follow, and the difference becomes especially prominent when we come to QCD
(Sections 4 and 5).  In Appendix A, we give a detailed comparison between
these two approaches, which may help to explain the underlying reasons for
constructing these special lump functions and for the particular way that we
formulate our lattice approximation.
\vskip26pt\noindent
{\it 2.4.  Link Functions}
\vskip4pt
For $\,m\,$ odd,  $\,L_m(x - j\ell)\,$ peaks at $\,x = (j + {1\over 2})\ell\,$,
which is the mid-point of the link connecting the $\,j^{th}\,$ and $\,(j +
1)^{th}\,$ sites, located at $\,x = j\ell\,$ and $\,(j + 1)\ell\,$.  It
is convenient to associate these functions with links, as will be discussed in
later sections; therefore, the lump functions of odd order will also be
called ``link functions''.  Here, in one space-dimension, this difference is
not
an essential one since the mid-point of any link is also the lattice site in
the
dual structure.  We shall illustrate the use of these link-functions by
considering the special case $\,m = 3\,$. \vskip4pt
For the scalar field $\,\phi(x, t)\,$, instead of (2.21), write

$$\phi(x, t)\;=\;\sum\,\chi_j(t)\,S(x - j\ell) \eqno(2.28)$$
\vskip4pt\noindent
where $\,S(x) = L_3(x)\,$ is given by (2.4).  Substituting this form into the
Lagrangian (2.20) and varying $\,\chi_j(t)\,$, we find, similar to
(2.22)-(2.26),

$$\chi_j(t)\;=\;\sum_K\;\left[\,{30\over N\ell(16 + 13 \cos \theta + \cos^2
\theta)}\,\right]^{1\over 2}\,e^{i(j + {1\over
2})\theta}\;Q_K^0(t)\eqno(2.29)$$
\vskip4pt\noindent
where $\,Q_K^0(t)\,$ is the new normal mode satisfying

$$\ddot Q_K^0\;=\;- (v^2(\theta) + \mu^2)\,Q_K^0\eqno(2.30)$$\noindent
with
$$v^2(\theta)\;=\;{20(1 - \cos \theta) (2 + \cos \theta)\over \ell^2(16 + 13
\cos \theta + \cos^2 \theta)} \eqno(2.31)$$
\vskip4pt\noindent
and $\,\theta\,$ given by (2.8)-(2.9), as before.  Correspondingly, (2.28)
becomes

$$\phi(x, t)\;=\;\sum_K\,Q_K^0(t)\,g_0(K\,|\,x)\eqno(2.32)$$\noindent
where
$$g_0(K\,|\,x)\;=\;\left[\,{30\over N\ell(16 + 13 \cos \theta + \cos^2
\theta)}\,\right]^{1\over 2}\;\sum_j\,e^{i(j + {1\over 2})\theta}\,S(x -
j\ell)\,. \eqno(2.33)$$
\vskip4pt\noindent
For notational clarity, we denote the normalized zeroth-band Bloch function for
the link functions by $\,g_0(K\,|\,x)\,$, instead of $\,f_0(K\,|\,x)\,$.
\vskip4pt
{}From (2.4), we see that $\,S({\ell\over 2}) = {3\over 4}\,$ and $\,S(-
{\ell\over 2}) = S({3\ell\over 2}) = {1\over 8}\,$.  Thus, the value
$\,\overline
\phi_j(t) \equiv \phi(x,t) \,$ at the mid-point of each link (i.e., at $\,x =
(j
+ {1\over 2})\,\ell\,$) is related to $\,\chi_j(t)\,$ introduced in (2.28) by

$$\overline\phi_j\;=\;{1\over 8}\,(6\,\chi_j + \chi_{j-1} +
\chi_{j+1})\,.\eqno(2.34)$$
\noindent
The inverse relation is
$$\chi_j\;=\;\sum_\alpha\;\sum_{j'=1}^N\,e^{i(j-j')\alpha}\;\;4\left[\,N(3 +
\cos
\alpha)\,\right]^{-1}\,\overline\phi_{j'}
\eqno(2.35)$$\noindent
where
$$\alpha\;=\;2\pi\,n/N$$\noindent
and its sum extends over $\,n = 1, 2,\,\cdot\cdot\,,N\,$.  Both $\,\chi_j\,$
and $\,\overline\phi_j\,$ satisfy the periodic boundary condition $\,\chi_{j+N}
= \chi_j\,$ and $\,\overline\phi_{j+N} = \overline\phi_j\,$.
\vskip4pt
In the usual lattice calculation, one tends to use $\,\overline\phi_j\,$ as the
primary variables.  To see how closely correlated the corresponding
$\,\chi_j\,$ are, let us consider the special case $\,N \to \infty\,$ and

$$\overline\phi_j\;=\;\cases{1&for $j = 0$\cr
0&otherwise\.\cr}\eqno(2.36)$$
\noindent
The corresponding $\,\chi_j\,$ is
$$\,\chi_j\;=\;\sqrt 2\;Z^{|\,j\,|}\;\cong (-)^j\,\sqrt 2\;e^{-1.763\,|\,j\,|}
\eqno(2.37)$$
\vskip4pt\noindent
where $\,Z = - 3 + 2\sqrt 2\,$.  Thus, $\,\chi_j\,$ decreases rapidly with
increasing $\,|\,j\,|\,$.
\vskip4pt
By following the steps (2.16)-(2.18), we can start with the $\,0^{th}\,$ band
link-function $\,g_0(K\,|\,x)\,$ and construct a complete set of orthonormal
Bloch functions $\,\{\,g_n(K\,|\,x)\,\}\,$, satisfying

$$\int_0^{N\ell}\,g_n(K\,|\,x)^*\,g_{n'}(K'\,|\,x)\,dx\;=\;\delta_{KK'}\,\delta_{nn'}\,.
\eqno(2.38)$$
\vskip4pt\noindent
The expansion of $\,\phi(x, t)\,$ in terms of all bands would, of course,
give the correct continuum solution.
\vskip4pt
For $\,|\,\theta\,| \le \pi\,$, like  $\,w^2(\theta)\,$,
$\,v^2(\theta)\,$ is larger than the exact continuum eigenvalue
$\,K^2 = (\theta/\ell)^2\,$.  This is again expected since the expansion
 (2.28) may also be regarded as a constraint imposed on the continuum problem.
 In addition,
we have

$$\delta(\theta)\;\equiv\;w^2(\theta) - v^2(\theta)\;=\;{2(1 - \cos \theta)^2
(8
+ 7 \cos \theta)\over \ell^2(2 + \cos \theta) (16 + 13 \cos \theta + \cos^2
\theta)}\;>\;0\,, \eqno(2.39)$$\vskip4pt\noindent
a relation that will be of use later on.
\vskip26pt\noindent {\it 2.5.  Generalized Lump Functions}
\vskip4pt
The generalized lump functions are defined by

$$L_{m, n}(x)\;\equiv\;\int_{-\infty}^\infty  {dk\over 2\pi}\;{(2 \sin
{k\over 2})^m (\cos {k\over 2})^n\over k^m}\;\times\;\cases{e^{ikx/\ell}&for
$\,m + n\,$ even\cr
e^{ik(x - {1\over \ell})/\ell}&for $\,m + n\,$ odd\cr}\eqno(2.40)$$
\vskip4pt\noindent
where $\,m\,$ and $\,n\,$ are both positive integers.  When $\,n = 0\,$,
$\,L_{m, n}(x)\,$ reduces to $\,L_m(x)\,$ given by (2.1).  These
generalized functions are nonzero only within $\,m + n\,$ lattice cells.
Similar
to (2.2)-(2.3) and (2.5)-(2.6), we have for $\,m + n\,$ even

$$\ell\;{d\over dx}\;L_{m, n}(x)\;=\;- L_{m - 1, n}(x) + L_{m - 1, n}(x +
\ell)\,,\eqno(2.41)$$
$$\sum_{j = - (m + n - 2)/2}^{(m + n)/2}\;L_{m, n}(x - j\ell)\;=\;{\rm
constant}$$
\noindent
and for $\,m + n\,$ odd

$$\ell\;{d\over dx}\;L_{m, n}(x)\;=\;L_{m - 1, n}(x) - L_{m - 1, n}(x -
\ell)\,,\eqno(2.42)$$
$$\sum_{j = - (m + n - 1)/2}^{(m + n - 1)/2}\;L_{m, n}(x - j\ell)\;=\;{\rm
constant}\,.\eqno(2.43)$$
\vskip4pt
The following example for $\,m + n = 4\,$ illustrates the variety of ways that
we may interpolate from a discrete set of lattice values $\,\{ \phi_j\}\,$ to a
continuous function $\,\phi(x)\,$ in the continuum.  Introduce a lump function
of mixed orders:

$$L(x)\;\equiv\;3\,L_{4, 0}(x) - 2\,L_{3, 1}( x)\,. \eqno(2.44)$$
\noindent
We find,
$$L(x)\;=\;\cases{(1 - {|\,x\,|\over\ell})\,(1 + {|\,x\,|\over\ell} - {3\over
2}\,{|\,x\,|^2\over\ell^2})&for $\,|\,x\,| < \ell$\cr
{1\over 2}\,(1 - {|\,x\,|\over\ell})\,(2 - {|\,x\,|\over\ell})^2&for $\ell <
|\,x\,| < 2\ell$\cr
0&for $|\,x\,| > 2\ell\,$.\cr}\eqno(2.45)$$
\vskip4pt\noindent
Because $\,L(0) = 1\,$ and $\,L(\pm \ell) = 0\,$, the expansion

$$\phi(x)\;=\;\sum_j\,\phi_j\,L(x - j\ell) \eqno(2.46)$$
\vskip4pt\noindent
gives $\,\phi(x_j) = \phi_j\,$ at {\it all} lattice sites $\,x_j = j\ell\,$,
even though each $\,\phi_j\,$ influences the value of $\,\phi(x)\,$ over four
nearby lattice cells (instead of two, as in the expansion (2.21) using
$\,\Delta(x)\,$).  The above lump function $\,L(x)\,$ is not always positive;
it is $\,1\,$ at $\,x = 0\,$, decreases to 0 at $\,x = \pm \ell\,$, reaches a
minimum $\,- 2/27\,$ at $\,x = \pm {4\over 3}\,\ell\,$, then increases to 0 at
$\,x = \pm 2\ell\,$ and remains 0 for $\,|\,x\,| > 2\ell\,$.  The corresponding
normalized $\,0^{th}\,$ band Bloch function is, for $\,0 < x < \ell\,$,
$$\eqalign{f_0(K\,|\,x)\;=&\;\left[ {70\over N\ell(62 + 17 \cos \theta - 10
\cos^2
\theta + \cos^3 \theta}\right]^{1\over 2}\cr
&\times\;\left[\,e^{-i\theta}\,L(x + \ell)
+ L(x) + e^{i\theta}\,L(x - \ell) + e^{2i\theta}\,L(x - 2\ell)\,\right]\cr}
\eqno(2.47)$$
\vskip4pt\noindent
with $\,\theta\,$ given by (2.8)-(2.9).  From (2.47), one can construct a
complete set of orthonormal Bloch functions $\,\{\,f_n(K\,|\,x)\,\}\,$, as
before.
\vfill\eject

\centerline{\bf 3.  QED (Free Field)}
\vskip14pt
The Lagrangian density of a free electromagnetic field $\,V_\mu(\vec r, t)\,$
is

$${\cal L}\;=\;- {1\over 4}\;F_{\mu\nu} F_{\mu\nu} \eqno(3.1)$$
\vskip4pt\noindent
 where $\,F_{\mu\nu} = {\partial V_\nu\over\partial x_\mu} - {\partial
V_\mu\over
\partial x_\nu}\,$, and the repeated subscripts $\,\mu\,$ and $\,\nu\,$ are
summed over
1, 2, 3 and 4 (i.e., $\,x, y, z\,$ and $\,it\,$). We follow the standard
procedure of
deriving the continuum quantum Hamiltonian by first fixing the gauge.  The next
step is to choose an appropriate complete set of orthonormal $\,c\,$ number
Bloch functions, in terms of which the quantized field operators can be
expanded.  The restriction to the $\,0^{th}\,$ band Bloch functions constitutes
the lattice theory in the present approach, as in (2.27) and (2.25).  A
critical point for the gauge theory lies in being able to preserve the
continuum gauge-fixing condition (valid at all continuous space-time
coordinates) within the lattice approximation.  Because of (1.2), the lattice
gauge theory maintains the local character of the original field theory.  The
compatibility between the continuum gauge-fixing and the band decomposition
makes it possible to carry out the transition from lattice to continuum in a
straightforward and systematic way.  As we shall discuss, the result is a new
lattice gauge formulation, which retains the noncompact nature of the continuum
gauge theory, and thereby may serve as its first approximation.
\vskip20pt\noindent
{\it 3.1.  Time-Axial Gauge}
\vskip10pt
In the time-axial gauge,
$$V_4\;=\;i\,V_0\;=\;0\/. \eqno(3.2)$$
\vskip8pt
Embed a cubic lattice of $\,{\cal N} = N^3\,$ sites in the continuum.  Consider
first the unit cubic cell (of volume $\,\ell^3\,$) which has the origin (0, 0,
0) and the lattice-site $\,(1, 1, 1)\,\ell\,$ as two of its eight vertices.
Altogether the cell borders on 12 links.  Assign the values of $\,V_x\,$ on the
four
$\,x\,$ links (all parallel to the $\,x$-axis) to be the constants $\,a, a',
a''\,$ and $\,a'''\,$.  Likewise, on each of the $\,y\,$ and $\,z\,$ links
$\,V_y\,$ and $\,V_z\,$ assume constant values $\,b, \cdot\cdot\,, b'''\,$ and
$\,c, \cdot\cdot\,, c'''\,$, as shown in Figure 1.  Inside the cell,
let $\,V_x\,$ be a multilinear function of $\,y\,$ and $\,z\,$:

$$V_x\;=\;a + (a' - a)\,{y\over\ell} + (a'' - a)\,{z\over\ell} + (a''' - a'' -
a' + a)\;{yz\over\ell^2}\,;\eqno(3.3)$$\noindent
similarly,
$$V_y\;=\;b + (b' - b)\,{z\over\ell} + (b'' - b)\,{x\over\ell} + (b''' - b'' -
b' + b)\,{zx\over\ell^2}\eqno(3.4)$$\noindent
and
$$V_z\;=\;c + (c' - c)\,{x\over\ell} + (c'' - c)\,{y\over\ell} + (c''' - c'' -
c' + c)\;{xy\over\ell^2}\,.\eqno(3.5)$$\vskip8pt
The magnetic field $\,\vec B = \vec\nabla \times \vec V\,$ inside the
cell is given by

$$B_x\;=\;\ell^{-1}\left[\,(1 - {x\over \ell})\,(c'' - c - b' + b) +
{x\over\ell}\,(c''' - c'  - b''' + b'')\,\right]\,,
\eqno(3.6)$$
\vskip4pt\noindent
and similar expressions for $\,B_y\,$ and $\,B_z\,$.  Thus,
on the $\,(y, z)\,$ plaquette at $\,x = 0\,$, $\,B_x\,$ is a constant.
{}From Figure 1, we see that this constant value $\,\ell^{-1}(c'' - b' - c +
b)\,$
is equal to $\,\ell^{-1}\,$  times the
algebraic sum of link-values along the counter-clockwise direction, with the
signs determined by the arrow directions shown in the figure.  Likewise, on the
$\,(y, z)\,$ plaquette at $\,x = \ell\,$, $\,B_x\,$ is again a constant,
$\,\ell^{-1}(c''' - b''' - c' + b'')\,$.  Inside the cell, $\,B_x\,$ depends
linearly on $\,x\,$.  Identical considerations can be applied to $\,B_y\,$ and
$\,B_z\,$.
\vskip4pt
Next, we consider other cubic cells, each of which (say, the $\,j^{th}\,$ cell)
can be arrived at through a space-translation, moving the corner of the above
unit cell from the origin $\,(0, 0, 0)\,$ to the $\,j^{th}\,$ lattice site at
$\,\vec r_j = (j_1, j_2, j_3)\ell\,$.  At the same time, replace the
link-values $\,a, b\,$ and $\,c\,$ by $\,a_j, b_j\,$ and $\,c_j\,$.  The entire
$\,V_\mu$-function in the continuum space can then be obtained by following the
same multilinear interpolation procedure (3.3)-(3.5).  This will lead to the
$\,0^{th}\,$ band approximation in the new lattice gauge formulation.  In this
way, the $\,a_j, b_j\,$ and $\,c_j\,$ are defined once and for all on each
link, so that in passing from one cell to another, there is no discontinuity
of either the tangential components of $\,\vec V\,$ or the {\it the\  normal\
 components\  of \  $\,\vec B\,$}.  (See (3.6) above.)
\vskip4pt
As we shall see, by using the lump and link functions introduced in the
previous
 section, we can cast the result in terms of Bloch functions consisting
of products of (2.7).  We write

$$V_a\;=\;\sum_{\vec K}\;q_a(\vec K)\,g(K_a\,|\,x_a)\,\prod_{b\ne
a}\,f(K_b\,|\,x_b) \eqno(3.7)$$
\vskip4pt\noindent
where the subscript $\,a\,$ or $\,b\,$ denotes the space-components $\,1, 2\,$
and $\,3\,$ (or $\,x, y\,$ and $\,z\,$), $\,\vec K\,$ denotes the Bloch wave
number, which is to be summed over the Brillouin zone with its components
$\,K_a\,$ satisfying (2.8), (2.9) and (2.11); i.e.

$$K_a\,N\ell\;=\;2\pi \times {\rm integer} \eqno(3.8)$$\noindent
and the components of $\,\vec\theta \equiv \vec K\ell\,$ bounded by
$$- \pi\;\le\;\theta_a\;\le\;\pi\,. \eqno(3.9)$$\noindent
The functions $\,g(K_a\,|\,x_a)\,$ and $\,f(K_b\,|\,x_b)\,$ are

$$g(K_a\,|\,x_a)\;=\;\left({1\over N\ell}\right)^{1\over 2}\,e^{{i\over
2}\,\theta_a}\,\sum_{n = - \infty}^\infty e^{in\theta_a}\;C(x_a - n\ell)
\eqno(3.10)$$\noindent
and
$$f(K_b\,|\,x_b)\;=\;\left({3\over N\ell(2 + \cos \theta_b)}\right)^{1\over
2}\,
\sum_{n = - \infty}^\infty e^{in\theta_b}\;\Delta(x_b -
n\ell)\eqno(3.11)$$
\vskip4pt\noindent
where $\,C(x)\,$ and $\,\Delta(x)\,$ are the lump functions $\,L_1(x)\,$
and $\,L_2(x)\,$ introduced in the previous section. By referring to (2.4), we
see that (3.7) reproduces (3.3)-(3.5).
\vskip4pt
 Consider again the
$\,j^{th}\,$ cubic cell.  By using (2.4), (3.7) and (3.10)-(3.11) we can write
down immediately the values of $\,V_x, V_y\,$ and $\,V_z\,$ on the three links
connected to the lattice site at $\,\vec r_j\,$.  Thus,

$$a_j\;=\;\sum_{\vec K}\;{1\over \sqrt\Omega}\;e^{i\vec K\cdot\vec
r_j}\,e^{i{1\over 2}\,\theta_1}\;\left({3\over 2 + \cos
\theta_2}\right)^{1\over 2}\,\left({3\over 2 + \cos
\theta_3}\right)^{1\over 2}\,q_1(\vec K)\,,$$
$$b_j\;=\;\sum_{\vec K}\;{1\over \sqrt\Omega}\;e^{i\vec K\cdot\vec
r_j}\,e^{i{1\over 2}\,\theta_2}\;\left({3\over 2 + \cos
\theta_3}\right)^{1\over 2}\,\left({3\over 2 + \cos
\theta_1}\right)^{1\over 2}\,q_2(\vec K)\,,\eqno(3.12)$$
$$c_j\;=\;\sum_{\vec K}\;{1\over \sqrt\Omega}\;e^{i\vec K\cdot\vec
r_j}\,e^{i{1\over 2}\,\theta_3}\;\left({3\over 2 + \cos
\theta_1}\right)^{1\over 2}\,\left({3\over 2 + \cos
\theta_2}\right)^{1\over 2}\,q_3(\vec K)$$
\noindent
where $\,\Omega = (N\ell)^3\,$ is the total volume.  (In this connection, it
may be helpful to compare the conventional discretized Abelian lattice gauge
theory to the present formulation based on using Boch wave functions, of
which the zeroth-band ones are simple continuum interpolations of discrete
values
$\,a_j\,, b_j\,$ and
$\,c_j\,$.  A discussion is given in Appendix A.)
\vskip4pt
Substituting (3.3) and (3.7) into (3.1), we find the Lagrangian (in the
$\,0^{th}\,$ band approximation) to be

$$\int\,{\cal L}\,d^3r\;=\;\sum_{\vec K}\;\left[\,\dot q^\dagger(\vec K)\,\dot
q(\vec K) - q^\dagger(\vec K)\,(w^2 - w\,\tilde w)\,q(\vec K)\,\right]
\eqno(3.13)$$
\vskip4pt\noindent
where $\,q(\vec K)\,$ and $\,w = w(\vec \theta)\,$ are $\,3 \times 1\,$ column
matrices whose components are $\,q_a(\vec K)\,$ and
$$w_a(\vec\theta)\;=\;{2\over \ell}\;\sin {\theta_a\over 2} \left({3\over 2 +
\cos \theta_a}\right)^{1\over 2}\eqno(3.14)$$
\vskip4pt\noindent
with $\,\vec\theta = \vec K\ell\,$ as before.  Sometimes these column matrices
will also be represented by vectors $\,\vec q(\vec K)\,$ and $\,\vec
w(\vec\theta)\,$.  In (3.13), ~ denotes the transpose and

$$w^2\;=\;\sum_{a = 1}^3\,w_a(\vec\theta)^2\;=\;\sum_{a = 1}^3\;{6(1 - \cos
\theta_a)\over \ell^2(2 + \cos \theta_a)}\,,\eqno(3.15)$$
\vskip4pt\noindent
which is the three-dimensional generalization of (2.24).  Throughout the
paper,  $\,\sum_{\vec K}'\,$ denotes the sum over half of the $\,\vec K$-space
within the Brillouin zone.  The corresponding Hamiltonian is

$$H\;=\;\sum_{\vec K}\,\left[\,p^\dagger(\vec K)\,p(\vec K) + q^\dagger(\vec
K)\,(w^2 - w\,\tilde w)\,q(\vec K)\,\right] \eqno(3.16)$$
\vskip4pt\noindent
with $\,p(\vec K) = \dot q(\vec K)\,$ satisfying the equal-time commutation
relation

$$\left[\,p_a(\vec K), q_{a'}(\vec K\,')\,\right]\;=\;\delta_{\vec K \vec
K'}\,\delta_{aa'}\,.$$
\vskip8pt
Both the Lagrangian and the Hamiltonian are invariant under the
time-independent gauge transformation
$$a_j\;\to\;a_j + \phi_{j'} - \phi_{j}\,,$$
$$b_j\;\to\;b_j + \phi_{j''} - \phi_{j}\,, \eqno(3.17)$$
$$c_j\;\to\;c_j + \phi_{j'''} - \phi_{j}\,,$$
\noindent
where the subscripts refer to the sites
$$j\;=\;(j_1, j_2, j_3)\,, \hskip4em j'\;=\;(j_1+1, j_2, j_3)\,,$$
$$j''\;=\;(j_1, j_2+1, j_3)\;\;{\rm and}\;\;j'''\;=\;(j_1, j_2, j_3+1)\,.
\eqno(3.18)$$
\vskip4pt\noindent
This property can be readily established by noting that $\,\dot{\vec V}\,$ is
unchanged since the transformation is independent of time and $\,\vec B =
\vec\nabla \times \vec V\,$ is also not affected, on account of (3.6).  As in
the standard treatment in the time-axial gauge, the above symmetry necessitates
the following constraint on the state vector $\,| >\,$:

$${\cal J}_j\;| >\;=\;0 \eqno(3.19)$$
\vskip4pt\noindent
where $\,{\cal J}_j\,$ is the generator associated with the transformation
(3.17).  There are altogether $\,{\cal N} - 1\,$ such constraints. (Note that
the special case $\,\phi_1 = \phi_2 = \,\cdot\cdot\cdot\, = \phi_{\cal N}\,$
produces no change in $\,a_j,\;b_j\,$ and $\,c_j\,$; therefore (3.17) for all
$\,j\,$ only imposes $\,{\cal N} - 1\,$ constraints.)
\vskip4pt
  In the
Hilbert space representation in which the basis vectors are chosen to be the
eigenstates $\,|\,q >\,$ of the operators
$\,q_a(\vec K)\,$, the state vector takes the form $\,< q\,|\,>\,$ and (3.19)
becomes
$${\cal J}_j < q\,|\,>\;=\;0 \eqno(3.20)$$\noindent
with $\,{\cal J}_j\,$ given by the differential operator

$${\cal J}_j\;=\;- i \sum_{\vec K}\,\left( \prod_{b = 1}^3 \sqrt{2 + \cos
\theta_b\over  3N}\;\right)\;e^{-i \vec K \cdot \vec r_j}\;\sum_{a =
1}^3\,w_a\;{\partial\over\partial q_a(\vec K)}\,. \eqno(3.21)$$\noindent
Hence, (3.20) is equivalent to, for $\,\vec K \ne 0\,$,
$$\sum_{a = 1}^3\,w_a\;{\partial\over\partial q_a(\vec K)}\;< q\,|\,>\;=\;0\,.
\eqno(3.22)$$
\vskip6pt
At a given $\,\vec K\,$, introduce a right-hand basis of three orthonormal
vectors $\,\hat e_1(\vec K)\,$, $\,\hat e_2(\vec K)\,$ and $\,\hat w(\vec K)
\equiv \vec w/\sqrt {w^2}\,$.  Define the generalized coordinate for the
longitudinal mode to be

$$Q_\ell(\vec K)\;\equiv\;\hat w(\vec K) \cdot \vec q\,(\vec K)
\eqno(3.23)$$\noindent
and that for the transverse mode
$$Q_t(\vec K)\;=\;\hat e_t(\vec K) \cdot \vec q\,(\vec K) \eqno(3.24)$$
\vskip4pt\noindent
where $\,t = 1, 2\,$ and $\,\vec q,\,\vec w\,$ are the vectorial forms of the
column matrices $\,q\,$ and $\,w\,$.  The constraint (3.22) implies
$\,< q\,|\,>\,$ to be independent of all $\,Q_\ell(\vec K)\,$.  Thus, in the
$\,0^{th}\,$ band approximation, the spectrum of the free electromagnetic field
is given by two transverse modes for each $\,\vec K\,$ in the Brillouin zone
with
$\,w^2(\vec\theta)\,$ as the square of the energy.  (Use of
 the subscript $\,\ell\,$ notation to denote ``longitudinal'' occurs only in
this subsection and Appendix A.)
\vskip4pt
It is straightforward to construct higher band Bloch functions through
(2.16)-(2.18).  Expanding $\,V_a\,$ in terms of all bands, the spectrum changes
from $\,w^2(\vec\theta)\,$ to the exact continuum result $\,\vec K^2 =
\vec\theta\,^2/\ell^2\,$ for $\,\vec K\,$ within the Brillouin zone, as
expected.
\vskip26pt\noindent
{\bf Remarks:}
\vskip6pt
The inverse relation of (3.23)-(3.24),

$$\vec q\,(\vec K)\;=\;\hat w(\vec K)\;Q_\ell(\vec K) + \sum_t\,\hat e_t(\vec
K)
\,Q_t(\vec K)\,, \eqno(3.25)$$
\noindent
enables us to decompose (3.7) into a sum of two parts:
$$\vec V(\vec r, t)\;=\;\vec V_\ell(\vec r, t) + \vec V\,'(\vec r, t)
\eqno(3.26)$$\noindent
with their components given by
$${(V_\ell)}_a\;=\;\sum_{\vec K}\,\hat w\,(\vec K)_a\,Q_\ell(\vec
K)\,g(K_a\,|\,x_a)\,\prod_{b \ne a}\,f(K_b\,|\,x_b)$$\noindent
\line {and \hfil (3.27)}
$$V_a'\;=\;\sum_{\vec K} \sum_t\,\hat e_t\,(\vec K)_a\,Q_t(\vec
K)\,g(K_a\,|\,x_a)\,
\prod_{b \ne a}\,f(K_b\,|\,x_b)\,.$$
\vskip4pt\noindent
{}From (2.2) and (2.4), we have
$$\ell\;{d\over dx}\;\Delta(x)\;=\;- C(x) + C(x + \ell) \eqno(3.28)$$
\vskip4pt\noindent
which leads to the following relation between  $\,g(K\,|\,x)\,$ and
$\,f(K\,|\,x)\,$, defined by (3.10)-(3.11):
$$\ell\;{d\over dx_a}\;f(K_a\,|\,x_a)\;=\;i\,w_a\,g(K_a\,|\,x_a)
\eqno(3.29)$$
\vskip4pt\noindent
with $\,w_a\,$ given by (3.14).  Thus, $\,\vec\nabla \times \vec V_\ell = 0\,$
at all points in the continuum space, since

$$\vec V_\ell(\vec r, t)\;=\;\vec\nabla\;\chi\,(\vec r, t)
\eqno(3.30)$$\noindent
where
$$\chi\;=\;\sum_{\vec K}\,\chi_{\vec K}\,\prod_a f(K_a\,|\,x_a)$$\noindent
\line {and \hfil (3.31)}
$$\chi_{\vec K}\;=\;- i\,Q_\ell(\vec K)\,/\,\sqrt{w^2}\,.$$
\vskip4pt\noindent
On the other hand, because $\,dg\,(K_a\,|\,x_a)\,/\,dx_a\,$ is very different
from
$\,f(K_a\,|\,x_a)\,$, one can readily verify

$$\vec\nabla\,\cdot\,\vec V\,'\;\ne\;0\,. \eqno(3.32)$$
\vskip4pt\noindent
Indeed, should we require $\,\vec V\,'\,$ to be divergence free, the roles of
$\,f\,$ and $\,g\,$ would have to be interchanged in (3.29), as will be
discussed in the next section.
\vskip4pt
The state vector $\,< q\,|\,>\,$ introduced in (3.20) is a functional of
$\,\vec V\,$ (i.e., of $\,\vec V_\ell\,$ and $\,\vec V')\,$.  The constraint
(3.22) makes
$\,< q\,|\, >\,$ to be a functional only of $\,\vec V\,'(\vec r, t)\,$,
independent of the longitudinal component $\,\vec V_\ell(\vec r, t)\,$.
However,
$\,\vec\nabla
\cdot \vec V\,' \ne 0\,$ reminds us that the familiar difference between the
time-axial gauge and the Coulomb gauge persists throughout the band structure.
\vskip4pt
We note that in the decomposition (3.25), the component
$\,\hat e_t(\vec K)\,Q_t(\vec K)\,$ is, of course, perpendicular (i.e.,
transverse) to the longitudinal component $\,\hat w(\vec K)\,Q_\ell(\vec
K)\,$.  Because $\,\vec K\,$ lies within the Brillouin zone, this is only the
lattice transversality condition; in contrast, (3.30) and (3.32) refer to
relations in which $\,\vec r\,$ is a continuum coordinate.
\vskip20pt\noindent
{\it 3.2.  Coulomb Gauge} \vskip6pt Let $\,A_\mu(\vec r, t)\,$ be the
electromagnetic field in the Coulomb gauge.  Its spatial part $\,\vec A(\vec r,
t)\,$ is divergence-free; i.e.,

$$\vec\nabla\,\cdot\,\vec A\;=\;0\,. \eqno(3.33)$$\noindent
The free-field Hamiltonian is

$$H\;=\;\int\;{1\over 2}\,(\vec {\rm \Pi}\,^2 + \vec B\,^2)\,d^3r \eqno(3.34)$$
\vskip4pt\noindent
where $\,\vec B = \vec\nabla \times \vec A\,$ denotes the magnetic field and
$\,- \vec {\rm \Pi}\,$ is the electric field which also satisfies the
transversality condition

$$\vec\nabla\;\cdot\;\vec {\rm \Pi}\;=\;0\,. \eqno(3.35)$$\noindent
In addition, we have the equal-time commutation relation

$$\left[\,{\rm \Pi}_a(\vec r, t), A_b(\vec r\,', t)\,\right]\;=\;-
i\,(\delta_{ab} -\nabla^{-2}\,\nabla_a\,\nabla_b)\;\delta^3(\vec r_a - \vec
r_b)\,. \eqno(3.36)$$
\vskip4pt\noindent
The central problem that confronts us is the construction of the appropriate
$\,0^{th}\,$ band Bloch functions, so that (3.33) and (3.35) can be valid at
{\it all} continuum points in the lattice approximation.
\vskip8pt
As mentioned at the end of the preceding section, in order to have a
divergence-free field, the roles of $\,f\,$ and $\,g\,$ in (3.29) have to be
interchanged.  Thus, we replace the link function (3.10) by (2.33).  We write,
for the $\,0^{th}\,$ band function expansion,

$$A_a(\vec r, t)\;=\;\sum_{\vec K}\;q_a(\vec K)\,F_a(\vec K\,|\,\vec r\,)$$
\line {and \hfil (3.37)}
$${\rm \Pi}_a(\vec r, t)\;=\;\sum_{\vec K}\;p_a(- \vec K)\,F_a(\vec K\,|\,\vec
r\,)$$
\noindent
where
$$F_a(\vec K\,|\,\vec r\,)\;=\;c_3(\theta_a)\,{\cal G}(K_a\,|\,x_a)\;\prod_{b
\ne a}\,c_2(\theta_b)\,{\cal F}(K_b\,|\,x_b)\eqno(3.38)$$\noindent
in which, similar to (2.7), but with $\,m = 2\,$ and 3,
$${\cal F}(K_a\,|\,x_a)\;=\;\sum_{n = -
\infty}^\infty\,e^{i\,n\,\theta_a}\,\Delta(x_a - n\ell)\,,$$\noindent
\line {and \hfil (3.39)}
$${\cal G}(K_a\,|\,x_a)\;=\;e^{i{1\over 2}\,\theta_a}\;\sum_{n = -
\infty}^\infty\,e^{i\,n\,\theta_a}\,S(x_a - n\ell)\,.$$\noindent
The factors
$$c_2(\theta_a)\;=\;\left[\,{3\over N\ell(2 + \cos \theta_a)}\,\right]^{1\over
2}\,,$$
\line {and \hfil (3.40)}
$$c_3(\theta_a)\;=\;\left[\,{30\over N\ell(16 + 13 \cos
\theta_a + \cos^2 \theta_a)}\,\right]^{1\over 2}\,,$$
\vskip4pt\noindent
refer to the normalization constants $\,c_m\,$ in (2.15) with the subscripts
$\,m = 2\,$ and 3 denoting the order of the lump functions in $\,{\cal F}\,$
and  $\,{\cal G}\,$.  We note that apart from normalization factors,
$\,{\cal F}\,$ and  $\,{\cal G}\,$ are the same functions as $\,f\,$ of (3.11)
and $\,g_0\,$ of (2.33).
\vskip6pt
{}From (2.3)-(2.4), it follows that
$$\ell\;{dS(x)\over dx}\;=\;\Delta(x) - \Delta(x - \ell) \eqno(3.41)$$\noindent
and therefore
$$\ell\;{d\over dx_a}\;{\cal G}(K_a\,|\,x_a)\;=\;2i\,\sin {\theta_a\over 2}\;
{\cal F}(K_a\,|\,x_a)\,. \eqno(3.42)$$
\vskip6pt\noindent
The transversality conditions (3.33) and (3.35) become simply
$$\vec q\,(\vec K)\;\cdot\;\vec v(\vec K)\;=\;0$$
\line {and \hfil (3.43)}
$$\vec p\,(\vec K)\;\cdot\;\vec v(\vec K)\;=\;0$$\noindent
where the components of $\,\vec v(\vec K)\,$ are

$$v_a(\vec K)\;=\;{2\over\ell}\;\sin {\theta_a\over 2}\;\left[ {10\,(2 + \cos
\theta_a)\over 16 + 13 \cos \theta_a + \cos^2 \theta_a}\right]^{1\over 2}\,.
\eqno(3.44)$$\noindent
Because
$$F_a(\vec K\,|\,\vec r\,)^*\;=\;F_a(- \vec K\,|\,\vec r\,)\,,
\eqno(3.45)$$\noindent
we have
$$q_a(\vec K)\;=\;q_a(- \vec K)^\dagger \hskip2em {\rm and}\hskip2em
p_a(\vec K)\;=\;p_a(- \vec K)^\dagger\,. \eqno(3.46)$$\vskip8pt
At a given $\,\vec K\,$, form a right-hand basis of three real orthonormal
vectors
$\,\hat\epsilon_1(\vec K), \hat\epsilon_2(\vec K)\,$ and $\,\hat v(\vec K)
\equiv \vec v(\vec K)\,/\,\sqrt {\vec v(\vec K)^2}\;$ so that

$$\,\hat\epsilon_1(\vec K) = \hat\epsilon_2(\vec K) \times \hat v(\vec K)\,.
\eqno(3.47)$$
\vskip4pt\noindent
Thus, if $\,\hat\epsilon_2(- \vec K) = \hat\epsilon_2(\vec K)\,$ then
$\,\hat\epsilon_1(- \vec K) = - \hat\epsilon_1(\vec K)\,$, since $\,\hat
v(-\vec
K) = - \hat v(\vec K)\,$, in accordance with (3.44).  Write

$$q_a(\vec K)\;=\;\sum_t\,\hat\epsilon_t(\vec K)_a\,Q_t(\vec K)$$
\line {and \hfil (3.48)}
$$p_a(\vec K)\;=\;\sum_t\,\hat\epsilon_t(\vec K)_a\,P_t(\vec K)$$\noindent
where the sum is over $\,t = 1, 2\,$.  The commutation relation (3.36), when
applied to the $\,0^{th}\,$ band components of $\,\vec\Pi\,$ and $\,\vec A\,$,
gives

$$\left[\,P_t(\vec K), Q_{t'}(\vec K\,')\,\right] \;=\;-i \delta_{\vec K \vec
K'}
\delta_{t t'}\,, \eqno(3.49)$$\noindent
or in its equivalent form
$$\left[\,p_a(\vec K), q_{a'}(\vec K\,')\,\right] \;=\;-i \delta_{\vec K \vec
K'}
\left(\delta_{aa'} - \hat v_a(\vec K)\,\hat v_{a'}(\vec K)\right)\,.
\eqno(3.50)$$
\vskip8pt
In the lattice approximation (3.37) and in terms of the column matrices

$$p(\vec K)\;=\;\left(\matrix{p_1(\vec K)\cr
p_2(\vec K)\cr
p_3(\vec K)\cr}\right) \hskip2em{\rm and}\hskip2em  q(\vec
K)\;=\;\left(\matrix{q_1(\vec K)\cr q_2(\vec K)\cr q_3(\vec
K)\cr}\right)\,,\eqno(3.51)$$
\noindent
the Hamiltonian (3.34) becomes, on account of (3.28) and (3.43),
$$H\;=\;\sum_{\vec K}\;[\,p^\dagger\,p + q^\dagger\,\Delta q\,]\eqno(3.52)$$
\vskip2pt\noindent
where the matrix $\,\Delta = \left(\Delta_{ab}(\vec K)\right)\,$ is diagonal,
with its diagonal elements

$$\Delta_{aa}\;=\;w^2 - \delta_a\,, \eqno(3.53)$$\noindent
$\,w^2\,$ given by (3.15) and, similar to (2.39), $\,\delta_a = w_a^2 -
v_a^2\,$; i.e.,

$$\delta_a\;=\;{2(1 - \cos \theta_a)^2 (8 + 7 \cos \theta_a)\over \ell^2\,(2 +
\cos \theta_a) (16 + 13 \cos \theta_a + \cos^2 \theta_a)}\,. \eqno(3.54)$$
\vskip6pt\noindent
As before, $\,\theta_a = K_a\,\ell\,$.  At a given $\,\vec\theta\,$, although
$\,\Delta\,$ is a $\,3 \times 3\,$ matrix, because of the orthogonality
condition (3.43), there are only two transverse modes whose eigen-frequencies
$\,\omega^2 = \omega_\pm^2\,$ are determined by the secular equation (cf. Eq.
(2.16) of I)
$${v_1^2\over \Delta_{11} - \omega^2} + {v_2^2\over \Delta_{22} - \omega^2} +
{v_3^2\over \Delta_{33} - \omega^2}\;=\;0\,. \eqno(3.55)$$
\noindent
The solution is
$$\eqalign {\omega_\pm^2(\vec\theta)\;=\;&w^2 - {1\over
2v^2}\,[\,v_1^2(\delta_2 + \delta_3) + v_2^2(\delta_3 + \delta_1) +
v_3^2(\delta_1 + \delta_2)\,]\cr
& \pm {1\over 2v^2}\,[\,\delta_1^2(v_2^2 + v_3^2)^2 + \delta_2^2(v_3^2 +
v_1^2)^2 + \delta_3^2(v_1^2 + v_2^2)^2\cr
&\hskip1em + 2\delta_1\delta_2(v_1^2 v_2^2 - v_3^2 v^2) +
2\delta_2\delta_3(v_2^2 v_3^2 - v_1^2 v^2) + 2\delta_3\delta_1(v_3^2 v_1^2 -
v_2^2 v^2)\,]^{1\over 2}\cr} \eqno(3.56)$$\noindent
where
$$v^2\;=\;\sum_{a=1}^3\,v_a(\vec\theta)^2\;=\;\sum_{a=1}^3\;{20(1 - \cos
\theta_a) (2 + \cos \theta_a)\over\ell^2\,(16 + 13 \cos \theta_a + \cos^2
\theta_a)}\,.\eqno(3.57)$$
\vskip8pt
Because $\,\delta_a > 0\,$, each of the diagonal elements $\,\Delta_{aa}\,$
lies
within $\,v^2\,$ and $\,w^2\,$.  The two eigenvalues $\,\omega_\pm^2\,$ have to
be
sandwiched between $\,\Delta_{11}\,$, $\,\Delta_{22}\,$ and $\,\Delta_{33}\,$;
consequently, they must also satisfy

$$v^2\;<\;\omega_\pm^2\;<\;w^2\,. \eqno(3.58)$$
\vskip4pt\noindent
Using the spherical coordinates $\,v_1 = v\, \sin \alpha\,\cos \beta\,$, $\,v_2
= v\,
\sin \alpha\,\sin \beta\,$ and $\,v_3 = v \,\cos \alpha\,$, we may express
(3.56)
in an alternative form:

$$\omega_\pm^2(\vec\theta)\;=\; w^2 + {1\over 2}\,\left[\,\lambda_{11} +
\lambda_{22} \pm \sqrt {(\lambda_{11} - \lambda_{22})^2 +
4\lambda_{12}^2}\;\right]\eqno(3.59)$$\noindent
where
$$\lambda_{11}\;=\;- \delta_1\;\sin^2 \beta - \delta_2 \cos^2 \beta\,,$$
$$\lambda_{22}\;=\;- (\delta_1\;\cos^2 \beta + \delta_2 \sin^2 \beta)\;\cos^2
\alpha - \delta_3 \sin^2 \alpha\,,$$\noindent
and
$$\lambda_{12}\;=\;(\delta_1 - \delta_2)\;\sin \beta\;\cos \beta\;\cos
\alpha\,.$$
\vskip4pt\noindent
In the long wavelength region $\,|\,\theta_a\,| << 1\,$, $\,\omega_\pm^2\,$
deviates from $\,w^2\,$ of (3.15) by  $\,O(\theta^4)\,$, and (3.59)
becomes

$$\omega_\pm^2\;=\;w^2 - {1\over 24\ell^2}\;\left[ s_2 - {3s_3\over s_1} \mp
\sqrt {\left(s_2 - {s_3\over s_1}\right) \left(s_2 - {9s_3\over
s_1}\right)}\;\right] + O(\theta^6) \eqno(3.60)$$
\vskip4pt\noindent
where $\,s_1 = \theta_1^2 + \theta_2^2 + \theta_3^2\,,\; s_2 =
\theta_2^2\,\theta_3^2 + \theta_3^2\,\theta_1^2 + \theta_1^2\,\theta_2^2\,$
and $\,s_3 = \theta_1^2\,\theta_2^2\, \theta_3^2\,$.
\vskip6pt
The higher-band Bloch functions can be constructed through a generalization of
(2.16)-(2.18).  (See (3.73)-(3.79) below).  As we shall see, these higher-band
Bloch functions are very close to the corresponding Fourier series of high wave
numbers.  By expanding $\,\vec A\,$ in terms of all bands, one changes the
spectrum from the lattice approximation $\,\omega_\pm^2\,$ to the continuum
result.
\vskip26pt\noindent
{\bf Remarks:}
\vskip8pt\noindent
{\bf 1.}  By using (3.48), the $\,0^{th}\,$ band expansion (3.37) of $\,\vec
A\,$
in the Coulomb gauge can be written as (with the time variable suppressed)

$$\vec A(\vec r\,)\;=\;\sum_{\vec K} \sum_{t=1}^2\,Q_t(\vec K)\;\vec F\,^t(\vec
K\,|\,\vec r\,) \eqno(3.61)$$
\noindent
where $\vec F\,^t(\vec K\,|\,\vec r\,)\,$ is a vector function whose components
are

$$ F\,^t(\vec K\,|\,\vec r\,)_a\;=\;\hat\epsilon_t(\vec K)_a\;F_a(\vec
K\,|\,\vec r\,) \,.\eqno(3.62)$$
\vskip4pt\noindent
Because of (3.38)-(3.42) and (3.47), for any $\,\vec K\,$ within the Brillouin
zone

$$\vec\nabla\,\cdot\,\vec F\,^t(\vec K\,|\,\vec
r\,)\;=\;0\eqno(3.63)$$\noindent
everywhere in the coordinate space.  Thus, we have

$$\vec F\,^t(\vec K\,|\,\vec r\,)\;=\;\vec\nabla \times \vec G\,^t(\vec
K\,|\,\vec r\,)\eqno(3.64)$$
\vskip4pt\noindent
with the components of $\,\vec G\,^t(\vec K\,|\,\vec r\,)\,$ given by

$$G\,^t(\vec K\,|\,\vec r\,)_a\;=\;i \sum_{t'=1}^2 (v^2)^{-{1\over
2}}\,\eta^{tt'}\,\hat\epsilon_{t'}(\vec K)_a\;c_2(\theta_a)\;{\cal
F}(K_a\,|\,x_a)\;\prod_{b \ne a}\,c_3(\theta_b)\;{\cal
G}(K_b\,|\,x_b)\eqno(3.65)$$
\vskip4pt\noindent
and $\,(\eta^{tt'})\,$ denoting the $\,2 \times 2\,$ antisymmetric matrix whose
elements are $\,\eta^{11} = \eta^{22} = 0\,$ and $\,\eta^{12} = - \eta^{21} =
1\,$.  Although the independent variables in the lattice approximation are
restricted to the set $\,\{\,Q_t(\vec K)\,\}\,$, the $\,0^{th}\,$ band Bloch
functions $\,\vec F\,^t(\vec K\,|\,\vec r\,)\,$ satisfy the differential
relations (3.63)-(3.64) at all $\,\vec r\,$ in the continuum.
\vskip10pt\noindent
{\bf 2.}  From (2.1) and (2.4), we see that the Fourier series expansion of
(3.38)
is

$$F_a(\vec K\,|\,\vec r\,)\;=\;e^{i\vec K\cdot\vec r}\;\sum_{\vec m}\,{\cal
C}_a(\vec\theta\,|\,\vec m)\,e^{i\,2\pi\,\vec m\cdot\vec r/\ell}\eqno(3.66)$$
where
$${\cal C}_a(\vec\theta\,|\,\vec m)\;=\;\left( {2 \sin {\theta_a\over 2}\over
\theta_a + 2\pi\,m_a} \right)^3\, c_3(\theta_a)\,\prod_{b \ne a}\;\left( {2
\sin
{1 \over 2}\,\theta_b\over\theta_b + 2\pi\,m_b} \right)^2\;c_2(\theta_b)\,,
\eqno(3.67)$$
\vskip4pt\noindent
$\,m_a\,$ (the components of $\,\vec m\,$) are all integers, and $\,c_2, c_3\,$
are given by (3.40).
\vskip10pt\noindent
{\bf 3.}  To construct the higher-band Bloch functions in the Coulomb gauge, we
generalize the steps (2.12)-(2.16):  Define

$$\vec\psi_{\vec p}\,^t(\vec r\,)\;\equiv\;\Omega^{-{1\over 2}}\,\hat e^t(\vec
p\,)\,e^{i\vec p\cdot \vec r}\eqno(3.68)$$
\vskip4pt\noindent
where $\,t = 1, 2\,$ denote two sets of divergence-free functions, with $\,\hat
e^1\,$ and $\,\hat e^2\,$ both real unit vectors perpendicular to $\,\vec p\,$
and
to each other.  Resolve

$$\vec p\;=\;\vec K + (2\pi\,\vec n/\ell) \eqno(3.69)$$
\vskip4pt\noindent
with the components of $\,\vec n\,$ being any integer.  At a given $\,\vec K\,$
within the Brillouin zone, write (3.68) as

$$\vec\psi_{\vec n}\,^t(\vec r\,)\;=\;\Omega^{-{1\over 2}}\,\hat e^t(\vec
n\,)\,e^{i\vec K \cdot \vec r +i\,2\pi\,\vec n \cdot \vec r/\ell}\,.
\eqno(3.70)$$
\vskip8pt
Introduce the $\,2 \times 2\,$ matrix $\,T_{\vec n}\,$ whose elements are

$$T_{\vec n}^{tt'}(\vec K)\;\equiv\;\int \vec F\,^t(\vec K\,|\,\vec r\,) \cdot
\vec\psi_{\vec n}\,^{t'}(\vec r\,)^*\,d^3r \eqno(3.71)$$
\vskip4pt\noindent
where $\,\vec F^t(\vec K\,|\,\vec r\,)\,$ is given by (3.62).  The complete set
of orthonormal divergence-free Bloch functions $\,\{\,\vec F_{\vec n}^t(\vec
K\,|\,\vec r\,)\,\}\,$ can be constructed by requiring, for $\,\vec n = 0\,$,

$$\vec F_0^t(\vec K\,|\,\vec r\,)\;=\; \vec F\,^t(\vec K\,|\,\vec r\,)
\eqno(3.72)$$
\noindent
and for $\,\vec n \ne 0\,$
$$\vec F_{\vec n}^t(\vec K\,|\,\vec r\,)\;=\;\vec\psi_{\vec n}\,^t(\vec r\,) -
\sum_{t'=1}^2\,\left[\,(M_{\vec n})^{tt'}\,\vec F_0^{t'}(\vec K\,|\,\vec r\,) +
(N_{\vec n})^{tt'}\,\vec\psi_0\,^{t'}(\vec r\,)\,\right]\,. \eqno(3.73)$$
\vskip4pt\noindent
The orthogonality between $\,\vec F_0^t\,$ and $\,\vec F_{\vec n}^{t'}\,$
yields
the $\,2 \times 2\,$ matrix relation

$$T_{\vec n} - M_{\vec n}^\dagger - T_0\,N_{\vec n}^\dagger\;=\;0\,,
\eqno(3.74)$$
\vskip4pt\noindent
which together with the orthonormality between $\,\vec F_{\vec n}^t\,$ and
$\,\vec
F_{\vec m}^{t'}\,$ for nonzero $\,\vec n\,$ and $\,\vec m\,$ give

$$M_{\vec n}\,M_{\vec m}^\dagger\;=\;N_{\vec n}\,N_{\vec m}^\dagger\,.
\eqno(3.75)$$
\noindent
The solution of (3.75) is
$$N_{\vec n}\;=\;M_{\vec n}\,u \eqno(3.76)$$
\noindent
with $\,u\,$ an arbitrary $\,2 \times 2\,$ unitary matrix satisfying

$$u\,u^\dagger\;=\;1\,. \eqno(3.77)$$
\noindent
By using (3.74), we have
$$M_{\vec n}\;=\;T_{\vec n}^\dagger(1 + u\,T_0^\dagger)^{-1}$$
\line {and \hfil (3.78)}
$$N_{\vec n}\;=\;T_{\vec n}^\dagger(1 + u\,T_0^\dagger)^{-1}\,u\,.$$
\vskip4pt\noindent
It follows then (for all $\,\vec n\,$ and $\,\vec m\,$, including 0) that the
orthonormality relation

$$\int_{\Omega}\,d^3r\,\vec F_{\vec n}^t(\vec K\,|\,\vec r\,)^*\,\cdot\,\vec
F_{\vec m}^{t'}(\vec K'\,|\,\vec r\,)\;=\;\delta_{\vec n \vec
m}\,\delta_{tt'}\,\delta_{\vec K \vec K'}\,; \eqno(3.79)$$
\vskip4pt\noindent
in addition, every member of the set $\,\{\,\vec F_{\vec n}^t\,\}\,$ satisfies
the divergence-free condition

$$\vec\nabla \cdot \vec F_{\vec n}^t(\vec K\,|\,\vec r\,)\;=\;0\,.
\eqno(3.80)$$
\vskip4pt\noindent
If one wishes, the arbitrary $\,u\,$ can be set to 1; (3.78) becomes
simply

$$M_{\vec n}\;=\;N_{\vec n}\;=\;T_{\vec n}^\dagger(1
+ T_0^\dagger)^{-1}\,.\eqno(3.81)$$
\vskip8pt
Because of (3.67), the matrix elements

$$T_{\vec n}^{tt'}\;=\;\sum_{a=1}^3 \hat\epsilon^t(\vec K)_a \cdot
\hat e^{t'}(\vec n)_a\;{\cal C}_a(\vec\theta\,|\,\vec n\,) \eqno(3.82)$$
\vskip4pt\noindent
are all real.  The expansion of $\,\vec A(\vec r, t)\,$ and $\,\vec\Pi(\vec r,
t)\,$ in terms of all bands,

$$\vec A(\vec r, t)\;=\;\sum_{\vec n, \vec K, t}\,Q_{\vec n}^t(\vec K)\,\vec
F_{\vec n}^t(\vec K\,|\,\vec r\,)$$
\line{and \hfil (3.83)}
$$\vec\Pi(\vec r, t)\;=\;\sum_{\vec n, \vec K, t}\,P_{\vec n}^t(- \vec K)\,\vec
F_{\vec n}^t(\vec K\,|\,\vec r\,)\,,$$
\vskip4pt\noindent
changes the $\,0^{th}\,$ band spectrum from the lattice approximation
$\,\omega_\pm^2(\vec\theta)\,$ of (3.56) to the exact continuum result  $\,\vec
K^2 = (\vec\theta\,/\ell)^2\,$ within the Brillouin zone and, for $\,\vec n \ne
0\,$, $\,\left[\,\vec K + (2\pi\,\vec n\,/\ell)\,\right]^2\,$ outside the zone.
\vskip10pt\noindent
{\bf 4.}  The gauge-fixing conditions, $\,V_0(\vec r, t) = 0\,$ in the
time-axial gauge and
\vskip1pt\noindent
 $\,\vec\nabla \cdot \vec A(\vec r, t) = 0\,$ in the Coulomb
gauge, are valid at {\it all} $\,\vec r\,$ in the lattice approximation.  This
necessitates the difference between the $\,0^{th}\,$ band functions in these
two
gauges.  Of course, in either gauge, the Bloch functions of all bands are
complete.  The transformation matrix between these different band structures
can
therefore be readily derived.
\vskip10pt\noindent
{\bf 5.}  Similar considerations are also applicable to a generalized
Coulomb-like gauge where, instead of $\,\vec\nabla \cdot \vec A = \nabla_a\,A_a
= 0\,$, we have$^5$ at all continuum $\,\vec r\,$

$$\int (\vec r\,|\,\Gamma_a\,|\,\vec r\,')\;A_a(\vec r\,')\,d^3r'\;=\;0\,,
\eqno(3.84)$$
\vskip4pt\noindent
with $\,\Gamma_a\,$ a linear operator independent of $\,\vec A\,$.
Accordingly, we have to create  a new class of lump functions of
different orders
$\,m\,$, for which relations like (2.2)-(2.3) remain valid, but
with the differential operator
$\,\partial\,/\,\partial x_a\,$ replaced by $\,\Gamma_a\,$.  Details will be
given elsewhere.
\vskip4pt
  The extension to QCD in a Coulomb or Coulomb-like gauge
is straightforward since the group index is external to the band
decomposition, as will be discussed in the next section.  However, as
mentioned in the Introduction, the extension to QCD in the time-axial gauge is
more complicated: While the gauge condition
$\,A_4(\vec r, t) = 0\,$ being a linear equation offers no problem in the band
decomposition, the generalization of the constraint equation (3.19)

$${\cal J}_j\,|\,>\;=\;0$$\vskip4pt\noindent
becomes nonlinear in QCD.  That means in solving this constraint, there would
be additional coupling between different bands which must be taken into
account.
\vskip4pt
In the next two sections, we will discuss QCD, but only in Coulomb gauge.
\vfill\eject
\noindent
\centerline {\bf 4.  QUANTUM CHROMODYNAMICS}
\vskip12pt
Let $\,{\cal A}_a^\ell(\vec r, t)\,$ be the field in the Coulomb gauge, with
the
subscript $\,a = 1, 2, 3\,$ denoting its spatial components and the superscript
$\,\ell = 1, 2,\,\cdot\cdot,\,n^2 - 1\,$ its $\,SU(n)\,$ index.  The
conjugate momentum is $\,{\cal P}_a^\ell(\vec r, t)\,$.  Both satisfy the
divergence-free condition

$$\vec\nabla\,\cdot\,\vec {\cal A}\,^\ell(\vec r,
t)\;=\;\vec\nabla\,\cdot\,\vec
{\cal P}\,^\ell(\vec r, t)\;=\;0 \eqno(4.1)$$
\noindent
and the equal-time commutation relation

$$\left[\,{\cal P}_a^\ell(\vec r, t)\,,\;{\cal A}_{a'}^{\ell '}(\vec r\,',
t)\,\right]\;=\;-
i\,\delta^{\ell\ell '}(\delta_{ab} - \nabla^{-2}\,\nabla_a
\nabla_b)\,\delta^3(\vec r - \vec r\,')\,. \eqno(4.2)$$
\vskip4pt\noindent
The continuum QCD Hamiltonian (without quarks) is$^5$

$${\cal H}\;=\;{1\over 2}\,\int\,[\,{\cal J}^{-1}\,{\cal P}_a^\ell\;{\cal
J}\,{\cal P}_a^\ell +
{\cal B}_a^\ell\,{\cal B}_a^\ell\,]\;d^3r + {\cal H}_{\rm Coul} \eqno(4.3)$$
\noindent
where the Coulomb interaction is

$$\eqalign{{\cal H}_{\rm Coul}\;=\;{1\over 2}\;g_0^2\,{\cal
J}^{-1}\,\int\,&\sigma^\ell(\vec r\,)\,(\ell, \vec r\;|\,(\nabla_a\,{\cal
D}_a)^{-1} ( -
\nabla^2)\,(\nabla_b\,{\cal D}_b)^{-1}\,|\,\ell', \vec
r\,')\,\cr
&\cdot\,{\cal J}\,\sigma^{\ell'}(\vec r\,')\,d^3r\,d^3r'\,,\cr}\eqno(4.4)$$
with the charge density

$$\sigma^\ell\;=\;f^{\ell mn}\,{\cal A}_a^m\,{\cal P}_a^n\,; \eqno(4.5)$$
\noindent
the color magnetic field $\,{\cal B}_a^\ell\,$ is given by

$$\epsilon_{abc}\,{\cal B}_c^\ell\;=\;\nabla_a\,{\cal A}_b^\ell -
\nabla_b\,{\cal A}_a^\ell +
g_0\,f^{\ell mn}\,{\cal A}_a^m\,{\cal A}_b^n\,,\eqno(4.6)$$
\vskip4pt\noindent
$\,{\cal D}_a\,$ denotes the covariant derivative matrix whose elements are

$${\cal D}_a^{\ell m}\;=\;\delta^{\ell m}\,\nabla_a - g_0\,f^{\ell mn}\,{\cal
A}_a^n\,,
\eqno(4.7)$$
\vskip4pt\noindent
$\,{\cal J}\,$ is the the Jacobian$^6$, or the Faddeev-Popov determinant$^7$

$${\cal J}\;=\;{\rm det}\;|\,\nabla_a\,{\cal D}_a\,|\,, \eqno(4.8)$$
\vskip4pt\noindent
$\,\epsilon_{abc}\,$ is the usual antisymmetric tensor in the
three-dimensional space, $\,f^{\ell mn}\,$ is the antisymmetric structure
constant
of the
$\,SU(n)\,$ group algebra and $\,g_0\,$ is the bare coupling constant, related
to
the renormalized coupling $\,g\,$ by

$$g_0\;=\;g + \delta g\,,\eqno(4.9)$$
\vskip4pt\noindent
as will be discussed later.  All the above repeated indices are summed over
(but
not for the equations below).
\vskip20pt\noindent {\it 4.1.  $\,0^{th}\,$ Band Expansion} \vskip6pt
The set $\,\vec F_{\vec n}^t(\vec K | \vec r\,)\,$, given by (3.72)-(3.73),
forms a complete orthonormal functional basis that satisfies $\,\{\,\vec\nabla
\cdot
\vec F_{\vec n}^t(\vec K | \vec r\,)\,\} = 0\,$.  Consequently, we can always
expand the field operators $\,\vec{\cal A}\,^\ell(\vec r, t)\,$ and
$\,\vec{\cal P}\,^\ell(\vec r, t)\,$ in terms of these Bloch functions, similar
to
(3.83).  In the $\,0^{th}\,$ band approximation, we restrict the expansion only
to $\,\vec n =
0\,$.  For clarity, the $\,0^{th}\,$ band approximations of
$\,\vec{\cal A}\,^\ell(\vec r, t)\,$ and  $\,\vec{\cal P}\,^\ell(\vec r, t)\,$
will be denoted
by $\,\vec A\,^\ell(\vec r, t)\,$ and $\,\vec\Pi\,^\ell(\vec r, t)\,$, whose
components
are given by, analogous to (3.37),

$$A_a^\ell(\vec r, t)\;=\;\sum_{\vec K}\,q_a^\ell(\vec K)\;F_a(\vec
K\,|\,\vec r\,)\eqno(4.10)$$\noindent
and
$$\Pi_a^\ell(\vec r, t)\;=\;\sum_{\vec K}\,p_a^\ell(- \vec K)\;F_a(\vec
K\,|\,\vec
r\,)\eqno(4.11)$$
\vskip4pt\noindent
where $\,F_a(\vec K\,|\,\vec r\,)\,$ is given by (3.38).  As in (3.43) and
(3.50), the divergence-free condition (4.1) requires

$$\sum_{a=1}^3\,q_a^\ell(\vec K)\,v_a(\vec K)\;=\;\sum_{a=1}^3\,p_a^\ell(\vec
K)\,v_a(\vec K)\;=\;0\,,\eqno(4.12)$$\noindent
and the commutation relation (4.2) leads to
$$\left[\,p_a^\ell(\vec K), q_{a'}^{\ell'}(\vec K')\,\,\right]\;=\;-
i\,\delta^{\ell\ell'}\,\delta_{\vec K \vec K'}\left(\delta_{aa'} - \hat
v_a(\vec K)\,\hat v_{a'}(\vec K)\right)\,.\eqno(4.13)$$
\vskip4pt\noindent
The substitution of (4.10)-(4.11) into (4.3)-(4.4) gives the noncompact lattice
formulation of QCD; in addition, $\,g_0\,$ should be replaced by $\,g_\ell\,$,
the lattice coupling constant, as will be discussed in Section 5.  (See
(5.1)-(5.6)
below.)
\vskip8pt
Because of (3.38)-(3.39), the Bloch function $\,F_a(\vec K\,|\,\vec r\,)\,$
satisfies the lattice locality condition (1.2).  To appreciate its
consequences,
we express (4.10), the expansion of $\,\vec A\,^\ell(\vec r, t)\,$, in an
equivalent form:

$$A_x^\ell(\vec r, t)\;=\;\sum_j\,a_j^\ell(t)\,S(x - j_1\ell)\;\Delta(y -
j_2\ell)\;\Delta(z - j_3\ell)\,,$$
$$A_y^\ell(\vec r, t)\;=\;\sum_j\,b_j^\ell(t)\,\Delta(x - j_1\ell)\;S(y -
j_2\ell)\;\Delta(z - j_3\ell)\eqno(4.14)$$
\noindent and
$$A_z^\ell(\vec r, t)\;=\;\sum_j\,c_j^\ell(t)\,\Delta(x - j_1\ell)\;\Delta(y -
j_2\ell)\;S(z - j_3\ell)\,,$$
\vskip4pt\noindent
where $\,\Delta(x)\,$ and $\,S(x)\,$ are the lump functions given by (2.4), and
$\,j\,$ denotes the $\,j^{th}\,$ lattice site located at  $\,\vec r_j = (j_1,
j_2, j_3)\,\ell\,$.  On account of (3.41), the $\,\vec\nabla \cdot \vec A^\ell
=
0\,$ condition gives
$$a_j^\ell - a_{j'_-}^\ell + b_j^\ell - b_{j''_-}^\ell + c_j^\ell -
c_{j'''_-}^\ell\;=\;0 \eqno(4.15)$$
\noindent
where the subscripts refer to the sites
$$j\;=\;(j_1, j_2, j_3)\,,\hskip3em j'_-\;=\;(j_1-1, j_2, j_3)\,,$$
$$j''_-\;=\;(j_1, j_2-1, j_3)\;\;\;{\rm and}\;\;\;j'''_-\;=\;(j_1, j_2,
j_3-1)\,.\eqno(4.16)$$
\vskip4pt\noindent
Equation (4.15) is the familiar field-flux conservation relation on a lattice
(analogous to the Kirchhoff law).  In our case, this discrete form co-exists
with its continuum realization $\,\vec \nabla \cdot \vec A = 0\,$ at all
$\,\vec
r\,$.
\vskip4pt
(A comment on notation:  Whenever the letter $\,\ell\,$ appears as a {\it
superscript}, it denotes the color index.  Otherwise, it may designate the
lattice spacing, as in $\,x - j_1\ell\,$ above, or it may indicate quantities
associated with a lattice; e.g., the lattice coupling constant $\,g_\ell\,$
that will be introduced later.)
\vskip8pt
Figure 2 gives a simple example of a localized configuration with

$$A_x^\ell\;=\;S(x)\,\left[\,\Delta(y) - \Delta(y -
\ell)\,\right]\,\Delta(z)\,\delta^{\ell 1}\,,$$
$$A_y^\ell\;=\;- \left[\,\Delta(x) - \Delta(x -
\ell)\,\right]\,S(y)\,\Delta(z)\,\delta^{\ell 1}\,,\eqno(4.17)$$
\noindent and
$$A_z^\ell\;=\;0$$
\vskip4pt\noindent
(the factor $\,\delta^{\ell 1}\,$ can be replaced by any other distribution in
the $\,SU(n)\,$ index $\,\ell\,$).  As explained in the caption, by examining
the
circulation of link fluxes associated with the four links of the central
plaquette in Figure 2, one sees that (4.15) is fulfilled.  Through direct
differentiation, this configuration also satisfies
$\,\vec\nabla
\cdot \vec A^\ell = 0\,$ everywhere; it can be readily verifed that there are
altogether
$\,2 \times 9\,$ cubic cells with nonzero $\,\vec A^\ell\,$.  The number inside
each of the nine plaquettes (all located at $\,z = 0\,$) in Figure 2 denotes
the
corresponding integral
$\,\ell^{-1} \int B_z^1\,dx dy\,$ over the plaquette.
\vskip24pt\noindent {\it 4.2.  Lattice Variables in the Coordinate Space}
\vskip6pt
In the Bloch wave-number space, the generalized coordinates $\,q_a^\ell(\vec
K)\,$ in the expansion (4.10) satisfy

$$q_a^\ell(\vec K)\;=\;q_a^\ell(- \vec K)^\dagger\eqno(4.18)$$
\vskip4pt\noindent
and the transversality condition (4.12).  At a given superscript $\,\ell\,$ and
for each pair $\,\pm \vec K \ne 0\,$, we can construct four independent
Hermitian variables from the two combinations
$$Q_t^\ell(\vec K)\;\pm\;Q_t^\ell(\vec K)^\dagger\eqno(4.19)$$
\noindent
each with the index $\,t\,$ = 1 or 2, where
$$Q_t^\ell(\vec K)\;\equiv\;\sum_{a=1}^3\,\hat\epsilon_t(\vec
K)_a\,q_a^\ell(\vec K)\eqno(4.20)$$
\vskip4pt\noindent
and $\,\hat\epsilon_t(\vec K)\,$ given by (3.47).  Thus, for a finite lattice
of
$\,{\cal N} = N^3\,$ unit cubic cells with periodic boundary conditions, there
are $\,2({\cal N} - 1)\,$ independent Hermitian variables for $\,\vec K \ne
0\,$, and for $\,\vec K = 0\,$

$$3\;\;{\rm constant}\;\;A_a^\ell\;\;{\rm solutions}\,, \eqno(4.21)$$\noindent
making a total of
$$2\,{\cal N} + 1\eqno(4.22)$$
\vskip4pt\noindent
independent Hermitian variables in the Coulomb gauge, all explicitly exhibited.
Here, as well as throughout this and the next subsections, we keep the group
superscript
$\,\ell\,$ fixed; therefore in the counting of degrees of freedom we do not
include the trivial factor $\,n^2 -1\,$ due to the $\,SU(n)\,$ generators.
\vskip6pt
However, the same problem appears in a different form, if one wishes to use
 the coordinate space lattice variables.  We
note that the expansion (4.14) employs
$\,3\,{\cal N}\,$ Hermitian variables
$$\{\,a_j^\ell, b_j^\ell, c_j^\ell\,\}\,.\eqno(4.23)$$
\vskip4pt\noindent
Because of (4.15), there are $\,{\cal N} - 1\,$ constraints (in which the
subtraction of 1 is on account of the sum of (4.15) over all $\,j\,$ being 0
identically); this leads to the same total number $\,3\,{\cal N} - ({\cal
N} - 1) = 2{\cal N} + 1\,$ independent variables, given by (4.22).
\vskip4pt
Excluding the three constant $\,\vec K = 0\,$ solutions (mentioned in (4.21)),
we may write
$$\vec A^\ell\;=\;\vec\nabla \times \vec I\,^\ell \eqno(4.24)$$
\noindent
where the spatial components of $\,\vec I\,^\ell\,$ are given by

$$I_x^\ell\;=\;\ell\;\sum_j\,\xi_j^\ell\;\Delta(x - j_1\ell)\;S(y -
j_2\ell)\;S(z - j_3\ell)\,,$$
$$I_y^\ell\;=\;\ell\;\sum_j\,\eta_j^\ell\;S(x - j_1\ell)\;\Delta(y -
j_2\ell)\;S(z - j_3\ell)\,,\eqno(4.25)$$
$$I_z^\ell\;=\;\ell\;\sum_j\,\zeta_j^\ell\;S(x - j_1\ell)\;S(y -
j_2\ell)\;\Delta(z - j_3\ell)\,,$$
\vskip4pt\noindent
with the factor $\,\ell\,$ denoting the lattice spacing, as before.  Clearly,
$\,\vec A^\ell\,$ and, therefore, also (4.23) are invariant under

$$\vec I\,^\ell\;\to\;\vec I\,^\ell + \vec\nabla\,\chi^\ell\,.
\eqno(4.26)$$\noindent
For a lattice interpretation of these continuum relations, we specify

$$\chi^\ell\;=\;\ell^2\,\sum_j\;\chi_j^\ell\;S(x - j_1\ell)\;S(y -
j_2\ell)\;S(z - j_3\ell)\,.\eqno(4.27)$$\noindent
Because of (3.41), (4.24) can be expressed in terms of the discrete variables:

$$a_j^\ell\;=\;\zeta_j^\ell - \zeta_{j''_-}^\ell - \eta_j^\ell +
\eta_{j'''_-}^\ell\,,$$
$$b_j^\ell\;=\;\xi_j^\ell - \xi_{j'''_-}^\ell - \zeta_j^\ell +
\zeta_{j'_-}^\ell\,,\eqno(4.28)$$
$$c_j^\ell\;=\;\eta_j^\ell - \eta_{j'_-}^\ell - \xi_j^\ell +
\xi_{j''_-}^\ell\,;$$\noindent
similarly, the transformation (4.26) takes the discrete form

$$\xi_j^\ell\;\to\;\xi_j^\ell + \chi_j^\ell - \chi_{j'_-}^\ell\,,$$
$$\eta_j^\ell\;\to\;\eta_j^\ell + \chi_j^\ell -
\chi_{j''_-}^\ell\,,\eqno(4.29)$$
$$\zeta_j^\ell\;\to\;\zeta_j^\ell + \chi_j^\ell - \chi_{j'''_-}^\ell\,,$$
\vskip4pt\noindent
where the lattice sites $\,j,\; j_-',\;j_-''\,$ and $\,j_-'''\,$ are given by
(4.16).
\vskip6pt
As remarked before, we may identify $\,a_j^\ell,\,b_j^\ell\,$ and
$\,c_j^\ell\,$ as ``link-fluxes'' which flow along the three links connecting
the site $\,j\,$ to sites $\,j',\,j''\,$ and $\,j'''\,$, given by (3.18).
Consider the cubic cell in which $\,x - j_1\ell,\, y- j_2\ell\,$ and $\,z -
j_3\ell\,$ are all between 0 and $\,\ell\,$.  Label its plaquette shared by
links $\,\overline{jj}''\,$ and  $\,\overline{jj}'''\,$ as $\,X_j\,$; likewise
the plaquette $\,Y_j\,$ is shared by links $\,\overline{jj}'''\,$ and
$\,\overline{jj}'\,$, and the plaquette $\,Z_j\,$ by $\,\overline{jj}'\,$ and
$\,\overline{jj}''\,$, as shown in Figure 3(a).  Identify $\,\xi_j^\ell\,$,
$\,\eta_j^\ell\,$ and $\,\zeta_j^\ell\,$ as the circulating
``plaquette-fluxes''
which flow counterclockwise along the edges of plaquettes $\,X_j\,, Y_j\,$ and
$\,Z_j\,$, as illustrated by Figure 3(b).  (For example, $\,\xi_j^\ell\,$
resembles the current loop represented by a magnetic moment pointing in the
$\,x$-direction, that is normal to $\,X_j\,$. The equation $\,\vec A = \vec
\nabla \times \vec I\,$ resembles $\,\vec J = \vec \nabla \times \vec M\,$
from electromagnetism.)
\vskip6pt
The continuum relation $\,\vec A^\ell = \vec\nabla \times \vec I^\ell\,$
implies $\,\vec \nabla \cdot \vec A^\ell = 0\,$.  The discrete expression
(4.28)
carries the same implication: Each link is shared by four plaquettes; through
superposition of these plaquette-fluxes, the net flux along each link is given
by (4.28), which automatically satisfies Kirchhoff's law (4.15). To summarize:
in accordance with (4.21)-(4.22), there are $\,2{\cal N} + 1\,$ independent
link-flux variables $\,\{ a_j^\ell, b_j^\ell, c_j^\ell \}\,$; excluding the
three constant $\,A_a^\ell\,$ solutions, the remaining $\,2{\cal N} + 1 - 3 =
2{\cal N} - 2\,$ independent link-flux variables can all be generated
by the plaquette-fluxes through (4.28).  In
other words, among the $\,3{\cal N}\,$ plaquette-flux variables
$\,\{\,\xi_j^\ell,\,\eta_j^\ell,\,\zeta_j^\ell\,\}\,$, there are
$$3{\cal N} - (2{\cal N} - 2)\;=\;{\cal N} + 2 \eqno(4.30)$$
\noindent
redundant ones that imply no net link-flux.  (See Section 4.3 below.)  The
problem of eliminating the redundant variables can then be shifted from
link-fluxes to plaquette-fluxes.
\vskip8pt
In the lattice formulation of QCD presented in this section, there is a precise
connection relating the continuum equations (4.1), (4.24) and (4.26) with their
discrete counterparts given by (4.15), (4.28) and (4.29).  The discrete
realizations can be further developed through the interplay between the lattice
and its dual structure as follows:
\vskip2pt
In the dual lattice, let links
$\,\overline{x}_j,\,\overline{y}_j,\,\overline{z}_j\,$ be duals to plaquettes
$\,X_j,\,Y_j,\,Z_j\,$ in the original lattice.  The plaquette-fluxes
$\,\xi_j^\ell,\,\eta_j^\ell,\,$ and $\,\zeta_j^\ell\,$ can also be regarded as
``currents'' in the dual lattice, all flowing along the respective links.  In
the same dual lattice, assign to each lattice site $\,j\,$ a ``potential''
$\,\chi_j^\ell\,$; transformation (4.29) generates a corresponding current
change along each link by an amount equal to the potential difference between
its two end points.  Such a change does not alter the link-fluxes
$\,(a_j^\ell,\,b_j^\ell,\,c_j^\ell)\,$ in the original lattice.  These extra
degrees of freedom enable us to eliminate $\,{\cal N} - 1\,$ of the $\,{\cal N}
+ 2\,$ redundant variables referred to in (4.30); the three remaining redundant
degrees of freedom may be represented by adding arbitrary constants to
$\,\xi_j^\ell,\,\eta_j^\ell\,$ and $\,\zeta_j^\ell\,$, as we shall see.
\vskip24pt\noindent {\it 4.3.  Elimination of Redundant Plaquette-Flux
Variables}
\vskip6pt
The lattice variables in the Bloch wave number space can be readily expressed
in
terms of the  $\,2{\cal N} + 1\,$ physical momentum-related ones, given by
(4.20)-(4.22).  However, for applications to confined configurations, such as
glueballs, etc., the use of lattice coordinate space variables gives a more
direct physical description.  In this section, we give an explicit procedure
for the elimination of  $\,{\cal N} + 2\,$ redundant variables among the
 $\,3{\cal N}\,$ plaquette-fluxes
$\,\xi_j^\ell,\,\eta_j^\ell\,$ and $\,\zeta_j^\ell\,$, in accordance with
(4.30).  Again, the group superscript $\,\ell\,$ is kept fixed throughout this
section.
\vskip6pt
In the dual lattice construct a ``tree'' called $\,T\,$, by connecting
arbitrarily $\,{\cal N} - 1\,$ links (without any closed cycles).  Start from
one of the end lattice sites of $\,T\,$ and label it 0.  The tree connects 0 to
another lattice site, say 1, through one of its links.  Continue this way to
move along the tree through all its links.  Since each new link leads to a new
lattice site, the
$\,{\cal N} - 1\,$ links of $\,T\,$ must lead from the site 0 to $\,{\cal N} -
1\,$ other lattice sites; i.e., the tree $\,T\,$ connects all $\,{\cal N}\,$
sites in the dual lattice.
\vskip6pt
Embed the dual lattice $\,L_D\,$ within the original cubic lattice $\,L\,$;
each
of the links in $\,L_D\,$ penetrates through a plaquette in $\,L\,$.  Assign
arbitrarily a distribution of  $\,3{\cal N}\,$ plaquette-fluxes
($\,\xi_j^\ell,\,\eta_j^\ell,\,$ and $\,\zeta_j^\ell\,$) in $\,L\,$, each
corresponding to a link-current in $\,L_D\,$.  Again, start from the end site 0
of
$\,T\,$ in $\,L_D\,$; by assigning the potential difference
$\,\chi_0^\ell - \chi_1^\ell\,$ between the sites 0 and 1 to be the negative of
the link current along $\,\overline{0\,1}\,$, we can change that link-current
(and therefore also the corresponding plaquette-flux in $\,L\,$) to zero
through
transformation (4.29).  Proceed the same way along the tree $\,T\,$; since
there
are $\,{\cal N} - 1\,$ arbitrary potential differences between its $\,{\cal
N}\,$
lattice sites, we can transform all the link-currents along $\,T\,$ to become
zero in $\,L_D\,$; i.e., all the corresponding
$${\cal N} - 1 \eqno(4.31)$$
\vskip4pt\noindent
plaquette-fluxes in $\,L\,$ can be set to zero through (4.29), without
affecting
any of the link-fluxes $\,(a_j^\ell,\,b_j^\ell,\,c_j^\ell)\,$.  Since the
lattice sites in $\,L_D\,$ are all located at the centers of the cubic cells in
$\,L\,$, the entire tree $\,T\,$ does not touch the surfaces of the original
lattice $\,L\,$.  In other words, none of the $\,{\cal N} - 1\,$
plaquette-fluxes that have been set to zero lie on the surfaces of $\,L\,$.
\vskip6pt
Without the boundary conditions, the entire cubic lattice $\,L\,$ would have
six
square surfaces.  Take any one of these surfaces, say $\,S\,$; it would have
four edges.  Assign an overall plaquette-flux $\,I^\ell\,$ flowing along the
edges of $\,S\,$; because of periodic boundary conditions, we see that the
edges facing each other become one, but with   $\,I^\ell\,$ flowing
along opposite directions, therefore cancelling each other.  Furthermore,
the six surfaces of $\,L\,$ reduce to three different ones.  An overall
plaquette-flux on any of these three surfaces produces no link-fluxes.  Thus,
from the tree $\,T\,$, we can add three links in the dual lattice $\,L_D\,$,
with each of the three different surfaces of $\,L\,$ penetrated once: Assign
zero also to the link-currents along these three additional links in $\,L_D\,$,
therefore making the three corresponding plaquette-fluxes in $\,L\,$ also
zero.  (This is a specific way of eliminating the three remaining redundant
degrees of freedom by adding constants to $\,\xi_j^\ell,\,\eta_j^\ell\,$ and
$\,\zeta_j^\ell\,$, mentioned at the end of Section 4.2.) Together with (4.31),
we have succeeded in setting   $\,{\cal N} - 1 + 3 = {\cal N} + 2\,$
plaquette-fluxes to zero, without affecting the link-flux distributions in
$\,L\,$.  This then removes the redundancy mentioned in (4.30).  The remaining
$\,2{\cal N} - 2\,$ nonzero plaquette-fluxes
($\,\xi_j^\ell,\,\eta_j^\ell,\,\zeta_j^\ell\,$) and the three constant
$\,A_a^\ell\,$ of (4.21) give the  $\,2{\cal N} + 1\,$ independent lattice
coordinate space variables.
\vskip6pt
An alternative procedure to eliminate the redundant lattice variables in the
coordinate space is given in Appendix B.
\vfill\eject
\noindent
\centerline {\bf 5.  LATTICE COUPLING CONSTANT}
\vskip12pt\noindent
{\it 5.1.  Lattice QCD Hamiltonian}
\vskip6pt
For the continuum QCD, the Hamiltonian $\,{\cal H}\,$, (4.3)-(4.4) in the
Coulomb gauge, is written in terms of
$\,\vec{\cal A}^\ell(\vec r, t)\,$ and its conjugate momentum $\,\vec{\cal
P}^\ell(\vec r, t)\,$, which satisfy (4.1) and (4.2).  In our lattice
formulation,  $\,\vec{\cal A}^\ell(\vec r, t)\,$ and $\,\vec{\cal P}^\ell(\vec
r, t)\,$ are replaced by  $\,\vec A^\ell(\vec r, t)\,$ and $\,\vec
\Pi^\ell(\vec r, t)\,$, given by (4.10)-(4.11), which denote the same field
operators, but restricted to the $\,0^{th}$-band Bloch functions.  At a given
lattice size $\,\ell\,$ and for a finite total volume $\,\Omega =
(N\ell)^3\,$, the system has only a finite number of variables, as discussed
in the previous section.  Correspondingly, the continuum Hamiltonian $\,{\cal
H}\,$ reduces to the lattice Hamiltonian

$$H\;=\;{1\over 2}\,\int\,[\,J^{-1}\,\Pi_a^\ell\,
J\,\Pi_a^\ell + B_a^\ell\, B_a^\ell\,]\;d^3r + H_{\rm
Coul} \eqno(5.1)$$
\noindent where, in place of (4.4)-(4.8), the lattice Coulomb interaction is

$$\eqalign{H_{\rm Coul}\;=\;{1\over 2}\;g_\ell^2\,
J^{-1}\,\int\,&\sigma_L^\ell(\vec r\,)\,(\ell, \vec r\;|\,(\nabla_a\,
D_a)^{-1} ( -
\nabla^2)\,(\nabla_b\,D_b)^{-1}\,|\,\ell', \vec r\,')\,\cr &\cdot\,
J\,\sigma_L^{\ell'}(\vec r\,')\,d^3r\,d^3r'\,,\cr}\eqno(5.2)$$
\vskip4pt\noindent
with the charge density now being
$$\sigma_L^\ell\;=\;f^{\ell mn}\,A_a^m\,\Pi_a^n\,, \eqno(5.3)$$
\noindent the color magnetic field $\, B_a^\ell\,$  given by

$$\epsilon_{abc}\,B_c^\ell\;=\;\nabla_a\,A_b^\ell -
\nabla_b\,A_a^\ell + g_\ell\,f^{\ell mn}\,A_a^m\,
A_b^n\,,\eqno(5.4)$$
\vskip4pt\noindent
$\,D_a\,$ denoting the covariant derivative matrix whose elements are

$$D_a^{\ell m}\;=\;\delta^{\ell m}\,\nabla_a - g_\ell\,f^{\ell mn}\,
A_a^n\,,
\eqno(5.5)$$
\vskip4pt\noindent
and $\,J\,$  the $\,0^{th}$-band Faddeev-Popov determinant (Jacobian)

$$J\;=\;{\rm det}\;|\,\nabla_a\,D_a\,|\,. \eqno(5.6)$$
\vskip4pt\noindent
As in (4.1)-(4.8), all superscripts $\,\ell, m, n\,$ refer to the group
indices,
from 1 to $\,n^2 - 1\,$ for $\,SU(n)\,$ and all repeated indices are summed
over.  (Notice that the {\it subscript} $\,\ell\,$ in $\,g_\ell\,$ denotes
``lattice''.)
\vskip6pt
The relation between the continuum and the lattice Hamiltonians $\,{\cal H}\,$
and $\,H\,$ may also be summarized as follows: Regard the continuum QCD
Hamiltonian $\,{\cal H}\,$ as a functional $\,{\cal F}\,$ of $\,\vec {\cal
A}^\ell\,,\vec {\cal P}^\ell\,$ and the bare coupling $\,g_0\,$.  We write

$${\cal H}\;=\;{\cal F}\,(\vec {\cal A}^\ell, \vec {\cal P}^\ell, g_0)
\eqno(5.7)$$
\vskip4pt\noindent
in accordance with (4.3)-(4.8).  The lattice QCD Hamiltonian $\,H\,$ denotes
the same functional $\,{\cal F}\,$, but with $\,\vec {\cal A}^a, \vec {\cal
P}^a\,$ and $\,g_0\,$ replaced by $\,\vec A^a, \vec \Pi^a\,$ and
$\,g_\ell\,$; i.e.,

$$H\;=\;{\cal F}\,(\vec A^a, \vec \Pi^a, g_\ell)\,. \eqno(5.8)$$
\vskip4pt\noindent
Since the continuum QCD action is noncompact, so is the lattice QCD action.
Because $\,H\,$ consists of only a
finite degree of freedom, the lattice coupling constant $\,g_\ell\,$
does not have to be renormalized, unlike (4.9) for the continuum case.  In the
following, we shall examine the relation between the lattice coupling
$\,g_\ell\,$, the continuum bare coupling  $\,g_0\,$ and its renormalization.
\vskip20pt\noindent  {\it 5.2.  Bare and Renormalized Continuum Coupling
Constants} \vskip6pt
A convenient and often used definition of the renormalized coupling in the
continuum case is to express it in terms of the interaction energy associated
with two external static color charges $\,e_1^\ell\,$ and $\,e_2^\ell\,$,
positioned at a distance $\,R\,$ apart:  Replace the continuum charge density
(4.5) by

$$\sigma^\ell(\vec r\,)\;=\;f^{\ell mn}\,{\cal A}_a^m(\vec r\,)\,{\cal
P}_a^n(\vec r\,) + \sigma_{\rm ext}^\ell(\vec r\,) \eqno(5.9)$$\noindent
where
$$\sigma_{\rm ext}^\ell(\vec r\,)\;=\;e_1^\ell\,\delta^3(\vec r - \vec
R_1) + e_2^\ell\,\delta^3(\vec r - \vec R_2)\,, \eqno(5.10)$$
\vskip4pt\noindent
with $\,e_1^\ell\,$ and $\,e_2^\ell\,$ located at $\,\vec R_1\,$ and $\,\vec
R_2\,$, and both being $\,(n^2 - 1)$-dimensional vectors in the group space.
Let
$\,E_{12}\,$ denote the part of the energy that is {\it proportional} to the
product
$\,e_1^\ell\,$ times
$\,e_2^\ell\,$.  Write

$$E_{12}\;\equiv\;{g_R^2\over 4\pi}\;{e_1^\ell\,e_2^\ell\over
R} \eqno(5.11)$$
\vskip4pt\noindent
where $\,R = |\,\vec R_1 - \vec R_2\,|\,$.  At a given $\,R\,$, the coefficient
$\,g_R^2\,$ can serve as the square of the {\it renormalized} coupling.  In
perturbation series$^8$ (assuming $\,g_R^2 << 1\,$)

$$g_R^2\;=\;g_0^2\,\left[ 1 + {11ng_0^2\over 24\pi^2}\;\left(ln\, \Lambda R
+ \gamma\right)\,\right] + O(g_0^6) \eqno(5.12)$$
\vskip4pt\noindent
where $\,\Lambda\,$ is the ultraviolet momentum cutoff and $\,\gamma =
0.5772157\,$ is the Euler constant.  (The reason for separating out the
constant $\,\gamma\,$ from
$\,ln\,\Lambda R\,$ is connected with (5.64) below.) The inverse relation of
(5.12) may be written as

$$g_0^2\;=\;g_R^2\,\left[ 1 + {11ng_R^2\over 24\pi^2}\;(ln \,\Lambda R +
\gamma) \right]^{-1}  + O(g_R^6)\,. \eqno(5.13)$$
\vskip4pt
Thus, at two distances $\,R\,$ and $\,R'\,$, to the same perturbative order the
corresponding renormalized couplings $\,g_R\,$ and $\,g_{R'}\,$ satisfy the
familiar asymptotic freedom relation$^8$

$${1\over g_{R'}^2} - {1\over g_R^2}\;=\;{11n\over 24\pi^2} \; ln
\,R/{R'}\,.
\eqno(5.14)$$
\vskip4pt\noindent
It follows then, for $\,R' < R\,$,  $\,g_{R'}^2 < g_R^2\,$; in
particular, when
$\,R'/R \to 0\,$, $\,g_{R'}^2 \to 0\,$.
\vskip20pt\noindent  {\it 5.3.  Lattice Coupling $\,g_\ell\,$ As the
Renormalized Continuum Coupling Constant}
\vskip6pt
Next, we consider the lattice Hamiltonian  (5.1)-(5.2).
Replace the lattice charge density (5.3) by

$$\sigma_L^\ell(\vec r\,)\;=\;f^{\ell mn}\,A_a^m(\vec r\,)\,\Pi_a^n(\vec r\,)
+ \sigma_{\rm ext}^\ell(\vec r\,) \eqno(5.15)$$
\vskip4pt\noindent
with $\,\sigma_{\rm ext}^\ell(\vec r\,)\,$ given by the external charge
density identical to (5.10), and then evaluate the same interaction energy
$\,E_{12}\,$ that is proportional to $\,e_1^\ell\,$ times
$\,e_2^\ell\,$.  Write the lattice result as

$$E_{12}\;=\;{(g_R^2)_L\over 4\pi}\;{e_1^\ell\,e_2^\ell\over R}
\eqno(5.16)$$
\vskip4pt\noindent
where, unlike (5.11)-(5.12), $\,(g_R^2)_L\,$ is a finite function of the
lattice coupling $\,g_\ell^2\,$.  Because $\,H\,$ does not have the symmetry of
the continuum, but only that of a cubic lattice, $\,(g_R^2)_L\,$ depends also
on $\,\vec R_1, \vec R_2\,$ and the lattice size $\,\ell\,$.  We may select
any $\,\vec R_1\,$ and  $\,\vec R_2\,$, and set
$$(g_R^2)_L\;=\;g_R^2\,; \eqno(5.17)$$
\vskip4pt\noindent
i.e., we choose the lattice coupling $\,g_\ell\,$ so that the interaction
energy $\,E_{12}(\vec R_1, \vec R_2)\,$ calculated from the continuum
Hamiltonian
$\,{\cal H}\,$ and that from its $\,0^{th}$-band approximation $\,H\,$ are
{\it exactly\ equal} for a particular pair of position vectors  $\,\vec R_1\,$
and $\,\vec R_2\,$.  Because of the cubic lattice symmetry, this equality
extends to all pairs of $\,\vec R_1\,, \vec R_2\,$ which can be reached from
the selected ones through a lattice translation and a cubic rotation.
\vskip4pt
The precise meaning of this equality can be stated in an alternative way:
Express the continuum coupling constant renormalization formula (4.9) in the
form

$$g_0\;=\;g_\ell + \delta g\,; \eqno(5.18)$$
\vskip4pt\noindent
i.e., instead of using (5.13), regard the lattice coupling $\,g_\ell\,$ as
the {\it renormalized} coupling for the {\it continuum case}.  Expand $\,\vec
{\cal A}^\ell\,$ and $\,\vec {\cal P}^\ell\,$ in terms of the complete set of
Bloch functions $\,\{\,\vec F_n^t(\vec K\,|\,\vec r\,)\,\}\,$, given by
(3.72)-(3.73).  Decompose

$$\vec{\cal A}^\ell\;=\;\vec A^\ell + \delta\vec{\cal A}^\ell \hskip2em {\rm
and} \hskip2em \vec{\cal P}^\ell\;=\;\vec \Pi^\ell + \delta\vec{\cal P}^\ell
\eqno(5.19)$$
\vskip4pt\noindent
where $\,\delta\vec{\cal A}^\ell\,$ and $\,\delta\vec{\cal P}^\ell\,$ consist
of all terms that depend on $\,\vec n \ne 0$-band Bloch functions.
Substituting (5.18)-(5.19) into the functional (5.7), we can express the
continuum
$\,{\cal H}\,$ as a sum

$${\cal H}\;=\;H + \delta{\cal H}\,, \eqno(5.20)$$
\vskip4pt\noindent
where $\,\delta{\cal H}\,$ contains all terms that depend on
$\,\delta\vec{\cal A}^\ell\,$, $\,\delta\vec{\cal P}^\ell\,$ and $\,\delta
g\,$.  The condition for determining $\,\delta g\,$ is: For $\,e_1^\ell\,$ and
$\,e_2^\ell\,$ located at the selected pair of position vectors $\,\vec R_1\,$
and $\,\vec R_2\,$, the interaction energy
$\,E_{12}\,$ calculated from $\,{\cal H}\,$ equals that using
$\,H\,$; i.e., $\,\delta{\cal H}\,$ gives no correction.  At different $\,\vec
R_1\,$ and $\,\vec R_2\,$ there will be, in general, a correction due to
$\,\delta{\cal H}\,$.
\vskip4pt
Because of asymptotic freedom, there is a wide range of
$\,R\,$ and lattice size $\,\ell\,$, for which the equality (5.17) can lead to
$\,g_\ell^2 << 1\,$, in which case corrections due to $\,\delta{\cal H}\,$ can
be computed perturbatively in powers of $\,g_\ell^2\,$.   In
the present noncompact lattice formulation, there is {\it no}  need to take the
limit
$\,\ell \to 0\,$, since the continuum solution is independent of the
lattice size $\,\ell\,$.
\vskip4pt
In the following, we assume that $\,g_\ell^2\,$ is indeed small; this enables
us to relate perturbatively $\,g_\ell^2\,$ with $\,g_R^2\,$, defined by
(5.11).  It is convenient to write

$$g_R^2\;=\;g_\ell^2\,\left[ 1 + {11n\over
24\pi^2}\;g_\ell^2\,\left(ln\,{R\over
\ell} + \gamma + \lambda + \delta \right) \right] + O(g_\ell^6)
\eqno(5.21)$$\noindent
 where, besides the Euler constant $\,\gamma\,$, there is another numerical
constant $\,\lambda\,$ and a function
$$\delta\;=\;\delta\,(\ell, \vec R_1, \vec R_2) \eqno(5.22)$$\noindent
which satisfies
$$\lim_{\ell \to 0}\;\delta\,(\ell, \vec R_1, \vec R_2)\;=\;0\,.\eqno(5.23)$$
\noindent
Accordingly, (5.18) becomes

$$g_0^2\;=\;g_\ell^2 - {11n\over 24\pi^2}\;g_\ell^4(ln\, \Lambda \ell -
\lambda -
\delta) + O(g_\ell^6)\,. \eqno(5.24)$$
\vskip4pt\noindent
In the rest of the paper, we shall evaluate the constant $\,\lambda\,$ and
derive the formula for $\,\delta\,$.  As we shall see,

$$\lambda\;=\;3.3559\,. \eqno(5.25)$$
\vskip4pt\noindent
The expression for $\,\delta\,$ is given in Appendix C (see (C.33)).
\vskip20pt\noindent
{\it 5.4.  Power Series Expansion of the Lattice Hamiltonian}
\vskip6pt
In the $\,0^{th}$-band approximation, the Hamiltonian (5.1)-(5.2) can be
expanded into a power series of the lattice coupling constant $\,g_\ell\,$:

$$H\;=\;H_0 + H_{\rm int} \eqno(5.26)$$\noindent
where $\,H_0\,$ is $\,g_\ell\,$ independent and

$$H_{\rm int}\;=\;g_\ell\,H_1 + g_\ell^2\,H_2 + g_\ell^3\,H_3 + g_\ell^4\,H_4
+ \cdots \,.\eqno(5.27)$$
\vskip4pt\noindent
The field $\,A_a^\ell(\vec r\,)\,$ and its conjugate momentum
$\,\Pi_a^\ell(\vec r\,)\,$ are given by (4.10) and (4.11).  At any given time
$\,t\,$, they can also be written in terms of the creation and annihilation
operators:

$$A_a^\ell(\vec r, t) = \sum_{\vec K, t} \;{1\over\sqrt {2\omega_t(\vec
K)}}\;\left[ a_t^\ell(\vec K)\,\hat\epsilon_t(\vec K)_a\,F_a(\vec K\,|\,\vec
r\,) + a_t^\ell(\vec K)^\dagger\,\hat\epsilon_t(\vec K)_a\,F_a(\vec K\,|\,\vec
r\,)^* \right]
\eqno(5.28)$$\noindent
and
$$\Pi_a^\ell(\vec r, t) = -i\sum_{\vec K, t} \sqrt{\omega_t(\vec K)\over
2} \left[ a_t^\ell(\vec K)\,\hat\epsilon_t(\vec K)_a\,F_a(\vec K\,|\,\vec r\,)
- a_t^\ell(\vec K)^\dagger\,\hat\epsilon_t(\vec K)_a\,F_a(\vec K\,|\,\vec
r\,)^* \right] \eqno(5.29)$$
\vskip4pt\noindent
where $\,\hat\epsilon_t(\vec K)\,$ and $\,\omega_t(\vec K)\,$, with $\,t = 1,
2\,$ (or $\,+\,, -\,$) are given by (3.47) and (3.56), $\,a_t^\ell(\vec K)\,$
and $\,a_t^\ell(\vec K)^\dagger\,$ satisfy the usual commutation relation

$$\left[\,a_t^\ell(\vec K), a_{t'}^{\ell'}(\vec
K')^\dagger\,\right]\;=\;\delta^{\ell\ell'}\,\delta_{tt'}\,\delta_{\vec K, \vec
K'}
\,. \eqno(5.30)$$
\vskip4pt\noindent
As before, $\,\vec K\,$ and $\,\vec K'\,$ are Bloch wave number vectors within
the Brillouin zone.  The operators $\,a_t^\ell(\vec K)\,$ and $\,a_t^\ell(\vec
K)^\dagger\,$ are related to  $\,q_a^\ell(\vec K)\,$ and  $\,p_a^\ell(\vec
K)\,$
of (4.10) and (4.11) by

$$q_a^\ell(\vec K)\;=\;\sum_t \;{1\over\sqrt {2\omega_t(\vec
K)}}\;\left[\,a_t^\ell(\vec K)\,\hat\epsilon_t(\vec K)_a + a_t^\ell(- \vec
K)^\dagger\,\hat\epsilon_t(- \vec K)_a\right]$$\noindent
\line {and \hfil (5.31)}
$$p_a^\ell(-\vec K)\;=\;- i \sum_t \;\sqrt{\omega_t(\vec K)\over
2}\;\left[\,a_t^\ell(\vec K)\,\hat\epsilon_t(\vec K)_a - a_t^\ell(- \vec
K)^\dagger\,\hat\epsilon_t(- \vec K)_a\,\right]\,.$$
\vskip4pt\noindent
In terms of these annihilation and creation operators, the zeroth-order
Hamiltonian in (5.27) becomes, apart from an additive constant which may be
set to zero,

$$H_0\;=\;\sum_{\vec K, t, \ell}\,\omega_t(\vec K)\,a_t^\ell(\vec
K)^\dagger\,a_t^\ell(\vec K)\,. \eqno(5.32)$$
\noindent
The eigenvalues of $\,H_0\,$ are

$$E_0(N)\;=\;\sum_{\vec K, t, \ell}\,N_{\vec K, t}^\ell\,\omega_t(\vec K)
\eqno(5.33)$$
\vskip4pt\noindent
where $\,N_{\vec K, t}^\ell = 0, 1, 2, \cdots\,$ is the eigenvalue of the
occupation-number operator $\,a_t^\ell(\vec K)^\dagger\,a_t^\ell(\vec K)\,$.
Let the corresponding eigenstates be $\,|\,N >\,$, with $\,|\,0 >\,$ denoting
the ground state of $\,H_0\,$. Thus, we have

$$a_t^\ell(\vec K)\,|\,0 >\;=\;0 \eqno(5.34)$$
\vskip4pt\noindent
and $\,E_0(0) = 0\,$.  To $\,O(g_\ell^2)\,$, the shift in the ground state
energy $\,E_0(\vec R_1, \vec R_2)\,$ due to $\,H_{\rm int}\,$ of (5.26)-(5.27)
is given by the familiar perturbation formula

$$E_0(\vec R_1, \vec R_2)\;=\;< 0\,|\,H_{\rm int}\,|\,0 > - \sum_{N\ne 0}\;{|<
N\,|\,H_{\rm int}\,|\,0 >|^2\over E_0(N)} + O(H_{\rm int}^3)\,; \eqno(5.35)$$
\vskip4pt\noindent
the interaction energy $\,E_{12}\,$, defined by (5.14), refers to the part of
$\,E_0(\vec R_1, \vec R_2)\,$ which is proportional to the product
$\,e_1^\ell\,e_2^\ell\,$ and corresponds to the following difference:

$$E_{12}\;=\;E_0(\vec R_1, \vec R_2) - \lim_{R \to \infty} E_0(\vec R_1, \vec
R_2)\eqno(5.36)$$\noindent
where $\,R = |\,\vec R_2 - \vec R_1\,|\,$, as before.
\vskip4pt
In the following, $\,E_{12}\,$ will be expanded in powers of $\,H_{\rm int}\,$:

$$E_{12}\;=\;E_{12}^{(i)} + E_{12}^{(ii)} + \cdots \eqno(5.37)$$
\vskip4pt\noindent
where $\,E_{12}^{(i)}\,$ is first-order in $\,H_{\rm int}\,$ and
$\,E_{12}^{(ii)}\,$ second order in $\,H_{\rm int}\,$.  As we shall see,
$\,E_{12}^{(i)}\,$ contains $\,O(g_\ell^2)\,$, $\,O(g_\ell^4)\,$ and higher
order terms; likewise $\,E_{12}^{(ii)}\,$ contains $\,O(g_\ell^4)\,$ and
higher-order terms.  In this paper, we shall calculate $\,E_{12}\,$ only up to
$\,O(g_\ell^4)\,$.
\vskip20pt\noindent
{\it 5.5.  First Order in $\,H_{\rm int}\,$}
\vskip6pt
To first order in $\,H_{\rm int}\,$, (5.35) reduces to

$$E_0(\vec R_1, \vec R_2)\;=\;< 0\,|\,H_{\rm int}\,|\,0 >\,, \eqno(5.38)$$
\vskip4pt\noindent
in which only the part proportional to $\,e_1^\ell\,e_2^\ell\,$ gives
$\,E_{12}^{(i)}\,$.  Substituting the power series expansion (5.27) into
(5.38),
we see that since all odd $\,g_\ell$-power terms carry odd numbers of gauge
field operators, whose vacuum expectation values are zero, the result is an
even function of $\,g_\ell\,$.  Because $\,E_{12}\,$ is proportional to
 $\,e_1^\ell\,e_2^\ell\,$, for our purpose we can equate $\,H_{\rm int} =
H_{\rm Coul}\,$, which is given by (5.2).  Up to $\,O(g_\ell^4)\,$, we can
also set the Jacobian $\,J = 1\,$.  In power series of $\,g_\ell\,$, we expand

$$\eqalign{(\nabla_a D_a)^{-1} (- \nabla^2) (\nabla_b D_b)^{-1}\;=&\;-
\nabla^{-2} + 2g_\ell\,\nabla^{-2}\,\nabla_a\,A_a\,\nabla^{-2}\cr
&-
3g_\ell^2\,\nabla^{-2}\,\nabla_a\,A_a\,\nabla^{-2}\,\nabla_b\,A_b\,\nabla^{-2}
+ \cdots\cr} \eqno(5.39)$$
\vskip4pt\noindent
where $\,A_a\,$ denotes the $\,(n^2 - 1) \times (n^2 - 1)\,$ matrix whose
$\,(\ell, \ell')\,$ element is related to the $\,0^{th}$-band gauge field
$\,A_a^m(\vec r\,)\,$ by

$$(A_a)^{\ell\ell'}\;=\;f^{\ell m \ell'}\,A_a^m\,.\eqno(5.40)$$\noindent
Hence, we find
$$E_{12}^{(i)}\;=\;g_\ell^2\;{e_1^\ell\,e_2^\ell\over 4\pi R} +
g_\ell^4\,{\cal E}^{(i)} + O(g_\ell^6) \eqno(5.41)$$\noindent
where
$${\cal E}^{(i)}\;=\;- 3\, e_1^\ell\,e_2^{\ell'}\;< 0\,|\,(\ell, \vec
R_1\,|\,\nabla^{-2}\,\nabla_a\,A_a\,\nabla^{-2}\,\nabla_b\,A_b\,\nabla^{-2}\,|\,\ell',
\vec R_2)\,|\,0 >\,. \eqno(5.42)$$
\vskip8pt
{}From (5.28) and (5.34), it follows that

$$< 0\,| A_a^\ell(\vec r\,)\,A_b^m(\vec r\,')\,|\,0 > = \delta^{\ell
m} \sum_{\vec K, t} \left[ 2\omega_t(\vec K) \right]^{-1} \hat\epsilon_t(\vec
K)_a\,\hat\epsilon_t(\vec K)_b\,F_a(\vec K\,|\,\vec r\,) F_b(\vec K\,|\,\vec
r\,')^* \eqno(5.43)$$
\vskip4pt\noindent
where, as before, $\,F_a(\vec K\,|\,\vec r\,)\,$ is the $\,0^{th}$-band Bloch
function, given by (3.38).  In (5.43), the repeated subscripts are not to be
summed over.  Using the familiar expressions

$$(\vec R_1\,|\,\nabla^{-2}\,\nabla_a\,|\,\vec r\,)\;=\;\int - {d^3p\over
(2\pi)^3}\;{ip_a\over p^2}\;e^{i\vec p \cdot (\vec R_1 - \vec r\,)}\,,$$
$$(\vec r\,'\,|\,\nabla_b \nabla^{-2}\,|\,\vec R_2\,)\;=\;\int -
{d^3p'\over (2\pi)^3}\;{ip_b'\over p'^2}\;e^{i\vec p\,' \cdot (\vec r\,' - \vec
R_2)}\eqno(5.44)$$\noindent
and
$$(\vec r\,|\,\nabla^{-2}\,|\,\vec r\,'\,)\;=\;\int - {d^3q\over
(2\pi)^3}\;{1\over q^2}\,e^{i\vec q\cdot (\vec r - \vec
r\,')}$$
\vskip4pt\noindent
with $\,p, p'\,$ and $\,q\,$ denoting the magnitudes of $\,\vec p, \vec p\,'\,$
and $\,\vec q\,$, and combining these with (5.42)-(5.43), we derive$^9$

$$\eqalign{{\cal E}^{(i)}\;=\;3n\,e_1^\ell\,e_2^\ell\,\sum \int
e^{i\phi}\;&{p_a\,p_b'\over p^2\,p'^2\,q^2}\;[ 2\omega_t(\vec
K)]^{-1}\;\hat\epsilon_t(\vec K)_a\,\hat\epsilon_t(\vec K)_b\cr
& \times F_a(\vec K\,|\,\vec r\,)\,F_b(\vec K\,|\,\vec r\,')^*\,d^3r\,d^3r'\cr}
\eqno(5.45)$$\noindent
where
$$\phi\;=\;\vec p \cdot (\vec r - \vec R_1) + \vec q \cdot (\vec r\,' - \vec
r\,) + \vec p\,'(\vec R_2 - \vec r\,') \eqno(5.46)$$
\vskip4pt\noindent
and the summation extends over the polarization index $\,t = 1, 2\,$, $\,\vec
K\,$ within the Brillouin zone, $\,\vec p, \vec p\,'\,$ and $\,\vec q\,$ over
{\it all}  wave numbers, and the subscripts $\,a, b = x, y\,$ and $\,z\,$.
\vskip6pt
Because $\,e^{-i\vec K \cdot \vec r}\,$ times the Bloch wave function
 $\,F_a(\vec K\,|\,\vec r\,)\,$ has the periodicity of the lattice, the
integrand in (5.45) consists of a periodic function in $\,\vec r\,$ and $\,\vec
r\,'\,$ multiplied by a phase factor $\,e^{i\phi + i \vec K \cdot (\vec r -
\vec r\,')}\,$.  Resolve the phase into a sum of terms

$$\phi + \vec K \cdot (\vec r - \vec r\,')\;=\;(\vec p - \vec q + \vec K)
\cdot \vec r - (\vec p\,' - \vec q + \vec K) \cdot \vec r\,' + (\vec p\,'
\cdot \vec R_2 - \vec p \cdot \vec R_1)\,. \eqno(5.47)$$
\noindent
The integration over $\,\vec r\,$ and $\,\vec r\,'\,$ gives
$$\vec p - \vec q + \vec K\;=\;2\pi\,\vec m_1/\ell$$
\line {and \hfil (5.48)}
$$\vec p\,' - \vec q + \vec K\;=\;2\pi\,\vec m_2/\ell$$
\vskip4pt\noindent
where $\,\vec m_1\,$ and $\,\vec m_2\,$ are three-dimensional vectors whose
components are all integers.  Introduce

$$\vec k\;=\;{1\over 2}\,(\vec p\,' + \vec p\,)\,,$$
$$\vec R_{\rm cm}\;=\;{1\over 2}\,(\vec R_2 + \vec R_1)\eqno(5.49)$$
\noindent and
$$\vec R\;=\;\vec R_2 - \vec R_1\,.$$
\noindent
The last term in (5.47) can be written as

$$\vec p\,' \cdot \vec R_2 - \vec p \cdot \vec R_1\;=\;{2\pi\over \ell}\;(\vec
m_2 - \vec m_1) \cdot \vec R_{\rm cm} + \vec k \cdot \vec R\,. \eqno(5.50)$$
\vskip4pt\noindent
Consequently, $\,{\cal E}^{(i)}\,$ is a periodic function in $\,\vec R_{\rm
cm}\,$ with the periodicity of the lattice cell.
\vskip4pt
By using (5.48)-(5.49), we can convert the sum over $\,\vec p, \vec p\,'\,$
and $\,\vec q\,$ into an integration over $\,\vec k\,$ times a sum over $\,\vec
m_1\,$ and $\,\vec m_2\,$; this leads to

$${\cal E}^{(i)}\;=\;3n\,e_1^\ell e_2^\ell \int {d^3k\over
(2\pi)^3}\;{e^{i\vec k \cdot \vec R}\over k^2}\; I(\vec k) \eqno(5.51)$$
\noindent
where $\,I(\vec k)\,$ is a dimensionless function, given by

$$I(\vec k)\;=\;\sum_{\vec m_1, \vec m_2} \int_B {d^3K\over (2\pi)^3}\;
f_{\vec m_1 \vec m_2}(\vec K, \vec k)\;e^{i\,{2\pi\over\ell}\,(\vec m_2 - \vec
m_1) \cdot \vec R_{\rm cm}} \eqno(5.52)$$
\vskip4pt\noindent
with the $\,\vec k$-integration over the entire continuum momentum space, but
the
$\,\vec K$-integration only within the Brillouin zone as indicated by the
subscript $\,B\,$,

$$\eqalign{f_{\vec m_1 \vec m_2}(\vec K, \vec k)\;=\;k^2
\sum_{t,a,b}&\;{\Omega^{1\over 2}\,{\cal C}_a(\vec\theta\,|\,- \vec
 m_1)\;\Omega^{1\over 2}\,{\cal C}_b(\vec\theta\,|\,- \vec m_2)
\hat\epsilon_t(\vec K)_a\,\hat\epsilon_t(\vec K)_b\over 2\omega_t(\vec
K)\,(\vec k + \vec K - {\vec m_1 + \vec m_2\over\ell}\;\pi)^2}\cr
& \times {(\vec k + {\vec m_1 - \vec m_2\over\ell}\;\pi)_a\;(\vec k - {\vec m_1
- \vec m_2\over\ell}\;\pi)_b\over (\vec k + {\vec m_1 - \vec
m_2\over\ell}\;\pi)^2\;(\vec k - {\vec m_1 - \vec
m_2\over\ell}\;\pi)^2}\,,\cr}\eqno(5.53)$$
\vskip4pt\noindent
and $\,\Omega^{1\over 2}\,{\cal C}_a(\vec\theta\,|\,\vec m)\,$ given by (3.67)
and (3.40), which is both dimensionless and independent of $\,\Omega\,$.  From
(5.52), one sees directly that $\,I(\vec k)\,$ and therefore $\,{\cal
E}^{(i)}\,$ are both periodic in $\,\vec R_{\rm cm}\,$.
\vskip6pt
To evaluate the contribution of $\,{\cal E}^{(i)}\,$ to the constant
$\,\lambda\,$ in (5.21), we examine the limiting case $\,\ell/R \to 0\,$ and
call the result $\,\lambda^{(i)}\,$.  The difference between the limiting
result
and that for an arbitrary nonzero $\,\ell/R\,$ gives the corresponding
contribution to the remainder $\,\delta(\ell, \vec R_1, \vec R_2)\,$.  When
$\,\ell/R \to 0\,$, the integral (5.51) is dominated by the region $\,k\ell =
O(R^{-1})\,$.  Furthermore, because of the rapid oscillatory phase factor
$\,{\rm exp}\,\left[\,i\;{2\pi\over \ell}\,(\vec m_1 - \vec m_2) \cdot \vec
R_{\rm cm}\,\right]\,$ in (5.52), we need only consider the special case

$$\vec m_1\;=\;\vec m_2\;=\;\vec m\,; \eqno(5.54)$$
\vskip4pt\noindent
as a result, $\,I(\vec k)\,$ becomes independent of $\,\vec R_{\rm cm}\,$.
{}From (5.53), one can readily verify that as $\,\vec k\,$ and $\,\vec K\,$
both
$\,\to 0\,$, the sum (5.52) over $\,\vec m\,$ is controlled by the single term
$\,\vec m_1 = \vec m_2 = 0\,$, for which the limiting behavior of the
corresponding $\,f_{\vec m_1 \vec m_2}(\vec K, \vec k)\,$ is given by the
continuum function $\,f_c\,$:

$$f_{00}(\vec K, \vec k)\;\to\;f_c(\vec K, \vec k)\;\equiv\;{1\over 2K(\vec k
+ \vec K)^2}\;\sum_t\;{[\,\vec k \cdot \hat\epsilon_t(\vec K)\,]^2\over k^2}
\eqno(5.55)$$
\vskip4pt\noindent
where $\,K\,$ and $\,k\,$ refer to the magnitudes of $\,\vec K\,$ and $\,\vec
k\,,$ as before, and the sum over the polarization index $\,t\,$ yields

$$\sum_t\,\hat\epsilon_t(\vec K)_a\,\hat\epsilon_t(\vec K)_b\;=\;\delta_{ab} -
{K_a K_b\over K^2}\,. \eqno(5.56)$$
\vskip6pt
To facilitate the integration (5.52), we decompose $\,I(\vec k)\,$ into a sum
of two terms:

$$I(\vec k)\;=\;I_c(\vec k) + I'(\vec k) \eqno(5.57)$$\noindent
where
$$I_c(\vec k)\;=\;\int_B {d^3K\over (2\pi)^3}\;f_c(\vec K, \vec k)
\eqno(5.58)$$\noindent
and, on account of (5.54),
$$I'(\vec k)\;=\;\int_B {d^3K\over (2\pi)^3}\;\left[\,\sum_m f_{\vec m, \vec
m}(\vec K, \vec k) - f_c(\vec K, \vec k)\,\right]\,.\eqno(5.59)$$
\vskip6pt\noindent
When $\,\ell/R \to 0\,$, the integration $\,I_c(\vec k)\,$ can be evaluated
analytically.  As shown in Appendix C, the result is
$$I_c(\vec k)\;=\;{1\over 6\pi^2}\,(- ln\,k\ell + \lambda_c) \eqno(5.60)$$
\noindent
in which the constant $\,\lambda_c\,$ is

$$\lambda_c\;=\;{4\over 3} + ln \,\pi - {6\over\pi} \int_0^1 dx\,{ln(x^2 +
2)\over (x^2 + 1) \sqrt {x^2 + 2}}\;=\;1.67040\,. \eqno(5.61)$$
\vskip4pt\noindent
The second term $\,I'(\vec k)\,$ in (5.57) is well-behaved at small $\,\vec
k\,$; its value at $\,\vec k = 0\,$ can be written as

$$I'(0)\;=\;{1\over 6\pi^2}\;\lambda' \eqno(5.62)$$\noindent
with the constant $\,\lambda'\,$ given by
$$\lambda'\;=\;- 0.1496\,.  \eqno(5.63)$$
\vskip8pt
By using

$$\int\;{d^3k\over (2\pi)^3}\;{1\over k^2}\;(ln\, k\ell)\;e^{i\vec k \cdot \vec
R}\;=\;- {1\over 4\pi R}\;[\,(ln\, R/\ell) + \gamma\,]\,,
\eqno(5.64)$$\noindent
we derive, as $\,\ell/R \to 0\,$,

$${\cal E}^{(i)}\;=\;{e_1^\ell\,e_2^\ell\over 4\pi\,R}\,\left( {n\over
2\pi^2}\right)\,[\,(ln\,R/\ell) + \gamma + \lambda^{(i)}\,] \eqno(5.65)$$
\noindent
where $\,\gamma\,$ is the Euler constant and

$$\lambda^{(i)}\;=\; \lambda_c +  \lambda'\;=\;1.5208\,. \eqno(5.66)$$
\vskip20pt\noindent
{\it 5.6.  Second Order in $\,H_{\rm int}\,$}
\vskip6pt
In accordance with (5.35)-(5.37),

$$E_{12}^{(ii)}\;=\;- \sum_{N \ne 0}\;{|\,< 0\,|\,H_{\rm int}\,|\,N >\,|^2\over
E_0(N)} + \lim_{R \to \infty} \sum_{N \ne 0}\;{|\,< 0\,|\,H_{\rm int}\,|\,N
>\,|^2\over E_0(N)} \,, \eqno(5.67)$$
\vskip4pt\noindent
in which the relevant states $\,|\,N >\,$ are those of two gauge quanta, i.e.
$\,|\,N > = |\,2 >\,$ with

$$ |\,2 >\;=\;a_{t_1}^{\ell_1}(\vec K_1)^\dagger\;a_{t_2}^{\ell_2}(\vec
K_2)^\dagger\,|\,0 > \,. \eqno(5.68)$$
\vskip4pt\noindent
To $\,O(g_\ell^4)\,$, we can write, similar to (5.41),

$$E_{12}^{(ii)}\;=\;g_\ell^4\,{\cal E}^{(ii)} + O(g_\ell^6) \eqno(5.69)$$
\vskip4pt\noindent
and set, in (5.67), $\,H_{\rm int} = H_{\rm Coul}\,$ with $\,J = 1\,$ and
$\,D_a = \nabla_a\,$, so that the relevant part of $\,H_{\rm Coul}\,$ is

$$H_{\rm Coul}\;=\;g_\ell^2 \int d^3 r \int  d^3
r\,'\;\sigma_A^\ell(\vec r\,)\,\sigma_{\rm ext}^\ell(\vec r\,')\,(\vec r\,|\,-
\nabla^{-2}\,|\,\vec r\,') \eqno(5.70)$$
\vskip4pt\noindent
where $\,\sigma_A^\ell(\vec r\,) = f^{\ell mn}\,A_a^m(\vec r\,)\,\Pi_a^n(\vec
r\,)\,$ is the first term in (5.15).  By using (5.28)-(5.29), (5.68)-(5.69)
and the last equation in (5.44), we obtain

$$< 2\,|\,H_{\rm Coul}\,|\,0 >\;=\;g_\ell^2\,f^{\ell \ell_1 \ell_2} \sum
M_{t_1 t_2}(\vec q; \vec K_1, \vec K_2)\,q^{-2}\, (e_1^\ell\,e^{-i\vec q \cdot
\vec R_1} + e_2^\ell\,e^{-i\vec q \cdot\vec R_2}) \eqno(5.71)$$
\vskip4pt\noindent
where $\,\ell_1, t_1\,$ and $\,\ell_2, t_2\,$ denote the group and
polarization indices of the two-gauge quanta in $\,|\,2 >\,$, $\,\vec q\,$ is
the momentum variable in (5.44) for $\,\nabla^{-2}\,$, which is related to
the Bloch wave numbers $\,\vec K_1\,$ and $\,\vec K_2\,$ of the two quanta by

$$\vec q\;=\;\vec K_1 + \vec K_2 + {2\pi\over\ell}\;\vec m\,, \eqno(5.72)$$
\noindent
with the components of $\,\vec m\,$ all integers.  The factor $\,M_{t_1
t_2}\,$ is given by

$$\eqalign{M_{t_1 t_2}(\vec q; \vec K_1, \vec K_2)\;=\;&{i\,[\omega_{t_2}(\vec
K_2) - \omega_{t_1}(\vec K_1) ]\over 2\sqrt {\omega_{t_1}(\vec K_1)\;
\omega_{t_2}(\vec K_2)}}\;\sum_a \hat\epsilon_{t_1}(\vec
K_1)_a\,\hat\epsilon_{t_2}(\vec K_2)_a\cr
&\times \int d^3r\,F_a(\vec K_1\,|\,\vec r\,)^*\,F_a(\vec K_2\,|\,\vec
r\,)^*\,e^{i\vec q \cdot \vec r}\,.\cr} \eqno(5.73)$$
\noindent
We observe that when $\,t_1 = t_2 = t\,$, as $\,\vec K_1 + \vec K_2 \to 0\,$

$$M_{t_1 t_2}(\vec q; \vec K_1, \vec K_2)\;\to\;0 \eqno(5.74)$$
\vskip4pt\noindent
for all $\,\vec m\,$ including 0, since in this case $\,\omega_{t_2}(\vec K_2)
- \omega_{t_1}(\vec K_1) = \omega_t(\vec K_2) - \omega_t(\vec K_1) \to 0\,$.
Also, when $\,t_1 \ne t_2\,$ but $\,\vec m = 0\,$, as $\,\vec K_1 + \vec K_2
\to 0\,$ we have again

$$M_{t_1 t_2}(\vec q; \vec K_1, \vec K_2)\;\to\;0\,, \eqno(5.75)$$
\noindent
because
$$\int d^3r\,F_a(\vec K_1\,|\,\vec r\,)^*\,F_a(\vec
K_2\,|\,\vec r\,)^*\,\to\,\int d^3r\,F_a(\vec K_1\,|\,\vec r\,)^*\,F_a(\vec
K_1\,|\,\vec r\,)\;=\;1\,$$
\vskip4pt\noindent
which makes the subsequent sum over the polarization vectors in (5.73)

$$\sum_a \hat\epsilon_{t_1}(\vec K_1)_a\,\hat\epsilon_{t_2}(- \vec
K_1)_a\;=\;0\,.$$
\vskip4pt\noindent
Both properties (5.74) and (5.75) will be useful for calculating $\,{\cal
E}^{(ii)}\,$ in the limit $\,\ell/R \to 0\,$.
\vskip6pt
Substituting (5.71)-(5.73) into (5.67) and (5.69), we derive

$${\cal E}^{(ii)}\;=\;- n\,e_1^\ell\,e_2^\ell\,\sum e^{i\phi'}\;{M_{t_1
t_2}(q; \vec K_1, \vec K_2)^*\,M_{t_1 t_2}(q'; \vec K_1, \vec K_2)\over [
\omega_{t_1}(\vec K_1) + \omega_{t_2}(\vec K_2) ] q^2 q'^2} \eqno(5.76)$$
\noindent where
$$\phi'\;=\;\vec q \cdot \vec R_1 - \vec q\,' \cdot \vec R_2\,,$$
$$\vec q\;=\;\vec K_1 + \vec K_2 + {2\pi\,\vec m\over \ell}\,, \eqno(5.77)$$
$$\vec q\,'\;=\;\vec K_1 + \vec K_2 + {2\pi\,\vec m\,'\over \ell}\,,$$
\vskip4pt\noindent
and the sum is over the polarization indices $\,t_1\,$ and $\,t_2\,$, the two
Bloch wave number vectors $\,\vec K_1\,$ and $\,\vec K_2\,$, as well as the
vectors
$\,\vec m\,$ and $\,\vec m\,'\,$, both having integer components.  From (5.49)
and (5.77), we have

$$\phi'\;=\;{2\pi\over\ell}\;(\vec m - \vec m\,') \cdot \vec R_{\rm cm} +
\vec k \cdot \vec R \eqno(5.78)$$\noindent
where
$$\vec k\;=\; \vec K_1 + \vec K_2 + {\pi\over \ell} \;(\vec m + \vec
m\,')\,. \eqno(5.79)$$
\vskip4pt\noindent
Hence, $\,{\cal E}^{(ii)}\,$ is also a periodic function in $\,\vec R_{\rm
cm}\,$, the same as $\,{\cal E}^{(i)}\,$.  Likewise, as $\,\ell \to 0\,$, we
can set
$$\vec m\;=\;\vec m\,' \eqno(5.80)$$
\noindent
and  $\,{\cal E}^{(ii)}\,$ becomes independent of  $\,\vec R_{\rm cm}\,$.
\vskip6pt
To derive the asymptotic behavior of  $\,{\cal E}^{(ii)}\,$ in the limit
$\,\ell/R \to 0\,$, we follow steps
parallel to those for  $\,{\cal E}^{(i)}\,$.  Write

$${\cal E}^{(ii)}\;=\;- n\,e_1^\ell\,e_2^\ell\,\int {d^3k\over
(2\pi)^3}\;{e^{i\vec k \cdot \vec R}\over k^2}\; II(\vec k)\eqno(5.81)$$
\noindent
where
 $$II(\vec k)\;=\;\int_B\,{d^3K\over (2\pi)^3}\;h(\vec K, \vec k) \eqno(5.82)$$
\vskip4pt\noindent
with the $\,\vec k\,$ integration extending over the entire momentum space
but the $\,\vec K\,$ integration  only within the Brillouin zone.  The
function $\,h\,$ is given by

$$h(\vec K, \vec k)\;=\;{1\over k^2} \sum_{t_1, t_2}\;{M_{t_1 t_2}(\vec k; \vec
K, \vec K')^*\,M_{t_1 t_2}(\vec k; \vec K, \vec K')\over
 \omega_{t_1}(\vec K) + \omega_{t_2}(\vec K')} \eqno(5.83)$$
\vskip4pt\noindent
where $\,\vec K'\,$ is a dependent variable, given by $\,\vec K' = (\vec k -
\vec K)_{{\rm mod}\, 2\pi/\ell}\,$; i.e., like $\,\vec K,\; \vec K'\,$ also
lies
within the Brillouin zone, but with the difference $\,[\,\vec K' - (\vec k -
\vec K)\,]_a = (2\pi/\ell) \times\,$ integer. Here, as before, the subscript
$\,a = x, y\,$ and $\,z\,$.  We note that as
$\,\vec k
\to 0\,$, the limit
$\,h(\vec K,
\vec k)\,$ is well behaved, on account of (5.74)-(5.75).  However, after the
$\,\vec K\,$ integration, the function $\,II(\vec k)\,$ diverges as $\,\vec k
\to 0\,$.
\vskip6pt
As in (5.57), write
$$II(\vec k)\;=\;II_c(\vec k) + II'(\vec k) \eqno(5.84)$$\noindent
with
$$II_c(\vec k)\;=\;\int_B {d^3 K\over (2\pi)^3}\;h_c(\vec K, \vec k)
\eqno(5.85)$$\noindent
and
$$II'(\vec k)\;=\;\int_B {d^3 K\over (2\pi)^3}\;[\,h(\vec K, \vec k) -
h_c(\vec K, \vec k)\,]\,. \eqno(5.86)$$
\vskip4pt\noindent
Choose $\,h_c(\vec K, \vec k)\,$ to be the integrand of $\,II(\vec k)\,$ {\it
without} the $\,0^{th}$-band approximation, i.e.

$$\eqalign{h_c(\vec K, \vec k)\;=\;{1\over 4k^2}\;& {(K - |\,\vec k - \vec
K\,|)^2\over K\,|\,\vec k - \vec K\,|\,(K + |\,\vec k - \vec K\,|)}\cr
& \times \left(\delta_{ab} - {K_a K_b\over K^2}\right) \left(\delta_{ab} -
{(\vec k -
\vec K)_a\,(\vec k -
\vec K)_b\over (\vec k - \vec K)^2}\right)\,,\cr} \eqno(5.87)$$
\vskip4pt\noindent
with the constraint that $\,\vec K\,$ lie within the Brillouin zone imposed
as  a boundary condition on the integral (5.86).  One can readily verify that
$\,II'(\vec k)\,$ is free from divergence and obtain (see Appendix C for
details)

$$II(\vec k)\;=\;{n\over 24\pi^2}\;g_\ell^4\;\left(ln\;{1\over k\ell} +
\lambda^{(ii)}\right) \eqno(5.88)$$\noindent
where the constant
$$\lambda^{(ii)}\;=\;{23\over 3} - 11\,ln\,2 + \lambda_c + \lambda''
\eqno(5.89)$$\noindent
with $\,\lambda_c\,$ given by (5.60) and
$$\lambda''\;=\;24\pi^2\,II'(0)\;=\;\int_B {d^3\vec K\over
(2\pi)^3}\;\lim_{k \to 0}\,[\,h(\vec K, \vec k) - h_c(\vec K, \vec k)\,]\,.
\eqno(5.90)$$
\vskip4pt\noindent
Note that in the limit when the magnitude $\,k \to 0\,$, the integrand
remains dependent on the direction $\,\hat k = \vec k/k\,$.  (See (C.17),
(C.25) and (C.28) below.)  In Appendix C, we show that
$\,\lambda'' = 0.1238\,$ and therefore

$$\lambda^{(ii)}\;=\;1.8362\,. \eqno(5.91)$$
\vskip4pt\noindent
Substituting (5.88) into (5.81) and carrying out the integration over $\,\vec
k\,$, we find

$${\cal E}^{(ii)}\;=\;- {n\over 24\pi^2}\;\left( ln\;{R\over \ell} + \gamma +
\lambda^{(ii)}\right)\,. \eqno(5.92)$$
\vskip4pt\noindent
Combining (5.65)-(5.66) and (5.91)-(5.92), we establish the constant
$\,\lambda\,$ in (5.21) to be the one given by (5.25); i.e.,

$$\lambda\;=\;\lambda^{(i)} + \lambda^{(ii)}\;=\;3.3559\,. \eqno(5.93)$$

\vfill\eject

\centerline{\bf APPENDIX A}
\vskip18pt
In the following we wish to compare the lattice formulation given in this paper
to the conventional ones in which the lattice structure consists of only
discrete sites connected by links, but without being embedded within a
continuum.  An example of such a comparison for the case of a scalar field in
one dimension was given at the end of Subsection 2.3.  In this appendix we
extend the comparison to the gauge theory.
\vskip4pt
Let it be agreed that the aim is to discretize only the spatial coordinates
while time remains continuous.  This violates relativistic symmetries, but no
fixed spatial lattice can be invariant under the full Lorentz group anyway.
\vskip4pt
Take a simple cubic lattice of spacing $\,\ell\,$ and period $\,N\ell\,$ in
each orthogonal direction; the volume associated with the full period is
$\,\Omega = (N\ell)^3\,$.  Conventionally, one labels the sites of the lattice
as $\,j = (j_1, j_2, j_3)\,$ and attaches to each link a variable analogous to
the vector potential, say $\,a_j, b_j, c_j\,$ on the links leaving site $\,j\,$
in the $\,x-, y-, z-$directions (as in (3.12)).
\vskip4pt
Consider first an Abelian gauge theory.  The {\it conventional} lattice
approach would be to write a lattice Lagrangian for a discrete system
$$\eqalign{{\cal L}_d\;&=\;\sum_j {\ell^3\over 2}\;\left[(\dot a_j - d_{j'} +
d_j)^2 + (\dot b_j - d_{j''} + d_j)^2 + (\dot c_j - d_{j'''} + d_j)^2\right]\cr
&- \sum_j {\ell\over 2}\;\left[(a_j + b_{j'} - a_{j''} - b_j)^2 + (b_j +
c_{j''} - b_{j'''} - c_j)^2 + (c_j + a_{j'''} - c_{j'} + a_j)^2\right]\cr}
\eqno({\rm A}.1)$$\noindent
where, as in (3.18),
$$\eqalign{j'\;&=\;(j_1 + 1, j_2, j_3)\,,\cr
j''\;&=\;(j_1, j_2 + 1, j_3)\,,\cr
j'''\;&=\;(j_1, j_2, j_3 + 1)\,;\cr} \eqno({\rm A}.2)$$
\vskip4pt\noindent
the first term of (A.1) is ``electric'' and the second ``magnetic''.  The new
variable $\,d_j\,$ attached to site $\,j\,$ is analogous to the ``time''
component of the vector potential.
\vskip4pt
One notes that $\,{\cal L}_d\,$ is invariant under the {\it lattice} gauge
transformation

$$\eqalign{a_j\;&\to\;a_j + \lambda_{j'} - \lambda_j\,,\cr
b_j\;&\to\;b_j + \lambda_{j''} - \lambda_j\,,\cr
c_j\;&\to\;c_j + \lambda_{j'''} - \lambda_j\,,\cr
d_j\;&\to\;d_j + \dot\lambda_j\cr} \eqno({\rm A}.3)$$
\vskip4pt\noindent
where $\,\lambda_j\,$ is an arbitrary quantity attached to site $\,j\,$.  One
may take advantage of this to pass either to the lattice time-axial gauge where

$$d_j\;=\;0 \hskip4em ({\rm all}\;\;j) \eqno({\rm A}.4)$$
\noindent
or to the lattice Coulomb gauge where
$$a_j - a_{j'_-} + b_j - b_{j''_-} + c_j - c_{j'''_-}\;=\;0\,; \eqno({\rm
A}.5)$$
\noindent
here, as in (4.16),
$$\eqalign{j_-'\;&=\;(j_1 - 1, j_2, j_3)\,,\cr
j_-''\;&=\;(j_1, j_2 - 1, j_3)\,,\cr
j_-'''\;&=\;(j_1 + 1, j_2, j_3 - 1)\,.\cr}\eqno({\rm A}.6)$$
\vskip8pt
In time-axial gauge one may write the Lagrangian (A.1) in ``momentum space'':
$${\cal L}_d\;=\;\sum_{\vec K}\left(\dot q^\dagger \dot q - q^\dagger(u^2 -
u\tilde u)\,q\right) \eqno({\rm A}.7)$$
\vskip4pt\noindent
where $\,q = q(\vec K)\,$ and $\,u = u(\vec K)\,$
are $\,3 \times 1\,$ column matrices.  The components of $\,q\,$ are the
Fourier coefficients of $\,a_j, b_j, c_j\,$.  Those of $\,u\,$ are

$$u_a(\vec \theta)\;=\;{2\over\ell}\;\sin {\theta_a\over 2}\,, \eqno({\rm
A}.8)$$
\vskip4pt\noindent
with $\,\vec\theta = \vec K\ell\,$.  Evidently there are redundant degrees of
freedom, which one may eliminate by writing

$$Q_\ell(\vec K)\;=\;\hat u(\vec K) \cdot \vec q(\vec K)\eqno({\rm A}.9)$$
\noindent
where $\,(\hat u)_a = u_a\,/\,\sqrt {u^2}\,$, and
$$Q_t(\vec K)\;=\;\hat\epsilon_t(\vec K) \cdot \vec q\,(\vec K) \hskip3em
(t\;=\;1, 2) \eqno({\rm A}.10)$$
\noindent
with $\,\hat u, \hat\epsilon_1, \hat\epsilon_2\,$ forming a right-handed
orthonormal system.  Then (A.7) becomes
$${\cal L}_d\;=\;\sum_{\vec K}\left(\dot Q_\ell^\dagger\,\dot Q_\ell +
\sum_{t=1}^2\;(\dot Q_t^\dagger\,\dot Q_t -
u^2\,Q_t^\dagger\,Q_t)\right)\,.\eqno({\rm A}.11)$$
\noindent
In this gauge there is a supplementary condition on the state vector,
$${\partial\over\partial Q_\ell(\vec K)}\;<Q\,|\,>\;=\;0 \eqno({\rm A}.12)$$
\vskip4pt\noindent
in the absence of sources, derived from the lattice analog of Gauss' law.
Hence the variables $\,Q_\ell\,$ can be ignored and the Hamiltonian written as

$${\cal H}_d\;=\;\sum_{\vec K} \sum_{t=1}^2\,(P_t^\dagger P_t +
u^2\,Q_t^\dagger
\,Q_t) \eqno({\rm A}.13)$$
\noindent
where $\,P_t(\vec K)\,$ is the conjugate momentum of $\,Q_t(\vec K)\,$.
\vskip4pt
Alternatively, in lattice Coulomb gauge one has from the outset $\,Q_\ell =
0\,$,
all $\,\vec K \ne 0\,$, and the variables $\,d_j\,$ are decoupled from
$\,Q_t\,$.  In the absence of sources the $\,d_j\,$ vanish and the Hamiltonian
again takes the form (A.13).
\vskip4pt
All this proceeds smoothly, but one encounters difficulties in extending it to
a non-Abelian theory.  For example, how is one to represent the nonlinear term
in the magnetic field?
\vskip4pt
Two different approaches are possible.  The first is to go over to a compact
formulation$^2$ based not on $\,a_j, b_j, c_j\,$ but on $\,e^{i\ell a_j}\,$,
$\,e^{i\ell b_j}\,$, $\,e^{i\ell c_j}\,$.  The disadvantage is that
this formulation, for nonzero $\,\ell\,$, is no longer connected with the true
continuum theory based on

$${\cal L}\;=\;- \int d^3r\;{1\over 2}\,\left({\partial A_\nu\over\partial
x_\mu} - {\partial A_\mu\over\partial x_\nu}\right)^2 \eqno({\rm A}.14)$$
\noindent
where the $\,A_\mu\,$ are continuous fields with gauge transformation
$$A_\mu\;\to\;A_\mu + {\partial\Lambda\over\partial x_\mu}\,. \eqno({\rm
A}.15)$$
\vskip4pt\noindent
The ``compact'' theory derived by nonlinearizing (A.1) and (A.3) is an abstract
model of the original theory, not a concrete approximation to it.
\vskip4pt
The second approach, which we pursue in this paper, is to seek a noncompact
lattice theory that can serve as a first approximation in a systematic way
which can approach the exact continuum theory, without requiring $\,\ell\,$ to
approach zero.  Thus (A.14) and (A.15) will be our starting point, not (A.1)
and (A.3); the lattice variables $\,a_j, b_j, c_j\,$ will enter only as
parameters controlling the few degrees of freedom allowed to $\,\vec A\,$ in
the zeroth approximation.
\vskip4pt
A natural way to implement this idea is to make each component of $\,\vec A\,$
a constant along any link in the corresponding direction, and interpolate
linearly between links.  Referring to the lump functions defined above, of
which $\,L_1\,$ is a constant and $\,L_2\,$ is linear, we see that this amounts
to setting

$$\eqalign{\vec A(\vec r\,)\;=\;&\sum_j\,[\, a_j\,\hat x\,L_1(x -
j_1\ell)\,L_2(y - j_2\ell)\,L_2(z - j_3\ell)\cr
&+ b_j\,\hat y\,L_2(x -
j_1\ell)\,L_1(y - j_2\ell)\,L_2(z - j_3\ell)\cr
&+  c_j\,\hat z\,L_2(x - j_1\ell)\,L_2(y - j_2\ell)\,L_1(z - j_3\ell)\,]\,.\cr}
\eqno({\rm A}.16)$$
\vskip4pt\noindent
The idea is to substitute this expression into (A.14) so as to obtain a
Lagrangian depending on the ``effective'' lattice variables $\,a_j, b_j,
c_j\,$.  (Writing $\,\vec A(\vec r\,)\,, L_1(x)\,$ and $\,L_2(x)\,$ as $\,\vec
V(\vec r\,)\,, C(x)\,$ and $\,\Delta(x)\,$, we see that (A.16) is the same as
(3.7)).
\vskip4pt
For an Abelian theory this idea works nicely in time-axial gauge because the
time-independent gauge transformation (A.3) actually generates a
transformation of the form (A.15) on the continuous fields defined in (A.16).
As shown in (3.22), this makes it possible to implement the subsidiary state
condition derived from the {\it continuum} Gauss' law by eliminating some of
the
lattice variables similar to (A.13).  But this no longer works in a non-Abelian
theory: the non-Abelian version of (A.1) is not invariant under any
transformation of the type (A.3).
\vskip4pt
For the non-Abelian theory, we propose instead to work in Coulomb gauge (or
Coulomb-like gauge (3.84)).  We find that if we replace the function $\,L_1\,$
in (A.16) by $\,L_3\,$, then on account of (2.3) the condition (A.5) on the
lattice variables is sufficient to ensure the true Coulomb gauge condition

$$\vec\nabla \cdot \vec A\;=\;0 \eqno({\rm A}.17)$$
\vskip4pt\noindent
{\it at every point in space}.  Moreover, if the higher band functions are
built in the manner indicated by (3.73), then (A.17) can be satisfied {\it
independently within each band} at every point in space.  This remains true in
non-Abelian theory because the Coulomb gauge condition
is linear.  The details are carried out in Subsection 3.2 and Section
4.  Identical considerations can be applied to any gauge in which the
gauge-fixing condition has no nonlinear terms, as in (3.84). \vfill\eject
\centerline{\bf APPENDIX B}
\vskip18pt
In this appendix, we give another method for the removal of redundant lattice
coordinate space variables, alternative to the one given in Section 4.3.
\vskip6pt
The objective is to eliminate the $\,{\cal N} + 2\,$ redundant variables among
the set

$$\{\,\xi_j^\ell\,, \eta_j^\ell\,, \zeta_j^\ell\,\} \eqno({\rm B}.1)$$
\vskip4pt\noindent
of $\,3{\cal N}\,$ plaquette flux variables.  Again, the group superscript
$\,\ell\,$ will be kept fixed, and the factor of its $\,n^2 - 1\,$ degrees of
freedom due to $\,SU(n)\,$ will not be included.
\vskip12pt\noindent
{\bf Theorem.}\ \ \ \ For any plaquette flux distribution (B.1), through (4.29)
in which

$$\{\,\xi_j^\ell\,, \eta_j^\ell\,, \zeta_j^\ell\,\}\;\to\;\{\,\xi_j'^\ell\,,
\eta_j'^\ell\,, \zeta_j'^\ell\,\} \eqno({\rm B}.2)$$
\vskip4pt\noindent
the transformed plaquette fluxes can be made to satisfy the lattice
divergence-free equation:

$$\xi_{j'}'^\ell - \xi_j'^\ell + \eta_{j''}'^\ell - \eta_j'^\ell +
\zeta_{j'''}'^\ell - \zeta_j'^\ell\;=\;0
\eqno({\rm B}.3)$$
\vskip4pt\noindent
at all $\,j\,$, with the sites $\,j',\, j''\,$ and $\,j'''\,$ given by (3.18).
\vskip16pt\noindent
{\bf Proof.}\ \ \ \ Consider the cubic cell $\,\tau_j\,$ that has
$\,X_j,\,Y_j\,$  and $\,Z_j\,$ of Figure 3(a) as three of its six surfaces.
The
other three surfaces are (in the same notation)  $\,X_{j'},\,Y_{j''}\,$ and
$\,Z_{j'''}\,$.  Define the lattice differences

$$\partial_x\,(I_x^\ell)_j\;\equiv\;\xi_{j'}^\ell - \xi_j^\ell\,,$$
$$\partial_y\,(I_y^\ell)_j\;\equiv\;\eta_{j''}^\ell - \eta_j^\ell\,,
\eqno({\rm B}.4)$$
\noindent and
$$\partial_z\,(I_z^\ell)_j\;\equiv\;\zeta_{j'''}^\ell - \zeta_j^\ell$$
\vskip4pt\noindent
to be the net flux flowing out from $\,\tau_j\,$ along the $\,x, y\,$ and
$\,z\,$ axes.  The lattice divergence

$$\vec\partial\,\cdot\,\vec
I_j\,^\ell\;\equiv\;\sum_{a=x}^z\,\partial_a(I_a^\ell)_j \eqno({\rm B}.5)$$
\vskip4pt\noindent
represents the {\it total} flux leaving $\,\tau_j\,$.
\vskip6pt
Next, we turn to the dual lattice $\,L_D\,$:  The center of $\,\tau_j\,$ in
the original lattice $\,L\,$ becomes the lattice site $\,j\,$ in $\,L_D\,$.
Correspondingly, $\,\xi_j^\ell,\;\eta_j^\ell\,$ and $\,\zeta_j^\ell\,$ denote
link currents, and $\,\vec\partial\,\cdot\,\vec I_j\,^\ell\,$ is the net
current
leaving the site $\,j\,$ (with the components of $\,\vec I_j\,^\ell\,$ now
defined on the links $\,\overline{j'j},\;\overline{j''j}\,$ and $\,\overline
{j'''j}\,$ in $\,L_D\,$).  Likewise, introduce

$$\partial_x^2\,\chi_j^\ell\;\equiv\;\chi_{j'}^\ell - 2\chi_j^\ell +
\chi_{j'_-}^\ell\,,$$
$$\partial_y^2\,\chi_j^\ell\;\equiv\;\chi_{j''}^\ell - 2\chi_j^\ell +
\chi_{j''_-}^\ell\,,\eqno({\rm B}.6)$$
$$\partial_z^2\,\chi_j^\ell\;\equiv\;\chi_{j'''}^\ell - 2\chi_j^\ell +
\chi_{j'''_-}^\ell$$\noindent
and
$$\partial^2\,\chi_j^\ell\;\equiv\;\sum_{a=x}^z\,\partial_a^2\,\chi_j^\ell
\eqno({\rm B}.7)$$
\vskip4pt\noindent
to be the second-order lattice differences and the lattice Laplacian.  The
sites $\,j,\,j',\,\cdot\cdot\,j_-'''\,$ in $\,L_D\,$ are dual to the cubic
cells  $\,\tau_j,\,\tau_{j'},\,\cdot\cdot\,\tau_{j_-'''}\,$ in $\,L\,$
which, in turn, are related to the sites $\,j,\,j',\,\cdot\cdot\,j_-'''\,$ in
$\,L\,$, given by (3.18) and (4.16), in the way described above.
\vskip4pt
Equation (B.3) states that through the transformation (4.29), $\,\vec
I_j\,^\ell\,$ becomes  $\,\vec I_j'^\ell\,$ which satisfies $\,\vec\partial
\cdot
\vec I_j'^\ell = 0\,$; therefore

$$- \partial^2\,\chi_j^\ell\;=\;\vec\partial\,\cdot\,\vec
I_j\,^\ell\;\equiv\;\rho_j^\ell\,. \eqno({\rm B}.8)$$
\vskip4pt\noindent
We can readily obtain the solution through the lattice Green's function

$${\cal G}(\vec r\,)\;\equiv\;\sum_{\vec K \ne 0}\,{1\over 2{\cal N}}\,[\,3 -
\sum_a \cos \theta_a\,]^{-1}\,e^{i \vec K \cdot \vec r} \eqno({\rm B}.9)$$
\vskip4pt\noindent
where $\,\theta_a = K_a \ell\,$ as before, and the sum extends over the
Brillouin zone but with its origin excluded.  The result is

$$\chi_j^\ell\;=\;\sum_i\;{\cal G}(\vec r_j - \vec r_i)\,\rho_i^\ell\,.
\eqno({\rm B}.10)$$
\vskip4pt\noindent
The theorem is then established.
\vfill\eject
The function $\,{\cal G}(\vec r\,)\,$ has been well studied in the literature.
For example, at $\,\vec r = 0\,$ and $\,{\cal N} \to \infty\,$

$${\cal G}(0)\;=\;0.25273\,,\eqno({\rm B}.11)$$
\vskip4pt\noindent
which can also be expressed$^{10}$ in terms of the elliptic integral

$${\cal G}(0)\;=\;{2\over\pi^2}\;(18 + 12 \sqrt {2} - 10 \sqrt {3} - 7 \sqrt
{6}\,)\;\times\;K^2[\,(2 - \sqrt {3})\,(3 - \sqrt {2})\,] \eqno({\rm
B}.12)$$\noindent
where
$$K(k)\;=\;\int_0^{\pi\over 2} (1 - k^2 \sin^2 \phi)^{-{1\over 2}}\,d\phi\,.
\eqno({\rm B}.13)$$
\vskip10pt
Returning to the problem of eliminating redundant variables, we note that
(B.3) imposes $\,{\cal N} - 1\,$ conditions.  Since any plaquette flux
distribution can be transformed into one that satisfies (B.3), the resulting
divergence-free distribution consists only of

$$3{\cal N} -  ({\cal N} - 1)\;=\; 2{\cal N} + 1 \eqno({\rm B}.14)$$
\vskip4pt\noindent
independent plaquette-flux variables.  As explained at the end of
Section 4.3, we can arbitrarily set three surface plaquette fluxes of $\,L\,$
to
zero (one along each Cartesian direction) without affecting the link-flux
distribution ($\,a_j^\ell,\,b_j^\ell,\,c_j^\ell\,$).  This and (B.14) complete
the reduction process: the number of independent variables in $\,(\xi_j^\ell,\,
\eta_j^\ell,\,\zeta_j^\ell)\,$ becomes $\, 2{\cal N} + 1 - 3 =  2{\cal N} -
2\,$, in accordance with (4.30).
\vfill\eject
\centerline{\bf APPENDIX C}
\vskip18pt\noindent
{\it C.1.  The Integral $\,I_c(\vec k)\,$}
\vskip8pt
To calculate the integral $\,I_c(\vec k)\,$ given by (5.58) for $\,k\ell <<
1\,$,
we divide the Brillouin zone, $\,- {\pi\over \ell} \le K_a \le {\pi \over
\ell}\,$ (where $\,a = x, y, z)\,$, into two regions: the interior of a small
sphere
$\,s\,$ with its center at the origin $\,\vec K = 0\,$ and the volume
outside the sphere.  The radius of the sphere, $\,K_0\,$, is chosen such that
$\,k << K_0 << {\pi\over\ell}\,$.  Correspondingly, the integral (5.58) can be
separated into two terms:

$$I_c(k)\;=\;I_{\rm in} + I_{\rm out} \eqno({\rm C}.1)$$\noindent
where
$$I_{\rm in}\;=\;\int_{\rm in} {d^3K\over (2\pi)^3}\;f_c(\vec K, \vec k)
 \eqno({\rm C}.2)$$\noindent
and
$$I_{\rm out}\;=\;\int_{\rm out} {d^3K\over (2\pi)^3}\;f_c(\vec K, \vec k)
 \eqno({\rm C}.3)$$\vskip4pt\noindent
with $\,f_c(\vec K, \vec k)\,$ given by (5.55), the integral (C.2) referring to
the integration inside the sphere $\,s\,$ and (C.3) for the integration outside
$\,s\,$ but within the Brillouin zone.  Since $\,I_{\rm in}\,$ contains
$\,ln\,k\ell\,$, $\,k\,$ may not be set to zero in its integrand.  A
straightforward calculation in spherical coordinates yields:

$$I_{\rm in}\;=\;{1\over 6\pi^2}\;\left(ln\;{K_0\over k} + {1\over 3}\right)
\eqno({\rm C}.4)$$\vskip4pt\noindent
where we have dropped the terms that vanish in the limit $\,k \to 0\,$.  To the
same order, $\,k\,$ may be set to zero in the integrand of $\,I_{\rm out}\,$,
on account of the low momentum cutoff $\,(K > K_0)\,$.  We have

$$I_{\rm out}\;=\;\int_{\rm out} {d^3K\over (2\pi)^3}\;\lim_{k \to
0}\;f_c(\vec K, \vec k)\,. \eqno({\rm C}.5)$$\noindent
It follows from (5.55) and (5.56) that

$$\lim_{k \to 0}\; f_c(\vec K, \vec k)\;=\;{1\over 2K^3}\,\left[ 1 - {(\hat k
\cdot \vec K)^2\over K^2}\right]\,. \eqno({\rm C}.6)$$\vskip4pt\noindent
Because of cubic symmetry, we may replace $\,(\hat k
\cdot \vec K)^2\,/\,K^2\,$ by $\,{1\over 3}\,$ in the integration; (C.5)
becomes

$$I_{\rm out}\;=\;{1\over 3}\,\int_{\rm out}\;{d^3\kappa\over
(2\pi)^3}\;{1\over\kappa^3}  \eqno({\rm C}.7)$$\vskip4pt\noindent
in which we introduce the dimensionless vector $\,\vec\kappa \equiv
{\ell\over\pi}\;\vec K\,$, whose Cartesian coordinates will be denoted by
$\,(x, y, z)\,$, so that its magnitude is $\,\kappa = \sqrt{x^2 + y^2 +
z^2}\,$.  The integration domain of (C.7) in terms of $\,\vec\kappa\,$ is given
by $\,- 1 \le x, y, z \le 1\,$ and $\,\kappa > {\ell\over\pi}\,K_0\,$.  In
terms of spherical coordinates with $\,a\,$ and $\,b\,$ denoting the polar and
azimuthal angles, we obtain

$$I_{\rm out}\;=\;{1\over 24\pi^3} \int_0^{2\pi} db\,\int_0^\pi da\,\sin
a\,ln\,\kappa(a, b) - {1\over 6\pi^2}\;ln\;{K_0\ell\over\pi} \eqno({\rm
C}.8)$$\vskip4pt\noindent where $\,\kappa(a, b)\,$ is the radial coordinate of
a
point on the boundary of the Brillouin zone in the direction specified by the
angles $\,a\,$ and
$\,b\,$.  The Brillouin zone is a cube bounded by six faces:  $\,x = \pm 1\,$,
$\,y = \pm 1\,$ and $\,z = \pm 1\,$.  Because of cubic symmetry, we need only
integrate over one of these faces and then multiply the result by six.  Take
the surface at $\,z = 1\,$: we have

$$\eqalign{x\;=&\;\tan a\;\cos b\,,\cr
y\;=&\;\tan a\;\sin b\,,\cr}\eqno({\rm C}.9) $$\noindent
and
$$\kappa(a, b)\;=\;\sec a\,; \eqno({\rm C}.10)$$
\noindent
(C.8) reduces to
$$I_{\rm out}\;=\;{1\over 8\pi^3} \int_{-1}^1\,dx\,\int_{-1}^1\,dy\;\;{ln(x^2 +
y^2 + 1)\over (x^2 + y^2 + 1)^{3\over 2}} + {1\over 6\pi^2}\;ln\;{\pi\over
K_0\ell}\,. \eqno({\rm C}.11)$$\noindent
Carrying out the integration over $\,y\,$, we find

$$I_{\rm out}\;=\;{1\over 6\pi^2}\;\left( ln\;{\pi\over K_0\ell} + 1 - {6\over
\pi}\;\int_0^1 dx\;{ln(x^2 + 2)\over (x^2 + 1) \sqrt{x^2 + 2}} \right)\,.
\eqno({\rm C}.12)$$
\noindent
Combining (C.4) and (C.12), we establish (5.60) and (5.61).
\vfill\eject\noindent
{\it C.2.  The Integral $\,II_c(\vec k)$}
\vskip6pt
Following the same method as in the above section, we separate the integral
(5.85)
into two terms,

$$II_c(k)\;=\;II_{\rm in} + II_{\rm out} \eqno({\rm C}.13)$$\noindent
where, as in (C.1)-(C.3),
$$II_{\rm in}\;=\;\int_{\rm in} {d^3K\over (2\pi)^3}\;h_c(\vec K, \vec k)
\eqno({\rm C}.14)$$\noindent
is the integration inside the same small sphere  $\,s\,$ and
$$II_{\rm out}\;=\;\int_{\rm out} {d^3K\over (2\pi)^3}\;h_c(\vec K, \vec k)
\eqno({\rm C}.15)$$
\vskip4pt\noindent
is the integration outside $\,s\,$ but inside the Brillouin zone.  The
integrand $\,h_c(\vec K, \vec k)\,$ is given by (5.87).  In the limit $\,k \to
0\,$ we have

$$II_{\rm out}\;=\;\int_{\rm out} {d^3K\over (2\pi)^3}\;\lim_{k \to
0} \;h_c(\vec K, \vec k)\;=\;{1\over 4}\;\;I_{\rm out}
\eqno({\rm C}.16)$$
\noindent
since, on account of (5.87),
$$\lim_{k \to 0}\;h_c(\vec K, \vec k)\;=\;{1\over 4}\;{(\hat k \cdot \vec
K)^2\over K^5}\eqno({\rm C}.17)$$\vskip4pt\noindent
which can be replaced by $\,(12K^3)^{-1}\,$ because of cubic symmetry.
Therefore
$$II_{\rm out}\;=\;{1\over 24\pi^2}\;\left( ln\;{\pi\over K_0\ell} + 1 -
{6\over
\pi}\;\int_0^1 dx\;{ln(x^2 + 2)\over (x^2 + 1) \sqrt{x^2 + 2}} \right)
\eqno({\rm C}.18)$$
\noindent
where, as before, $\,K_0\,$ is the radius of the small sphere $\,s\,$.
\vskip6pt
To carry out the integration inside $\,s\,$ for $\,II_{\rm in}\,$, we find it
convenient to introduce the following prolate spheroidal coordinates for the
Bloch wave number vector $\,\vec K\,$:

$$K_x\;=\;{k\over 2}\;\sqrt {(\xi^2 - 1) (1 - \eta^2)}\;\cos \phi\,,$$
$$K_y\;=\;{k\over 2}\;\sqrt {(\xi^2 - 1) (1 - \eta^2)}\;\sin \phi\,,\eqno({\rm
C}.19)$$
$$K_z\;=\;{k\over 2}\;(\xi\eta + 1)$$
\vskip4pt\noindent
with $\,0 \le \phi \le 2\pi\,$, $\,- 1 \le \eta \le 1\,$ and $\,\xi \le 1\,$.
 To derive the $\,ln\;K_0/k\,$ term and the constant term of $\,II_{\rm in}\,$
in the limit $\,k \to 0\,$, we may approximate the spherical domain of the
integration by a prolate spheroid with foci ($\,0, 0, 0\,$) and ($\,0, 0, k\,$)
and with the length of its semi-major axis equal to  $\,K_0\,$, which
corresponds to the upper limit of $\,\xi < {2K_0\over k}\,$ in the
$\,\xi$-integration.  The error due to the approximation vanishes in the limit
$\,k \to 0\,$.  We find

$$\eqalign{II_{\rm in}\;&=\;{1\over 32\pi^2} \int_1^{2K_0/k} d\xi \int_{-1}^1
d\eta\;{\eta^2\over\xi}\;\left[ 1 + {(\xi^2 + \eta^2 - 2)^2\over (\xi^2 -
\eta^2)^2} \right]\;\cr
&=\;{1\over 24\pi^2}\;\left( ln\;{K_0\over k} + 8 -
11\,ln\,2\right)\,.\cr}\eqno({\rm C}.20)$$
\vskip4pt\noindent
Combining (C.18) and (C.20), we derive the expression of $\,II_c(k)\,$ when
$\,k\ell \to 0\,$; this leads to (5.88) and (5.89), except for the constant
$\,\lambda''\,$ which will be evaluated below.
\vskip20pt\noindent
{\it C.3.  The constants $\,\lambda'\,$ and $\,\lambda''\,$}
\vskip6pt
The expression for the constant $\,\lambda'\,$ follows from (5.53), (5.54),
(5.55), (5.59) and (5.62) and it reads:

$$\lambda'\;=\;2\pi^2 \int_B {d^3 K\over (2\pi)^3}\;\left[ \sum_{a,
t}\;{\hat\epsilon_t(\vec K)_a\,\hat\epsilon_t(\vec K)_a\over 2\omega_t(\vec
K)}\;\sum_{\vec m}\;{\Omega\,C_a(\vec
\theta\,|\,\vec m)^2\over
\left(\vec K + {2\pi\vec m\over\ell}\right)^2} - {1\over K^3}\right]\,.
\eqno({\rm C}.21)$$
\vskip4pt
To derive the expression for $\,\lambda''\,$, we separate the summation
$\,\sum_{t_1t_2}\,$ in $\,h(\vec K, \vec k)\,$ of (5.83) into a sum over the
same
polarization, $\,t_1 = t_2 = t\,$, and another sum over different polarizations
$\,t_1 \ne t_2\,$, i.e.

$$h(\vec K, \vec k)\;=\;h_1(\vec K,\vec k) + h_2(\vec K, \vec k) \eqno({\rm
C}.22)$$\noindent
where
$$h_1(\vec K, \vec k)\;=\;{1\over k^2}\;\sum_t\;{M_{tt}(\vec k; \vec K, \vec
K')^*\;M_{tt}(\vec k; \vec K, \vec K')\over \omega_t(\vec K) + \omega_t(\vec
K')} \eqno({\rm C}.23)$$\noindent
and
$$h_2(\vec K, \vec k)\;=\;{1\over k^2}\;\sum_{t_1 \ne t_2}\;{M_{t_1 t_2}(\vec
k;
\vec K, \vec K')^*\;M_{t_1 t_2}(\vec k; \vec K, \vec K')\over \omega_{t_1}(\vec
K) + \omega_{t_2}(\vec K')} \eqno({\rm C}.24)$$
\vskip4pt\noindent
with $\,\vec K'\,$ defined by the prescription following (5.83). The limit $\,k
\to 0\,$ of $\,h_1(\vec K, k)\,$ can be readily obtained, and we have

$$\lim_{k \to 0}\;h_1(\vec K, \vec k)\;=\;{1\over 4}\;\sum_t {(\hat k \cdot
\vec \nabla_{\vec K}\,\omega_t)^2\over \omega_t^3(\vec K)}\,. \eqno({\rm
C}.25)$$ \vskip4pt\noindent
In the same limit, using the Fourier expansion (3.66) and (3.67), we find

$$\int d^3r\,F_a(\vec K\,|\,\vec r\,)^*\,F_a(\vec K'\,|\,\vec
r\,)^*\,e^{i\vec k \cdot \vec r} = N^3 \sum_{\vec m} {\cal C}_a(\vec
\theta\,|\,\vec m)\,{\cal C}_a(\vec \theta - \vec k
\ell\,|\,\vec m) = 1 + O(k^2\ell^2) \eqno({\rm C}.26)$$\noindent
and therefore
$$M_{t_1t_2}(\vec k; \vec K, \vec K')\;=\;{i\,[\,\omega_{t_1}(\vec K) -
\omega_{t_2}(\vec K)\,]\over 2\sqrt {\omega_{t_1}(\vec K)\,
\omega_{t_2}(\vec K)}}\,\sum_a \hat\epsilon_{t_1}(\vec K)_a\;\vec k \cdot
\vec\nabla_{\vec K}\;\hat\epsilon_{t_2}(- \vec K)_a + O(k^2\ell^2)
\eqno({\rm C}.27)$$\vskip4pt\noindent
for $\,t_1 \ne t_2\,$, in accordance with (5.75).  Substituting (C.27) into
(C.24) we obtain

$$\eqalign{\lim_{k \to 0}\,h_2(\vec K, \vec k)\;=\;&{1\over 4}\,\sum_{t_1 \ne
t_2}\,{[\,\omega_{t_1}(\vec K) - \omega_{t_2}(\vec K)\,]^2\over
\omega_{t_1}(\vec
K)\,\omega_{t_2}(\vec K)\;[\,\omega_{t_1}(\vec K) + \omega_{t_2}(\vec K)\,]}\cr
&\times \sum_{a,b}\,[\,\hat\epsilon_{t_1}(\vec K)_a\;\hat k \cdot
\vec\nabla_{\vec K}\;\hat\epsilon_{t_2}(- \vec K)_a\,]\,
[ \,\hat\epsilon_{t_1}(\vec K)_b\;\hat k \cdot
\vec\nabla_{\vec K}\;\hat\epsilon_{t_2}(- \vec K)_b\,]\,.\cr} \eqno({\rm
C}.28)$$\vskip4pt\noindent
Following the definitions (5.90) and (5.86) we have

$$\lambda''\;=\;\lambda_1'' + \lambda_2''  \eqno({\rm C}.29)$$\noindent
where
$$\eqalign{\lambda_1''\;&=\;24\pi^2 \int {d^3K\over (2\pi)^3}\;\lim_{k \to 0}\;
[\,h_1(\vec K, \vec k) -  h_c(\vec K, \vec k)\,]\cr
&=\;2\pi^2 \int_B {d^3K\over (2\pi)^3}\;\left[
\sum_t\,{|\,\vec\nabla_{\vec K}\,\omega_t\,|^2\over 2\omega_t^3(\vec K)} -
{1\over K^3} \right]\cr} \eqno({\rm C}.30)$$\noindent
and
$$\eqalign{\lambda_2''\;&=\;2\pi^2 \int {d^3K\over (2\pi)^3}\;\sum_{t_1 \ne
t_2}\,{[\,\omega_{t_1}(\vec K) - \omega_{t_2}(\vec K)\,]^2\over
\omega_{t_1}(\vec
K)\,\omega_{t_2}(\vec K)\;[\,\omega_{t_1}(\vec K) + \omega_{t_2}(\vec K)\,]}\cr
&\times \sum_{a,b,c}\,\hat\epsilon_{t_1}(\vec K)_a\;{\partial\,
\hat\epsilon_{t_2}(- \vec K)_a\over \partial K_c}\;\;\hat\epsilon_{t_1}(\vec
K)_b\;\;{\partial\,\hat\epsilon_{t_2}(- \vec K)_b\over \partial K_c}\,.\cr}
\eqno({\rm C}.31)$$\vskip4pt\noindent
Both integrations (C.30) and (C.31) are well defined.  The result is
$\,\lambda'' = 0.1238\,$.
\vskip24pt\noindent
{\it  C.4.  The Remainder $\,\delta\,$}
\vskip6pt
	The function $\,\delta(\ell, \vec R_1, \vec R_2)\,$ is defined by
(5.21)-(5.23), which relates $\,g_R^2\,$ to $\,g_\ell^2\,$.  From
(5.16)-(5.17), we see that $\,g_R^2\,$ appears as the coefficient of
$\,e_1^\ell\,e_2^\ell\,/\,4\pi\,R\,$ in $\,E_{12}\,$; the latter can in turn be
written as (neglecting $\,O(g_\ell^6)\,$)

$$E_{12}\;=\;g_\ell^2\;{e_1^\ell\,e_2^\ell\over 4\pi\,R} + g_\ell^4({\cal
E}^{(i)} + {\cal E}^{(ii)})\,.\eqno({\rm C}.32)$$\noindent
It follows then
$$\delta(\ell, \vec R_1, \vec R_2)\;=\;{96\pi^3\,R\over
11n\,e_1^\ell\,e_2^\ell}\;({\cal E}^{(i)} + {\cal E}^{(ii)}) - ln\;{R\over
\ell} - \gamma - \lambda \eqno({\rm C}.33)$$\noindent
where $\,\gamma\,$ is the Euler constant, $\,\lambda\,$ is given by (5.25),

$${\cal E}^{(i)}\;=\;3n\,e_1^\ell\,e_2^\ell \int {d^3k\over
(2\pi)^3}\;{e^{i\vec
k \cdot \vec R}\over k^2}\;I(\vec k) \eqno({\rm C}.34)$$
\vskip4pt\noindent
with $\,I(\vec k)\,$ given by (5.52), and $\,{\cal E}^{(ii)}\,$  given by  a
similar expression (5.76).  When $\,\ell \to 0\,$, $\,\delta\,$ satisfies

$$\lim_{\ell \to 0} \delta(\ell, \vec R_1, \vec R_2)\;=\;0\,, \eqno({\rm
C}.35)$$
\vskip4pt\noindent
since that is the condition for the determination of $\,\lambda\,$.  In
general,
the dependence of $\,\delta\,$ on $\,\ell,\,\vec R_1\,$ and $\,\vec R_2\,$ is
not a simple one, though well defined through (C.33).
\vskip4pt
Because of (5.50) and (5.78), both  $\,{\cal E}^{(i)}\,$ and $\,{\cal
E}^{(ii)}\,$ are periodic in

$$\vec R_{\rm cm}\;=\;{1\over 2}\,(\vec R_1 + \vec R_2)$$
\vskip4pt\noindent
with the periodicity of the lattice.  It is convenient to average $\,\vec
R_{\rm
cm}\,$ over a unit lattice cell; this gives a $\,<\delta>_{\rm av}\,$ which
depends only on $\,\ell\,$ and $\,\vec R = \vec R_2 - \vec R_1\,$.  For
 $\,<\delta>_{\rm av}\,$, we may set

$$\vec m_1\;=\;\vec m_2 \hskip3em {\rm and} \hskip3em \vec m\;=\;\vec m'
\eqno({\rm C}.36)$$
\noindent
in (5.52)-(5.53) and (5.76)-(5.77).  Thus, the integrand in (C.34) can be
written as

$$I(\vec k)\;=\;\sum_{\vec m} \int_B\;{d^3K\over (2\pi)^3}\;f_{\vec m \vec
m}(\vec K, \vec k) \eqno({\rm C}.37)$$
\noindent where
$$f_{\vec m \vec m}(\vec K, \vec k)\;=\;k^{-2}\,\Omega^3 \sum_t\;{\left[\sum_a
{\cal C}_a(\vec\theta\,|\,\vec m)\,\hat\epsilon_t(\vec K)_a k_a\right]^2\over
2\omega_t(\vec K) \left(\vec k + \vec K - {2\pi\,\vec m\over \ell}\right)^2}\,.
\eqno({\rm C}.38)$$\noindent
\vskip4pt\noindent
Likewise, we have
$${\cal E}^{(ii)}\;=\;- n\,e_1^\ell\,e_2^\ell \int {d^3k\over
(2\pi)^3}\;{e^{i\vec k \cdot \vec R}\over k^2}\;II(\vec k) \eqno({\rm
C}.39)$$\noindent
where
$$II(\vec k)\;=\;\int_B {d^3 \vec K\over (2\pi)^3}\;h(\vec K, \vec k)
\eqno({\rm
C}.40)$$
\vskip4pt\noindent
and $\,h(\vec K, \vec k)\,$ is given by (5.83).  Substituting the above
$\,{\cal E}^{(i)}\,$ and $\,{\cal E}^{(ii)}\,$ into (C.33), we derive the
expression for $\,<\delta>_{\rm av}\,$.
 \vfill\eject
\centerline {\bf REFERENCES}\vskip16pt\noindent
1.	R. Friedberg, T. D. Lee and Y. Pang, J. Math. Phys. {\bf 35}(11), 5600
(1994).
\vskip4pt\noindent
2.	K. Wilson, Phys.Rev. D{\bf 10}, 2455 (1974); J. B. Kogut and L. Susskind,
\vskip1pt
\ \ Phys.Rev. D{\bf 11}, 395 (1975).  M. Creutz, Phys.Rev. D{\bf 21}, 2308
(1980).
\vskip4pt\noindent
3.	Lattice gauge theories have been extensively studied in the literature.  See
\vskip1pt
\ {\it Lattice\  93}, Nucl.Phys. (Proceedings Supplement) B{\bf 34} (1994),
eds.
\vskip1pt
\ \ T. Draper, S. Gottlieb, A. Soni and D. Toussaint, for recent reviews of the
 \vskip1pt
\ \ subject, and the references cited therein.   Noncompact lattice gauge
theory
\vskip1pt
\ \ has been discussed by K. Cahill, {\it ibid.}, 231; A. Patrascioiu, E.
Seiler and
\vskip1pt
\ \ I. Stamatescu, Phys.Lett. B{\bf 107}, 364 (1981); I. Stamatescu, U. Wolff
and
\vskip1pt
\ \ D. Zwanziger, Nucl.Phys. B{\bf 225} [FS9], 377 (1983); E. Seiler, I.
Stamatescu
\vskip1pt
\ \ and D. Zwanziger, Nucl.Phys. B{\bf 239}, 177 and 201 (1984); but their
\vskip1pt
\ \ formulations are quite different from the one given in this paper.
\vskip4pt\noindent
4. T. D. Lee, Phys.Lett. B{\bf 122}, 217 (1983); R. Friedberg and T. D. Lee,
\vskip1pt
\ \ Nucl.Phys. B{\bf 225}, 1 (1983).
\vskip4pt\noindent
5.	See N. H. Christ and T. D. Lee, Phys.Rev. D{\bf 22}, 939 (1980), and the
\vskip1pt
\ \ references cited therein.
\vskip4pt\noindent
6. T. D. Lee and C. N. Yang, Phys.Rev. {\bf 128}, 885 (1962).
\vskip4pt\noindent
7. L. D. Faddeev and V. N. Popov, Phys.Lett. {\bf 25}B, 29 (1967).
\vskip4pt\noindent
8.	D. J. Gross and F. Wilczek, Phys.Rev.Lett. {\bf 30}, 1343 (1973) and H.
Politzer,
\vskip1pt
\ \  {\it ibid.}, 1346.
\vskip4pt\noindent
9.	An uncommon feature here is the combined use of two different kinds of wave
\vskip1pt
\ \ number vectors,  with $\,\vec K\,$ related to the lattice structure, and
 $\,\vec p,\,\vec p\,',\,\vec q\,$
\vskip1pt
\ \  the usual continuum variables, as displayed
explicitly in (5.43)-(5.45).
\vskip1pt
\ \ Otherwise, the calculation is very similar to the
corresponding continuum
\vskip1pt
\ \ Coulomb gauge calculation.  Cf. V.N. Gribov, lecture at
the {\it 12th Winter
\vskip1pt
\ \ School of the Leningrad Nuclear Physics Institute} (1977),
\vskip1pt
\ \ S.D. Drell, in {\it A Festschrift for Maurice Goldhaber}, edited by G.
Feinberg,
\vskip1pt
\ \ A.W. Sunyar and J. Weneser (New York, New York Academy of Sciences,
\vskip1pt
\ \ 1980) and T. D. Lee, {\it Particle Physics and Introduction to Field
Theory},
 \vskip1pt
\ \ (Harwood Academic Publishers, 1981).
\vskip4pt\noindent
10.	G. N. Watson, Q.J.Math. {\bf 10}, 266 (1939).

\vfill\eject

\centerline {\bf FIGURE CAPTIONS} \vskip34pt\noindent
{\bf Figure 1.}\ \ \ \ Equations (3.3)-(3.5) give an electromagnetic field
configuration in the time-axial gauge within the unit cell in which $\,x, y,
z\,$ all lie between 0 and $\,\ell\,$.
\vskip20pt\noindent
{\bf Figure 2.}\ \ \ \ A picture of the localized QCD gauge field configuration
(4.17) in the Coulomb gauge.  All plaquettes are located on the $\,(x, y)\,$
plane at $\,z = 0\,$.  Move along the arrow direction of the central plaquette:
Start from its lower left lattice site and label it 0; then label consecutively
the other three lattice sites 1, 2 and 3.  The gluon field described by (4.17)
corresponds to a link-flux distribution (in the notation of (4.14))  $\,a_0^1
= b_1^1 = - a_3^1 = - b_0^1 = 1\,$ and all other $\,a_j^\ell\,,b_j^\ell\,$ and
$\,c_j^\ell\,$ zero.  Thus, Kirchhoff's law (4.15) is satisfied.  The
number inside each plaquette denotes the integral of
$\,\ell^{-1}\,\int\,B_j^1\,dx\,dy\,$ over the plaquette.  On account of the
factor $\,\Delta(z)\,$, the gauge field distribution (4.17) is nonzero over
two layers of 9 cubic cells each (making a total of 18 cells), extending from
the
9 plaquettes at $\,z = 0\,$ in the figure to $\,z = \pm \ell\,$.
\vskip20pt\noindent
{\bf Figure 3.}\ \ \ \ The lattice sites $\,j,\, j',\, j''\,$ and $\,j'''\,$
are defined by (3.18).  (a) shows the plaquettes $\,X_j\,, Y_j\,$ and $\,Z_j\,$
that are perpendicular to the $\,x, y\,$ and $\,z\,$ axes.
\vskip2pt\noindent
(b) gives the
plaquette-fluxes $\,\xi_j^\ell\,, \eta_j^\ell\,$ and $\,\zeta_j^\ell\,$
associated with these plaquettes.

\vfill\eject

\end